\documentclass[a4paper,10pt]{article}

\usepackage{jheppub} 

\usepackage{graphicx,floatflt,amssymb,epsfig}
\usepackage[normalem]{ulem}
\usepackage[utf8]{inputenc}
\usepackage[inline]{enumitem}
\usepackage{mathrsfs}
\usepackage{amssymb}
\usepackage{amsmath}
\usepackage{xcolor}
\usepackage{hyperref}
\usepackage{graphicx}
\usepackage{subfig}
\usepackage{multirow}
\usepackage{mathtools}
\usepackage{rotating}
\usepackage{lscape}
\usepackage{float}
\usepackage{pstricks}
\usepackage[toc,page]{appendix}
\usepackage{pifont}
\usepackage{braket}
\usepackage{bbold}
\usepackage{ragged2e}
\usepackage{slashed}
\usepackage{rotating}
\usepackage{tablefootnote}

\usepackage{tikz}
\usepackage{tikz-feynman}
\usetikzlibrary{arrows}
\usetikzlibrary{decorations.pathmorphing}

\renewcommand{\arraystretch}{1.5}

		\newcommand{\JP}{{J/\psi}}

		\newcommand{\bqa}{\begin{eqnarray}}
		\newcommand{\eqa}{\end{eqnarray}}
		\newcommand{\stateinequ}[3]{{}^{#1}{#2}_{#3}}
		
		\def\bfsigma{\mbox{\boldmath $\sigma$}}

		\def\bfsigma{\mbox{\boldmath $\sigma$}}

		\newcommand{\bmt}{\begin{pmatrix}}
			\newcommand{\emt}{\end{pmatrix}}
		\newcommand{\ba}{\begin{array}{c}}
			\newcommand{\ea}{\end{array}}
		\newcommand{\beq}{\begin{equation}}
		\newcommand{\eeq}{\end{equation}}
		\newcommand{\bea}{\begin{eqnarray}}
		\newcommand{\eea}{\end{eqnarray}}
		\newcommand{\nn}{\nonumber}
		\newcommand{\bi}{\begin{itemize}}
			\newcommand{\ei}{\end{itemize}}
		
		\newcommand{\baz}{\begin{array}{cc}}
			
			\newcommand{\mathsym}[1]{{}}

			\newcommand{\bt}{\begin{tabular}}
				\newcommand{\et}{\end{tabular}}

			\newcommand{\benu}{\begin{enumerate}}
				\newcommand{\eenu}{\end{enumerate}}
			
			\newcommand{\bav}{\begin{array}{cccc}}

				\allowdisplaybreaks

	\title{\boldmath Precision Study of Semileptonic and Non-Leptonic $B_c$ Decays to $\eta_c$ and P Wave Charmonia}
	
	\author[a] {Aritra Biswas,}
	\emailAdd{aritra1.biswas@gmail.com}
	
	\author[b] {Soumitra Nandi}
	\emailAdd{soumitra.nandi@iitg.ac.in}
	
	\author[b,c] {and Shantanu Sahoo}
	\emailAdd{shantanu\_sahoo@iitg.ac.in}
	
	\affiliation[a]{Theoretische Physik 1, Center for Particle Physics Siegen (CPPS), Universit\"at Siegen, Walter-Flex-Str. 3, 57068 Siegen, Germany}	
	\affiliation[b]{Indian Institute of Technology Guwahati, Department of Physics, Assam 781039, India}
    \affiliation[c]{University of Hyderabad, School of Physics, Hyderabad 500046, India}

\abstract{
We analyse semileptonic and non-leptonic decays of the $B_c$ meson into P wave charmonium states. The analytic expressions for the transition form factors are taken from NRQCD, while their normalisations and shapes are constrained directly using experimental data, without introducing any additional model-dependent inputs. Using radiative decay data of $\chi_{c0}$, $\chi_{c1}$, and $h_c$, we extract the derivatives of their radial wave functions at the origin and update the $B_c \to (\chi_{c0}, \chi_{c1}, h_c)$ form factors. We present predictions for the semileptonic branching fractions, the lepton flavour universality ratios $R(\chi_{c0})=0.185(3)$, $R(\chi_{c1})=0.147(26)$, and $R(h_c)=0.068(2)$, as well as selected non-leptonic $B_c$ decays and P-wave charmonium production in $e^+e^-$ annihilation and $Z$-boson decays.}

\arxivnumber{}

\begin{document}

\begin{flushright}
SI-HEP-2025-02
\end{flushright}

	\maketitle
\section{Introduction}

The $B_c$ meson has received considerable attention in recent years from both theoretical and experimental perspectives. Owing to its valence quark content of two heavy flavours, bottom and charm, its production is suppressed by a factor of $\alpha_s^2$ and is therefore most effectively studied at hadron colliders. First observed by the CDF collaboration in 1998 through the decay channel $B_c \to J/\psi \pi^+$~\cite{CDF:1998ihx}, the $B_c$ meson is now abundantly produced at the LHC, where its production cross section is expected to be of the order of tens of nanobarns~\cite{Chang:2003cr}, enabling detailed investigations of its properties and decay modes.

Decays of the $B_c$ meson into charmonium final states have attracted considerable interest in recent years, particularly in connection with possible lepton flavour universality violating hints of physics beyond the Standard Model. Semileptonic decays to the S-wave charmonium state $J/\psi$ play a central role through the LFUV observable $R(J/\psi)$, analogous to the well-known $R(D^{(*)})$ ratios. This quantity has been measured by LHCb~\cite{LHCb:2017vlu} and more recently by CMS~\cite{CMS:2024seh}, while the most precise and model-independent Standard Model prediction, $R(J/\psi)=0.2582(38)$, is provided by lattice QCD calculations of the HPQCD collaboration~\cite{Harrison:2020nrv}. Earlier Standard Model estimates in the literature were largely model dependent~\cite{Anisimov:1998uk,Kiselev:1999sc,Kiselev:2002vz,Ivanov:2006ni,Hernandez:2006gt,Qiao:2012vt,Wang:2012lrc,Rui:2016opu,Dutta:2017xmj,Watanabe:2017mip,Issadykov:2018myx,Leljak:2019eyw,Cohen:2018dgz}. Owing to large uncertainties, the CMS result is consistent with zero, and a naive combination of the LHCb uncertainties implies a tension with the Standard Model at the level of $\sim1.8\sigma$.

Although the $J/\psi$ channel is experimentally the cleanest, a reliable understanding of $B_c$ decays into P-wave charmonium states is essential, as they can constitute important backgrounds to S-wave analyses. In particular, the sizable branching ratio $\mathcal{B}(\chi_{c2}\to J/\psi\gamma)=(19.0\pm0.5)\%$~\cite{Zyla:2020zbs} implies that cascade decays such as $B_c^+\to \chi_{c2}(\to J/\psi\gamma)\pi^+$ can contaminate $B_c\to J/\psi\pi$ measurements. Experimentally, the first evidence for a $B_c$ decay into a P-wave charmonium was reported by LHCb through the observation of $B_c\to \chi_{c0}(\to K^+K^-)\pi^+$ with a significance of $4\sigma$~\cite{LHCb:2016utz}.

Semileptonic decays of the $B_c$ meson into P-wave charmonium states provide a theoretically clean and phenomenologically rich laboratory for testing heavy-quark dynamics, spin symmetry, and lepton flavour universality in the heavy-heavy sector. Unlike decays into S-wave charmonia, these modes probe a larger set of independent form factors and are sensitive to the internal orbital structure of the final-state charmonium, thereby offering complementary information on non-perturbative QCD effects. A reliable understanding of $B_c \to P$-wave semileptonic transitions is also essential in view of ongoing and future precision studies of $R(J/\psi)$, since cascade decays involving P-wave charmonia can constitute non-negligible backgrounds. With increasing $B_c$ production at the LHC and the lack of lattice or direct experimental determinations of the relevant form factors, precise and data-constrained predictions for these semileptonic decays are timely and well motivated.

A precise determination of the decay widths for $B_c$ transitions into P-wave charmonium states requires reliable knowledge of the corresponding form factors over the full kinematic $q^2$ range. At present, neither lattice QCD nor experimental determinations of these form factors are available. Existing results are therefore based on model-dependent approaches, including perturbative QCD~\cite{Wang:2012lrc,PhysRevD.90.114030,Sun:2008ew,Rui:2016opu,Liu:2023kxr,Rui:2018kqr}, non-relativistic and relativistic quark models~\cite{Hernandez:2006gt,Nobes:2000pm,Ebert:2003cn,Ivanov:2005fd,Ebert:2010zu,Nayak:2022gdo}, QCD sum rules~\cite{Colangelo:1992cx,Kiselev:2002vz,Leljak:2019eyw,Azizi:2009ny,Azizi:2013zta,Kiselev:1999sc}, light-front quark models~\cite{Wang:2008xt,Ke:2013yka}, and the NRQCD framework~\cite{Brambilla:2004jw,PhysRevD.51.1125,Bodwin:2006dm,Brambilla:2010cs,Kiselev:2001zb,Bell:2005gw,Bell:2006tz,Qiao:2011yz,Qiao:2012vt,Qiao:2012hp,Zhu:2017lqu,Zhu:2017lwi,Tao:2022yur}.

In the current article, we employ the NRQCD formalism with the corresponding relativistic corrections to the form factors for the decay of the $B_c$ meson to P-wave charmonia as computed in ref.~\cite{Zhu:2017lwi}. The article is organized as follows. Section~\ref{sec:Bctoetac_imp} focuses on the exclusive semileptonic $B_c \to \eta_c \ell\nu$ decay. Subsection~\ref{subsec:Bctoetacff_shape} presents the improved form factor shapes, while subsection~\ref{subsec:Bctoetac_obs} discuss the associated physical observables with this decay channel. Section~\ref{sec:BctoP_semilep} addresses the exclusive semileptonic $B_c \to P$ wave charmonium transitions. In subsection~\ref{subsec:NRQCD_theoP}, we discussed the NRQCD factorization framework and the associated transition form factors. Subsection~\ref{subsec:exp_th_Pwave} introduces the relevant experimental modes and the theoretical expressions used to fit the derivatives of the P-wave charmonia wave functions at the origin. The methodology and our fit results are presented in subsection~\ref{subsec:fit_res_Pwave}. Subsection~\ref{subsec:BctoPff_shape} is dedicated to the discussion of form factor shapes instrumental in the theoretical description of semileptonic $B_c\to h_c, \chi_{c0,1}$ decays, followed by subsection \ref{subsec:BctoP_obs} which provides predictions for the corresponding observables. Section~\ref{sec:BctoP_nonlep} extends the analysis to two-body non-leptonic decays involving a P-wave charmonium and a light pseudoscalar/vector meson. We also provide numerical estimates for a few production modes of the P-wave charmonia themselves in section~\ref{sec:Pwaveprod_Zdecay}. Finally, Section~\ref{sec:conclusions} summarizes our findings and conclusions. 
\section{Semileptonic $B_c \to \eta_c$ Decays}
\label{sec:Bctoetac_imp}
The effective Hamiltonian that describe the $b \to c \ell \nu$ transition within the SM, can be written as 
\begin{equation}
    \mathcal{H}_\text{eff.}= \frac{G_F V_{cb}}{\sqrt{2}} \bar{b} \gamma_\mu (1-\gamma_5)c \otimes \bar{\nu}_\ell \gamma^\mu (1-\gamma_5)\ell, 
    \label{eq:efflagrbtoc}
\end{equation}
where $G_F$ is the Fermi constant, and $V_{cb}$ is the CKM element. In the absence of QED corrections, $\mathcal{H}_\text{eff.}$ can be factorized into the hadronic and leptonic contributions. The hadronic contribution is proportional to the corresponding hadronic matrix element and is parametrized with form factors. 

The hadronic matrix element describing $B_c \to \eta_c \ell^- \bar{\nu}_\ell$ (pseudo-scalar to pseudo-scalar) decay is described by two form factors:
\begin{align}\label{eq:Bctoetac}
\langle \eta_c(p')|\bar{c}\gamma^{\mu} b|B_c(P)\rangle =&-i\bigg[f_0(q^2)\frac{m_{B_c}^2-m_{\eta_{c}}^2}{q^2}q^{\mu} +f_+(q^2)(P^{\mu}+p'^{\mu}-\frac{m_{B_c}^2-m_{\eta_{c}}^2}{q^2}q^{\mu})\bigg],
\end{align}
from which it follows that $f_0(0)=f_+(0)$, where $q^2=(P-p')^2$ is the dilepton invariant mass. Following the method given in~\cite{Korner:1989qb}, the total decay amplitude($\mathcal{M}$) for semileptonic $B_c \to \eta_c \ell \nu$ decay can be written as:
\begin{equation}
    \mathcal{M} = \frac{G_F V_{cb}}{\sqrt{2}}\left[\eta_{\lambda} H^p_{V,\lambda} L^{\lambda}_{\lambda_\ell}\right].
\end{equation}
where the $W$ boson polarization $\lambda \in (t, +, -, 0)$ and $\lambda_\ell=\pm 1/2$ is the helicity of lepton.  
The differential decay rate is then given by:
\begin{equation}
    \frac{d\Gamma (B_c \to \eta_c \ell^- \bar{\nu}_\ell)}{d q^2}=\frac{G_F^2|V_{cb}|^2 \sqrt{\lambda(m_{B_c}^2, m_{\eta_c}^2, q^2)}}{384 \pi^3 m_{B_c}^3}\left(1-\frac{m_\ell^2}{q^2}\right)^2\biggl[\bigl(1+\frac{m^2_\ell}{2 q^2}\bigr)|H^p_{V,0}(q^2)|^2 + \frac{3 m_\ell^2}{2 q^2}|H^p_{V,t}|^2\biggr],
\end{equation}
where, $\lambda(m_{B_c}^2, m_{\eta_c}^2, q^2)=((m_{B_c}+m_{\eta_c})^2-q^2)((m_{B_c}-m_{\eta_c})^2-q^2)$. In the expression above, the decay rate involves hadronic helicity amplitudes that are related to the form factors and are given by:
\begin{align}
    H^{p}_{V,0}(q^2)&=\sqrt{\frac{\lambda(m_{B_c}^2, m_{\eta_c}^2, q^2)}{q^2}} f_+(q^2),\\
    H^{p}_{V,t}(q^2)&=\frac{(m_{B_c}^2-m_{\eta_c}^2)}{\sqrt{q^2}} f_0(q^2).
\end{align}
To obtain an estimate for the shape of the differential decay width distribution over the entire allowed kinematical range, knowledge of the behavior of the non-perturbative aspects of the decay, i.e., the form factors over the corresponding range, is imperative. As of now, we do not have any reliable lattice estimates on the $B_c \to \eta_c$ form factors $f_{+,0}(q^2)$. In Ref.~\cite{Biswas:2023bqz}, we used $B_c \to J/\psi$ form factors and theoretical tools like heavy quark spin symmetry (HQSS) to estimate the form factors $f_{+,0}(q^2)$. In the subsequent section, we revisit the $B_c \to \eta_c$ decay and related form factors, focusing on their constrained shapes.
\subsection{Improved Form Factor Shape}
\label{subsec:Bctoetacff_shape}
To obtain this constrained shape, we use NRQCD estimates for $B_c \to \eta_c$ form-factors at $q^2=0$. Here we briefly discuss the methodology for our estimates on these transition form factors. Interested readers can look in ref.~\cite{Biswas:2023bqz} for details. In the NRQCD framework, transition matrix elements between two hadronic states are factorized into non-perturbative wave function parameters and a hard scattering kernel. For $B_c \to J/\psi(\eta_c)$, the transition matrix elements can then be written as:
\begin{equation}
    \langle J/\psi(\eta_c)|\bar{c}\gamma^{\mu}(1-\gamma_5) b|B_c\rangle\simeq \sum_{n=0} \psi(0)^R_{B_c} \,\psi(0)^R_{J/\psi(\eta_c)} T^n \,\,  ,
\end{equation}
where the hard kernels $T^n$ can be calculated perturbatively and the non-perturbative parameters $\psi(0)^R_{B_c}$, $\psi(0)^R_{J/\psi(\eta_c)}$ are the radial wave functions at the origin for the $b\bar{c}$ and $c\bar{c}$ systems. For a detailed discussion of the NRQCD formalism, see Ref.~\cite{PhysRevD.51.1125}. To estimate these form factors, we needed the radial wave functions, $\psi^R_{M}(0)$, where $ M$ stands for the respective meson. We use experimental input on the radiative decay of $J/\psi \to e^+ e^-$, $\eta_c \to \gamma \gamma$ and decay constant $f_{J/\psi}, f_{\eta_c}, f_{B_c}$ from the lattice QCD ref.~\cite{Donald:2012ga, Davies:2010ip,Colquhoun:2015oha} combined with form factor input from the LQCD~\cite{Harrison:2020gvo, Harrison:2020nrv} of $B_c \to J/\psi$ transition at $q^2=0$. With all inputs that are dealt with, the relevant parameters, the radial wave function; $\psi^R_{B_c}(0)$, $\psi^R_{J/\psi}(0)$, $\psi^R_{\eta_c}(0)$ at $r=0$ are extracted by a global $\chi^2$ analysis by maximizing the likelihood. With these extracted parameters we estimated the $B_c \to J/\psi(\eta_c)$ transition form factors. Our results has presented in the Ref.~\cite{Biswas:2023bqz} in table~7. In our present analysis, we used our estimated form factor for $B_c \to \eta_c$ transition at $q^2=0$ to constrain the previously estimated relevant form factor for this transition. For the completeness of the draft, we are presenting our estimate only for the $B_c \to \eta_c$ transition form factor below in eq.~\ref{eq:ffNRQCDinput}.
\begin{equation}
f_{+}^{B_c\to \eta_c}(q^2=0)\Bigr|_\text{NRQCD} = f_{0}^{B_c\to \eta_c}(q^2=0)\Bigr|_\text{NRQCD}= 0.64\pm 0.15 \, \,.
\label{eq:ffNRQCDinput}
\end{equation}
To constrain the shape of the form factor we also use generated pseudo data points for the form factors $f_+(q^2)$ and $f_0(q^2)$ at $q^2=6,~8,~q^2_\text{max}$ (in $GeV^2$) for the $B_c \to \eta_c$ transition taken from table~2 in ref.~\cite{Biswas:2023bqz}. These synthetic data points for the form factors are also presented here in table~\ref{tab:syndatfpf0} to illustrate the completeness of the draft.
\begin{table*}[t]
	\begin{center}
		\begin{tabular}{|*{6}{c|}}
			\hline
			\multicolumn{6}{|c|}{\bf Pseudo data points for $f^{B_c\to \eta_c}_{+,0}(q^2)$} \\
			\hline \hline
			$f_{+}(6)$ &  	$f_{+}(8)$ & $f_{+}(\text{$q^2_{max}$})$ & $f_{0}(6)$ & $f_{0}(8)$ & $f_{0}(\text{$q^2_{max}$})$ \\
            \hline
			$\text{0.72(28)}$& $\text{0.83(30)}$& $\text{1.01(30)}$& $\text{0.63(26)}$ & $\text{0.69(25)}$ &  $\text{0.79(24)}$ \\
			\hline
		\end{tabular}
		\caption{Synthetic data for the $B_c\rightarrow \eta_c$ form factors generated using eq.2.2 in Ref.~\cite{Biswas:2023bqz}. The fit results of the parameters $\Delta(1)'$ and $\Delta(1)''$ given in the third column of tab.~1 from Ref~\cite{Biswas:2023bqz}. Also, to generate the numbers, we have used inputs $\Delta_{+}(w=1)= \Delta_0(w=1)= 0.84(25)$ in eq.~2.6 and the respective correlation provided in table~15 in Ref~\cite{Biswas:2023bqz}.}
		\label{tab:syndatfpf0}
	\end{center}
\end{table*}
\paragraph{\underline{$z$-Expansion of the Form Factor}:}
In the case of semileptonic $B_c$ decays to $\eta_c$, $q^2$ ranging from $m_\ell^2$ to $(m_{B_c}-m_{\eta_c})^2$ but the form factor can be continued analytically in the $q^2$ complex plane. They have a pole at $q^2=(m_B+m_D)^2$ and various poles corresponding to $B_c$ resonances with the appropriate quantum numbers. We have adopted the model-independent approach proposed by Bourrely-Lellouch-Caprini (BCL) to parametrize the $q^2$ shapes of $f_{+,0}(q^2)$ \cite{Caprini}. In this parametrization, both the form factors are written as a polynomial series in powers of $q^2$.
Adopting these notation, we define 
\begin{eqnarray}
    t=q^2=(p-p^\prime)^2 \, , \, \,  t_+=(m_{B_c} + m_{D})^2 \, ,\, \, t_-=(m_{B_c}-m_{\eta_c})^2.
\end{eqnarray}
Then the form factor $f_i(q^2)$ is parametrized as follows:
\begin{equation}
	f_i(q^2) = \frac{1}{P(q^2)} \sum_{n=0}^{N} a^i_n z^n.
	\label{eqn:BCLeta}
\end{equation}
Here $a_i^n$ are the real parameters and $z \equiv z(q^2,t_0)$ maps the physical $q^2$ region on a unit circle,  $|z(q^2,t_0)|= 1$. $t_+$ denotes the beginning of the multi-particle cut.  Thus, this map sends the branch cut onto the unit circle and the rest of the first Riemann sheet onto the open unit circle $|z(q^2,t_0)|< 1$. Note that $z(t_+, t_0)= -1$ and $z(\infty, t_0)= +1$. The free parameter $t_0 < t_+$, the center of the circle in the $z$ plane, can be chosen for convenience $t_0 \in [0,t_+]$ that determines the point in the $q^2$-plane mapped to the origin in the $z$ plane, i.e. $z(t_0, t_0)=0$. The conformal transformation that maps $q^2$ to $z$ plane is defined as follows: 
{\small
\begin{equation}
     z(q^2,t_0)=\frac{\sqrt{t_+-q^2}-\sqrt{t_+-t_0}}{\sqrt{t_+-q^2}+\sqrt{t_+-t_0}}.
\end{equation}
}
We take $t_0$ to be $t_\text{opt.}$ and $t_\text{opt.}= t_+\left( 1-\sqrt{1-\frac{t_-}{t_+}}\right)$, so $q^2\in [0,q^2_\text{max}]$ is mapped to $z\in [+z_\text{max}, -z_\text{max}]$. With this prescription, the maximum physical value $|z_\text{max}|\approx 0.017 $ for $B_c \to \eta_c \ell \nu$ decay ensures faster convergence in the $z$-series expansion. $P(q^2)$ contains the the pole contribution and is defined as $P(q^2)=1-q^2/m_R^2$. $m_R$ represents the masses of the low-lying $B_c$ resonances. We use $m_R=6.331~\text{GeV}$~\cite{Mathur:2018epb} and $6.712~\text{GeV}$~\cite{Mathur:2018epb} for $f_+$ and $f_0$ respectively. In systematic expansions for the form factor, one can incorporate the constraints of unitarity as in the case of the Boyd-Grinstein-Lebed~(BGL)~\cite{Boyd:1994tt} parametrization. In the BGL parametrization form factor $f_i(q^2)$ is parametrized as follows:
\begin{equation}
	f_i(q^2) = \frac{1}{P_i(z) \, \phi_i(z)} \sum_{k=0}^{N} b_k^i z^k.
	\label{eqn:BGLetac}
\end{equation}
where $P_i(z)$ are the Blaschke factors that contain the sub-threshold pole information and $\phi_i(z)$ are the outer functions that can be found in~\cite{Boyd:1994tt, Boyd:1997kz, Ray:2023xjn}. Also, the parameter $z$ is related to the recoil variable $w$ as
\begin{equation}
    z=\frac{\sqrt{w+1}-\sqrt{2}}{\sqrt{w+1}+\sqrt{2}}
\end{equation}
$w$ is related to the momentum transfer $(q^2)$ to the dilepton system as $q^2=m_{B_c}^2+m_{\eta_c}^2-2 m_{B_c} m_{\eta_c} w$ and BGL coefficients $(b_k^i)$ follow the unitarity constraints:
\begin{equation}
    \sum_{k=0}^K b_k^2 \leq 1.
    \label{eq:BGLconst}
\end{equation}
\paragraph{\underline{Unitarity Constraint}:}
Unitarity bound discussed earlier for BGL parametrization can also be imposed onto the BCL coefficients mentioned in eq.~\ref{eqn:BCLeta} as discussed in ref.~\cite{Caprini}. By comparing the representation mentioned in eq.~\ref{eqn:BCLeta} and eq.~\ref{eq:BGLconst} we have,
\begin{equation}
    \sum_{n=0}^{N} a_n z^n= \Psi(z) \sum_{k=0}^K b_k z^k,
\end{equation}
 where $\Psi(z)$ can be calculated using the Blaschke function and the outer function for respective form factors, and this calculated function is as follows:
\begin{equation}
    \Psi(z)=\frac{m_{B_c^*}^2}{4(t_+-t_0)} \Phi(z) \frac{(1-z)^2((1-z^*)^2)}{(1-zz^*)^2}.
    \label{eq:BCLBGLequlty}
\end{equation}
We denote $z^*=z(m^2_{B_c^*},t_0)$ by the position of the pole in the variable $z$ and $\Phi(z)\equiv \phi(z)$ is the outer functions~\cite{Ray:2023xjn} relevant for the $f_+$ and $f_0$ from factors. 
\begin{align}
    \phi_{f_+}(z)=&\frac{8 r^2}{m_{B_c}} \sqrt{\frac{n_I}{3 \pi \tilde{\chi}^T_1(0)}} \frac{(1+z^2(1-z)^{1/2})}{[(1+r)(1-z)+2\sqrt{r}(1+z)]^4}\\
    \phi_{f_0}(z)=& r(1-r^2) \sqrt{\frac{ 8 n_I}{\pi \tilde{\chi}^L_{1^-}(0)}} \frac{(1-z^2)(1-z)^{1/2}}{[(1+r)(1-z)+2\sqrt{r}(1+z)]^4}
\end{align}
where $r=m_{\eta_c}/m_{B_c}$. 
The analytic function $\Psi(z)$ can be Taylor expend around $z=0$ as,
\begin{equation}
    \Psi(z)=\sum_{k=0}^\infty \eta_k \, z^k
\end{equation} 
Then the insertion in the eq.~\ref{eq:BCLBGLequlty} followed by the expansion around $z=0$ gives us
\begin{equation}
    b_n=\sum_{k=0}^{min[K,n]} \eta_{n-k} \, a_k, \, \, \, \, n \geq 0.
\end{equation}
and this leads to the inequility, expressed in terms of the coefficients $a_i$ and given by:
\begin{equation}
    \sum_{j,k=0}^K B_{jk} \, a_j \, a_k\leq 1  \quad \quad  \text{with} \quad \quad B_{jk}=\sum_{n=0}^\infty \eta_n \,\eta_{n+|j-k|}.
    \label{eq:BCLinequlty}
\end{equation}
As $|z|<<1$, this ensures the rapid convergence of the Taylor series with the coefficient $\eta_j$, a finite order summation upon the $j$ and $k$ can help us to compute $B_{jk}$. We computed $B_{jk}$ for $K=10$. For a more detailed discussion see ref.~\cite{Caprini}.  With $B_{jk}$ computed from eq.~\ref{eq:BCLinequlty}, we proceed to derive the unitarity bounds on the BCL coefficients, which are finally obtained as: 
\begin{itemize}
\label{eq:constraints_BCL}
    \item For $f_+$ and $N=2:$\\
    $0.00216 \, a_0^2 - 0.00054 \, a_0 \, a_1 + 0.00216 \, a_1^2 -0.00286 \, a_0 \, a_2 - 0.00054 \, a_1 \, a_2 +0.00216a_2^2 \leq 1$
    \item For $f_0$ and $N=1:$\\
    $0.02370 \, a_0^2 - 0.02422 \, a_0 \, a_1 + 0.02370 \, a_1^2 \leq 1$
\end{itemize}
Our goal is to constrain these BCL parameters in eq.~\ref{eqn:BCLeta} with the obtained unitary bound. To address the weak unitary bounds on the BCL coefficients, we conduct a revised BCL fit. Using the generated synthetic data points mentioned in the table~\ref{tab:syndatfpf0} and the NRQCD inputs mentioned in the eq.~\ref{eq:ffNRQCDinput} along with the calculated weak unitary bounds on the BCL coefficients, we perform a revised BCL fit~\footnote{Without unitarity bound, we also did the fit. Both (with and without unitarity bounds) fit results are consistent. Here, we present our results incorporating the unitarity condition that we derived.} and extract the corresponding BCL coefficients in the eq.~\ref{eq:constraints_BCL} using the QCD constraint $f_+(0)=f_0(0)$, which reduces the number of free parameters. Due to the inclusion of the $q^2=0$ input, which carries an uncertainty about $23\%$, significantly less than that of our earlier HQSS-based estimate. We expect an overall reduction in the uncertainties of the fitted BCL parameters. The extracted BCL coefficients are presented in table~\ref{tab:BCLcoeffwNRQCD}. In the present parametrization, the systematic uncertainty in the form factors must account for the truncation of the expansion in eq.~\ref{eqn:BCLeta}, taken at order $n=2$ for $f_+(q^2)$ and for $n=1$ for $f_0(q^2)$. Using the unitary bounds on the BCL coefficients, we can estimate the corresponding truncation errors. In the following, we will discuss this.
\begin{table*}[t]
	\begin{center}
		\begin{tabular}{|*{3}{c|}}
			\hline
			\textbf{BCL Coefficients} & \textbf{Fit Results} \\
			\hline \hline
			$a^{f_0}_1$  &  $\text{-2.4(11)}$ \\
			$a^{f_+}_0$  &  $\text{0.68(9)}$ \\
			$a^{f_+}_1$  &  $\text{-7.01(79)}$ \\
            $a^{f_+}_2$  &  $\text{20(14)}$ \\
			\hline
			$\textbf{dof}$  &  $3$ \\
			$\textbf{p-Value}$  & $0.85$ \\
            \hline 
		\end{tabular}
		\caption{Fit results for the BCL coefficients $a_n^i$ corresponding to the $B_c \to \eta_{c}$ transition form factors obtained from a fit to the synthetic data generated under HQSS displayed in table~\ref{tab:syndatfpf0}. We truncate the series at N=1 for $f_0$ and N=2 for $f_+$, since higher order coefficients are insensitive to the data. Correlation between the parameters presented in table~\ref{tab:corrBCLcoeff}.}
		\label{tab:BCLcoeffwNRQCD}
	\end{center}
\end{table*}
\paragraph{\underline{Systematic error due to truncation of the series}:}
The unitarity constraint eq.~\ref{eq:BCLinequlty} can be exploited to derive a conservative estimate of the truncation error for the associated $B_c \to \eta_c$ transition form factors. We adopt the prescription as mentioned in ref.~\cite{Caprini} for the systematic error on the form factor given by:
\begin{equation}
    \delta f_i(q^2)|_\text{syst.}= \frac{a_{n+1}^\text{max} \, \, |z^{n+1}|}{\bigl(1-\frac{q^2}{m_{B^*_c}^2}\bigr)}
    \label{eq:trunc_error}
\end{equation}
where, $a_{n+1}^\text{max}$ denotes the maximum value for the parameter $a_{n+1}$ allowed by the unitarity condition eq.~\ref{eq:BCLinequlty} for $a_n$ parameter fit, $n < N$. We calculate the maximum possible systematic error that can be received by the form factors at the physical region $|z_\text{max}|=0.017$. We obtain $a_3^\text{max}=4.517$ for $f_+$ and $a_2^\text{max}=7.041$ for $f_0$ from factor respectively. With these estimates for the non-zero value for $a_{n+1}^\text{max}$ in the eq.~\ref{eq:trunc_error} we estimate the numerator for each of the cases at the physical region $|z_\text{max}|$, where the truncation error is maximum. We find that the numerator of eq.~\ref{eq:trunc_error} is $<< 1 \%$ to the LO coefficient $a_0$ for both form factor expansions. As these systematically calculated truncation errors are very small, the results do not affect the form factor shape. Hence, we did not include this truncation error in the results we discuss later in this section. 

\begin{figure}[t]
	\centering
	\subfloat[]{\includegraphics[scale=0.35]{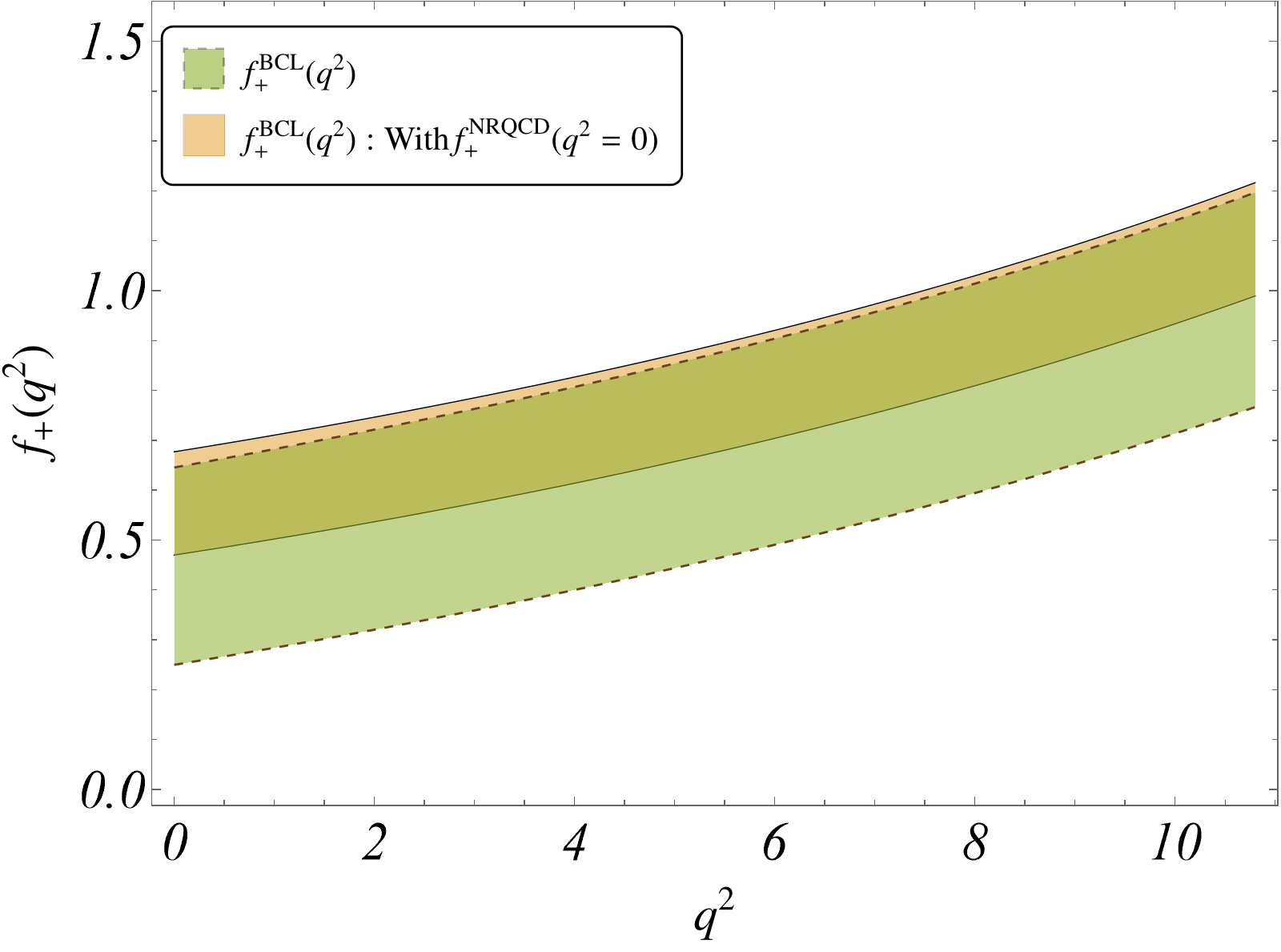}\label{fig:fp_Bcetac}}~~~~~~~
        \subfloat[]{\includegraphics[scale=0.35]{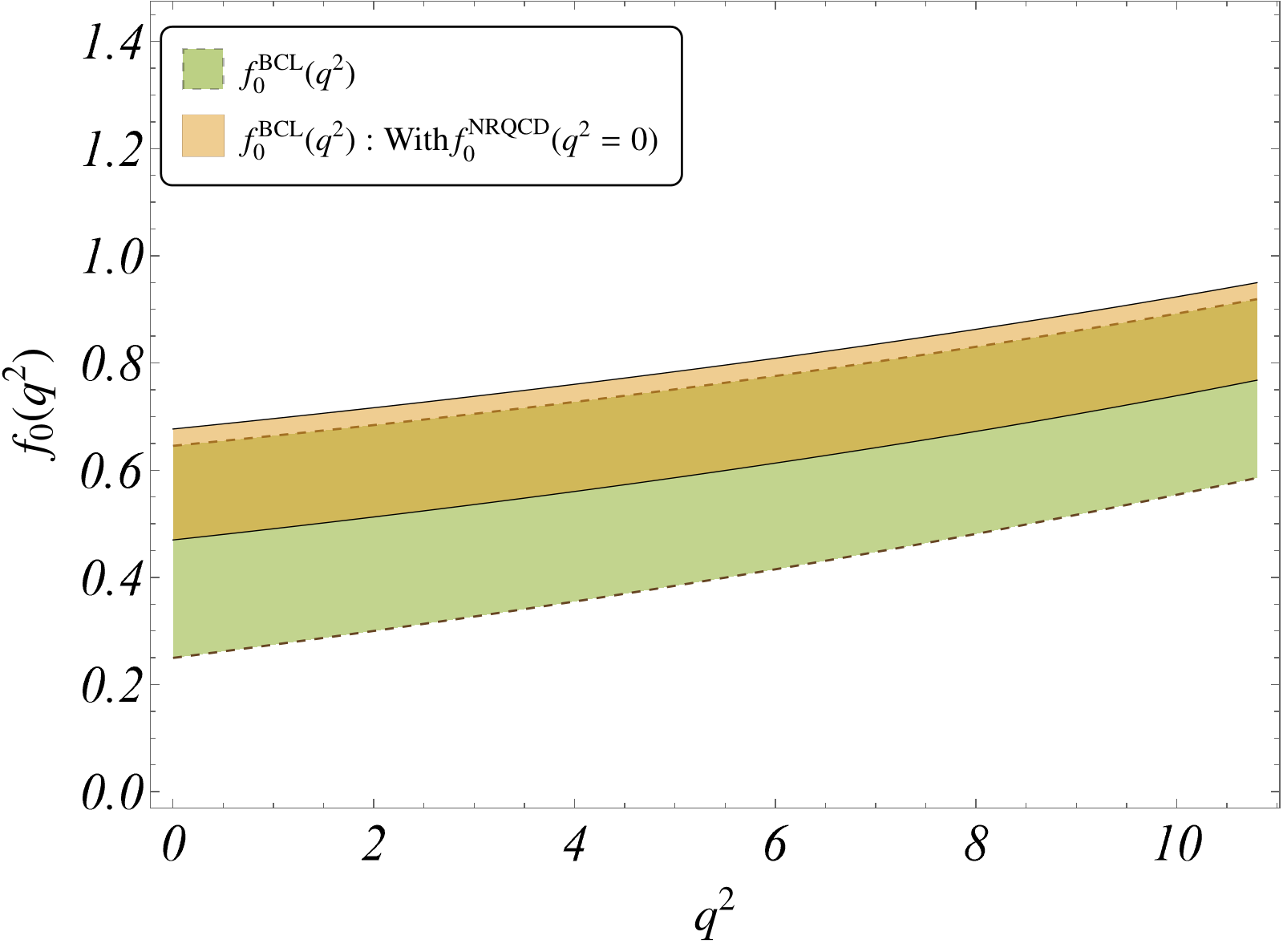}\label{fig:f0_Bcetac}}
	\caption{The brown band representing the form factors corresponding to the BCL parametrization using the fit results from table~\ref{tab:BCLcoeffwNRQCD} and the green band specifying the BCL parametrization using the fit results (without NRQCD input) from table~3 Ref~\cite{Biswas:2023bqz}. These bands represent the $1\sigma$ region for the corresponding form factors.}
	\label{fig:Bc2etacffplot}
\end{figure}

Using the central values of the BCL coefficients, along with their associated uncertainties and correlations, presented in table~\ref{tab:BCLcoeffwNRQCD} and \ref{tab:corrBCLcoeff} respectively, we extracted the shape of the form factors. The respective $q^2$ shapes are shown in figures~\ref{fig:fp_Bcetac} and \ref{fig:f0_Bcetac}. These form factors shape (obtained using the NRQCD input at $q^2=0$) are compared with our earlier estimates that did not include the NRQCD input. Our estimation is illustrated in figure~\ref{fig:Bc2etacffplot}.
Also, the estimates for the form factor using BCL parametrization at $q^2=0$ is as follows:
\begin{align}
f_{+}^{B_c\to \eta_c}(q^2=0) = f_{0}^{B_c\to \eta_c}(q^2=0) &=
\begin{cases}
0.45 \pm 0.20 \, ,\\
0.57 \pm 0.10\ \ \ \text{(With NRQCD input)}.
\end{cases}
\end{align}
Without the inclusion of the NRQCD from factor input at $q^2=0$, our estimate using the BCL parameterization has an uncertainty of about $44\%$. While we incorporated the NRQCD input and did the BCL parametrization for the form factors, we have $17\%$ uncertainty in our form factor estimate at $q^2=0$. We found that form factor estimates at $q^2=0$ from both methods are also consistent.
\subsection{Prediction for Observables}
\label{subsec:Bctoetac_obs}
\begin{figure}[t!]
	\centering
	\includegraphics[scale=0.42]{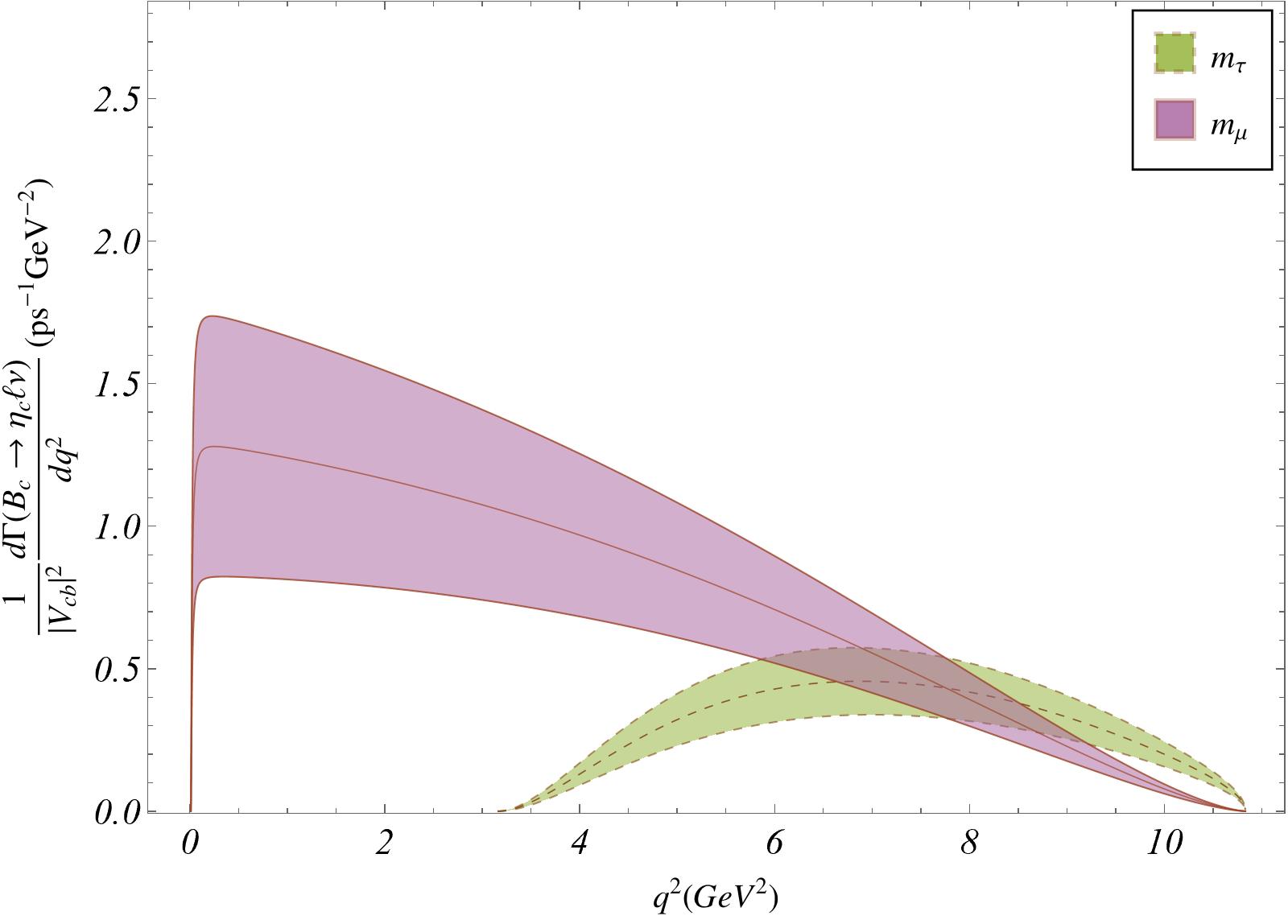}
	\caption{1$\sigma$ bands for the differential decay distributions, estimated using the fit results from table~\ref{tab:BCLcoeffwNRQCD}, are shown for the semileptonic $B_c\to\eta_c$ decays with a tau (green) or a muon (purple) in the final state.}
	\label{fig:Bc2etacdrates}
\end{figure}

Using the shape of the form factors obtained using the discussed methodology thus far, we constrain the shape of the decay distribution and estimate other related observables. These decay distributions for $\tau$ and $\mu(e)$ are shown in fig.~\ref{fig:Bc2etacdrates}. Our current uncertainty estimates remain relatively large. We anticipate significant improvements as more precise inputs on the form factors become available.
With $|V_{cb}| = 0.0403(5)$~\cite{Ray:2023xjn}, we update SM estimate for the semitauonic and semimuonic branching ratio :
\begin{equation}
	\mathcal{B}(B_c \to \eta_c \tau^- \bar{\nu}_\tau) =  (1.90 \pm 0.49)\times 10^{-3}, \ \ \ \   \mathcal{B}(B_c \to \eta_c \mu^- \bar{\nu}_\mu) = (6.51 \pm 1.97)\times 10^{-3}.~~~~
\end{equation}
The predicted values for the semileptonic branching fractions have an uncertainty of approximately $25-30\%$. Using the updated form factor estimates, our updated prediction for the LFU observable $R(\eta_c)$, is defined as
\begin{equation}
    R(\eta_c)=\frac{\mathcal{B}(B_c \to \eta_c \tau^- \bar{\nu}_\tau)}{\mathcal{B}(B_c \to \eta_c \ell^- \bar{\nu}_\ell)} \, ,
\end{equation}
is
\begin{align}\label{eq:Retac_ratio2}
	R(\eta_c)=0.292(14).
\end{align}
The uncertainty in this ratio is estimated to be around $5\%$. This is the most precise SM prediction of $R(\eta_c)$ to date, to the best of our knowledge. The reduction of error in $R(\eta_c)$ is due to a high positive correlation between the form factors in the numerator and the denominator, which we have explicitly checked~\footnote{The standard deviation of a ratio $R = A/B$ of two observables A and B is given by
	\begin{equation}
		\sigma_{R} = |R| \sqrt{\left(\frac{\sigma_{A}}{A}\right)^2 + \left(\frac{\sigma_{B}}{B}\right)^2 - 2 \rho_{AB} \ \frac{\sigma_{A}}{A} \frac{\sigma_{B}}{B} }
	\end{equation}
	where $\rho_{AB}$ is the correlations between the uncertainties of A and B. Here, $\sigma_A$ and $\sigma_B$ represent the standard deviations in the estimate of $A$ and $B$ respectively.}.
\section{Semileptonic $B_c \to \chi_{c(0,1)}, B_c\to h_c$ Decays}\label{sec:BctoP_semilep}
The decay rate distribution of semileptonic decays of $B_c$ meson to $P$-wave charmonia are similar to those of semileptonic decays of $B_c$ to $S$-wave charmonia ref.~\cite{Biswas:2023bqz} as this is also involve four fermion $b \to c \ell \nu$ charge current operator described as eq.~\ref{eq:efflagrbtoc}. P-wave orbitally excited charmonia include the spin singlet $h_c(^1P_1)$ (pseudo-vector), the scalar $\chi_{c0}(^3P_0)$ and the axial-vector $\chi_{c1}(^3P_1)$ states. The transition form factors involve in $B_c\to h_c (\chi_{c1})$ transition are defined as ~\cite{Rui:2018kqr,Li:2009tx}:
\begin{align}\label{eq:Bctohcchic1ff}
\langle h_c(\chi_{c1})(p,\epsilon^*)|\bar{c}\gamma^{\mu}b|B_c(P)\rangle=-i\bigg[&2m_{h_c(\chi_{c1})}A_0^{h_c(\chi_{c1})}\frac{\epsilon^*\cdot q}{q^2}q^{\mu} &\nn \\& +(m_{B_c} + m_{h_c(\chi_{c1})})A_1^{h_c(\chi_{c1})}(q^2)(\epsilon^{*\mu}-\frac{\epsilon^* \cdot q}{q^2}q^{\mu})& \nn \\
& -A_2^{h_c(\chi_{c1})}(q^2)\frac{\epsilon^*\cdot q}{m_{B_c}+m_{h_c(\chi_{c1})}}(P^{\mu}+p^{\mu}-\frac{m_{B_c}^2-m_{h_c(\chi_{c1})}^2}{q^2}q^{\mu})\bigg], \\
\langle h_c(\chi_{c1})(p,\epsilon^*)|\bar{c}\gamma^{\mu}\gamma^{5}b|B_c(P)\rangle =&\frac{2 V^{h_c(\chi_{c1})}(q^2)}{m_{B_c}+m_{h_c(\chi_{c1})}}\epsilon^{\mu \nu \rho\sigma}\epsilon^*_{\nu}p_{\rho}P_{\sigma},
\end{align}
where momentum transfer is defined as $q =P-p$ with the momentum for the $B_c$ meson being denoted by $P$ and that for the final state charmonium by $p$. The polarization vector for the final state charmonium is represented by $\epsilon^*$. In addition we have the following QCD relation between the form factors
\begin{equation}
    2 m_V A_0(0)= (m_{B_c}-m_V) A_1(0)-(m_{B_c}+m_V) A_2(0).
\end{equation}
As this kinematic relation has calculate from the first principle of the QCD we will be exploited this constrain into our numerical analysis. In this study we also explore the $B_c \to \chi_{c0}$ transition as well. The pseudo-scalar $B_c$ meson decays to the scalar $\chi_{c0}$ meson are describing by two form factors, and related to the transition matrix elements are given by~\cite{Rui:2018kqr}: 
\begin{align}\label{eq:Bctochic0ff}
\langle \chi_{c0}(p^\prime)|\bar{c}\gamma^{\mu}\gamma^{5}b|B_c(P)\rangle =&-i\bigg[F_0(q^2)\frac{m_{B_c}^2-m_{\chi_{c0}}^2}{q^2}q^{\mu} +F_+(q^2)(P^{\mu}+p^{\prime\mu}-\frac{m_{B_c}^2-m_{\chi_{c0}}^2}{q^2}q^{\mu})\bigg],
\end{align} 
where momentum transfer is defined as $q =P-p^\prime$ with the momentum for the $B_c$ meson being denoted by $P$ and that for the final state charmonium by $p^\prime$. The form factors also hold the QCD relation $F_+(0)=F_0(0)$ and will also be exploited in our numerical analysis. 

After calculating the total amplitude for the semileptonic decay $B_c \to \chi_{c0,1}(h_c)\, \ell \nu$, we can calculate the differential decay distribution by doing spin sum over the final states and the spin average over the initial state. Then the total decay width is obtained by integrating $d\Gamma/d q^2$ over the physical region $q^2$ ranging from $m_\ell^2<q^2< q^2_\text{max}=(m_{B_c}-m_V)^2$.
 The respective expressions for the decay rates for the final state vector charmonium can be expressed as follows:
\begin{align}
    \frac{d\Gamma (B_c \to V \ell^- \bar{\nu}_\ell)}{d q^2}&=\frac{G_F^2|V_{cb}|^2 \sqrt{\lambda(m_{B_c}^2, m_{V}^2, q^2)}}{384 \pi^3 m_{B_c}^3}\bigl(1-\frac{m_\ell^2}{q^2}\bigr)^2 \times \nn \\
    &\biggl[\Bigl(1+\frac{m^2_\ell}{2 q^2}\Bigr)\biggl(|H_{V,+}(q^2)|^2+|H_{V,-}(q^2)|^2+|H_{V,0}(q^2)|^2\biggr)+ \frac{3 m_\ell^2}{2 q^2}|H_{V,t}|^2\biggr],
\end{align}
where V can be a vector meson $\chi_{c1}$ or $h_c$. In the above equation the kellen-lambda function is expressed as $\lambda(m_{B_c}^2, m_{V}^2, q^2)= \left((m_{B_c}+m_V)^2-q^2\right) \left((m_{B_c}-m_V)^2-q^2\right)$, with $m_\ell$, $m_V$ are the masses of the respective lepton and the final state meson. The differential decay rate for $B_c \to \chi_{c0} \ell^- \bar{\nu}$ is then given by:
\begin{equation}
    \frac{d\Gamma (B_c \to \chi_{c0} \ell^- \bar{\nu}_\ell)}{d q^2}=\frac{G_F^2|V_{cb}|^2 \sqrt{\lambda(m_{B_c}^2, m_{\chi_{c0}}^2, q^2)}}{384 \pi^3 m_{B_c}^3}\left(1-\frac{m_\ell^2}{q^2}\right)^2\biggl[\bigl(1+\frac{m^2_\ell}{2 q^2}\bigr)|H^{s}_{V,0}(q^2)|^2 + \frac{3 m_\ell^2}{2 q^2}|H^{s}_{V,t}|^2\biggr],
\end{equation}
where, $\lambda(m_{B_c}^2, m_{\chi_{c0}}^2, q^2)=((m_{B_c}+m_{\chi_{c0}})^2-q^2)((m_{B_c}-m_{\chi_{c0}})^2-q^2)$. 
With the transition matrix element and kinematic variables (four momentum for the parent$(P^\mu)$ and daughter meson $(p^\mu)$ and four momentum transfer to the di-lepton ($q^\mu$), polarization vector of the off-shell $w$ and for the daughter vector meson) one can calculate the hadronic helicity amplitudes that are expressed in terms of QCD form factors. For $B_c \to \chi_{c1} (h_c) \ell^- \bar{\nu}_\ell$ hadronic helicity amplitudes that are related to $V$, $A_0$, $A_1$, $A_2$ form factors and are given by:
\begin{align}
    H_{V,\pm}(q^2)=&(m_{B_c}+m_V) A_1(q^2)\mp \frac{\sqrt{\lambda(m_{B_c}^2, m_{V}^2, q^2)}}{m_{B_c}+m_V} V(q^2),\\
    H_{V,0}(q^2)=&-\frac{(m_{B_c}+m_V)}{2 m_V\sqrt{q^2}}\Biggl[ (m_{B_c}^2-m_V^2-q^2) A_1(q^2) - \frac{\lambda(m_{B_c}^2, m_{V}^2, q^2)}{(m_{B_c}+m_V)^2} A_2(q^2) \Biggr],\\
    H_{V,t}(q^2)=&-\sqrt{\frac{\lambda(m_{B_c}^2, m_{V}^2, q^2)}{q^2}} A_0(q^2).
\end{align}
and for $B_c \to \chi_{c0} \ell^- \bar{\nu}_\ell$ hadronic helicity amplitudes that are related to the form factors and are given by:
\begin{align}
    H^{s}_{V,0}(q^2)&=\sqrt{\frac{\lambda(m_{B_c}^2, m_{\chi_{c0}}^2, q^2)}{q^2}} F_+(q^2),\\
    H^{s}_{V,t}(q^2)&=\frac{m_{B_c}^2-m_{\chi_{c0}}^2}{\sqrt{q^2}} F_0(q^2).
\end{align}
To obtain reliable predictions for the decay rates, the hadronic helicity amplitudes must be determined, which are in turn expressed in terms of the transition form factors introduced earlier. At present, there are no reliable determinations of the $B_c \to \chi_{c0,1}, h_c$ form factors from LQCD or light-cone sum rule (LCSR) approaches. As a result, existing estimates of these quantities necessarily rely on model-dependent calculations. 

In this work, we adopt the NRQCD factorization framework to compute the $B_c \to$ P-wave charmonium transition form factors at $q^2=0$. Crucially, while the functional structure of the form factors is dictated by NRQCD, the associated non-perturbative inputs are extracted directly from available experimental data. This allows us to follow a largely data-driven approach, minimising reliance on purely model-dependent assumptions. The resulting form factors provide a controlled and systematic basis for evaluating the corresponding decay widths. In the following subsection, we briefly outline the NRQCD framework employed in our analysis.

\subsection{Transition Form Factor in NRQCD Framework}\label{subsec:NRQCD_theoP}
Within the NRQCD factorization framework, the QCD form factors can be factorized as the product of NRQCD short-distance coefficients and long-distance matrix elements (LDMEs). The NRQCD LDMEs contain non-perturbative information from the soft and ultra-soft regions of the QCD amplitude. These LDMEs can be expended in terms of the relative velocity between the quarks inside the mesons. At the leading order of the velocity expansion, LDMEs defined as meson's wave function at the origin~\cite{PhysRevD.51.1125, Bell:2006tz,Zhu:2017lwi, Qiao:2011yz}. The short-distance contribution, on the other hand, corresponds to the contribution from the hard momentum region of the QCD amplitude. Hence, these short-distance coefficients can be calculated up to higher orders in the strong coupling constant $(\alpha_s)$. In the Ref.~\cite{Zhu:2017lwi}, the authors have explicitly expanded the LDME in the power of $\mathcal{O}(v^2)$ and calculated the LO of the $B_c \to$ P-wave charmonium form factors, as well as the Relativistic Correction (RC) to the form factors.  

We obtain the analytic expressions of the $B_c\to P$ wave charmonia form factors calculated in the framework of the NRQCD effective theory from ref.~\cite{Zhu:2017lwi}. In our analysis, we have incorporated the available LO results and the respective relativistic corrections to the form factors. Following ref~\cite{Zhu:2017lwi}, the form factors describing a generic $B_c\to$ P-wave charmonia decay, including all the corrections, can be expressed as:
{\small
	\begin{equation}
    \label{eq:pwaveff_totexp}
	f^i(q^2)=f^i_{LO}(q^2)\bigg[1+f^i_{RC1}(q^2)/f^i_{LO}(q^2)+f^i_{RC2}(q^2)/f^i_{LO}(q^2)+f^i_{RC^{\prime}1}(q^2)/f^i_{LO}(q^2)+f^i_{RC^{\prime}2}(q^2)/f^i_{LO}(q^2)\bigg]^{1/2}.
	\end{equation}
    }
The detail expressions of $f^i_{LO}(q^2)$, $f^i_{RC^{\prime}1}(q^2)$ and $f^i_{RC^{\prime}2}(q^2) $ can be seen from the ref. \cite{Zhu:2017lwi}. 
All these form factors $f_i(q^2)$ calculated in NRQCD is proportional to the product of wave functions $\psi_{B_c}(0) \psi^{\prime}_{X_{c\bar{c}}}(0)$. The wave function associated with the $B_c$ meson can be written as:
\begin{equation}
\psi_{B_c}(0)=\frac{1}{\sqrt{2 N_c}} \langle0|\chi_b^{\dagger}\psi_{c}|B_c\rangle.
\label{eq:nonpert_matrix}
\end{equation}
And the derivative of wave functions of the $P$-wave charmonia at the origin relevant for the calculations are defined as:
\begin{align}
\psi'(0)_{h_c }\epsilon^{*i}=&\frac{1}{\sqrt{2N_c}}\langle h_c(\epsilon^*)|
\psi^{\dagger}(-\frac{i}{2}{\overleftrightarrow{ { D}^i}})\chi|0\rangle,\\
\psi'(0)_{\chi_{c0} }=&\frac{1}{\sqrt{3}}\frac{1}{\sqrt{2N_c}}\langle \chi_{c0}|
\psi^{\dagger}(-\frac{i}{2}{\overleftrightarrow{ {\bold D}}}\cdot \bfsigma)\chi|0\rangle,\\
\psi'(0)_{\chi_{c1} }\varepsilon^{*i} =&\frac{1}{\sqrt{2}}\frac{1}{\sqrt{2N_c}}\langle \chi_{c1}(\varepsilon^*)|
\psi^{\dagger}(-\frac{i}{2}{\overleftrightarrow{ {\bold D}}}\times \bfsigma)^i\chi|0\rangle.
\end{align}
For a more detailed discussion, please refer to the ref.~\cite{Qiao:2012vt, Bell:2006tz, Zhu:2017lwi, PhysRevD.51.1125}. The magnitudes of the matrix elements for the higher dimensional operators instrumental for the relativistic corrections corresponding to the $P$-wave charmonium are given by:   
\begin{align}
&\langle h_c(\varepsilon^*)|
\psi^{\dagger}(-\frac{i}{2}{\overleftrightarrow{ { D}^i}})\left(-\frac{i}{2}  \overleftrightarrow {\bold D}\right)^2\chi|0\rangle \simeq |\bold{k}'|^2\langle h_c(\varepsilon^*)|
\psi^{\dagger}(-\frac{i}{2}{\overleftrightarrow{ { D}^i}})\chi|0\rangle.
\\
&\langle \chi_{c0}|
\psi^{\dagger}(-\frac{i}{2}{\overleftrightarrow{ {\bold D}}}\cdot \bfsigma)\left(-\frac{i}{2}  \overleftrightarrow {\bold D}\right)^2\chi|0\rangle
\simeq |\bold{k}'|^2\langle \chi_{c0}|
\psi^{\dagger}(-\frac{i}{2}{\overleftrightarrow{ {\bold D}}}\cdot \bfsigma)\chi|0\rangle.\\
&\langle \chi_{c1}(\varepsilon^*)|
\psi^{\dagger}(-\frac{i}{2}{\overleftrightarrow{ {\bold D}}}\times \bfsigma)^i\left(-\frac{i}{2}  \overleftrightarrow {\bold D}\right)^2\chi|0\rangle \simeq |\bold{k}'|^2\langle \chi_{c1}(\varepsilon^*)|
\psi^{\dagger}(-\frac{i}{2}{\overleftrightarrow{ {\bold D}}}\times \bfsigma)^i\chi|0\rangle.
\end{align}
In the above, $k^\prime$ represents half of the quark relative momentum inside the $B_c$ mesons. In mathematical terms, $k^\prime = m_{red}v^\prime = m_bm_cv^\prime/(m_b+m_c)$, where $v^\prime$ is the relative velocity of the heavy valence quarks inside the $B_c$ meson. and $m_{b,c}$ are the masses of the bottom and charm quarks. For a nice discussion regarding the theoretical picture leading up to the computation of the relativistic corrections, the interested reader is referred to ref.~\cite{Zhu:2017lqu, Zhu:2017lwi, PhysRevD.51.1125}. 

Following the discussion above, we can see that to provide  predictions for the $B_c\to P$ wave form factors, it is essential to determine the $B_c$ meson wave function at the origin along with the derivative of the $P$ wave charmonium wave functions at the origin. Following our earlier analysis~\cite{Biswas:2023bqz}, we were able to obtain $\psi_{B_c}(0)$ using lattice inputs on the $B_c \to J/\psi$ form factors~\cite{Harrison:2020gvo} and the decay constants $f_{B_c}$. Into the next section, we will describe the inputs required to extract the wave functions  $\psi'(0)_{\chi_{c0} }$, $\psi'(0)_{\chi_{c1} }$ and $\psi'(0)_{h_{c} }$, for $\chi_{c0}, \chi_{c1}$ and $h_c$ respectively. Therefore, the major uncertainties in the respective form factors, and hence in the decay rates, will come from the associated uncertainties in the wave functions. 

We are aware of the scheme sensitivity of the theoretical expressions used in our analysis. In order to keep our analysis scheme independent as far as possible, we average over the values of such scheme-dependent quantities in three schemes (Kinetic scheme, $\overline{MS}$ scheme, and Pole mass scheme). In our analysis, we use $m_c=1.34\pm 0.27$ (see ~\cite{Biswas:2023bqz}).

In this work, our focus is primarily on phenomenological studies of semileptonic $B_c$ meson decays to the scalar $\chi_{c0}$ and the axial-vector $\chi_{c1}$ and $h_c$ states. As no LQCD calculation of these transition form factors is available to date, it would be interesting to obtain the shapes of the form factors for further phenomenological studies and explore these decays and present predictions for measurable observables, such as semileptonic branching ratios, which can be verified once measurements of these decay modes become available. 

 We will conduct a phenomenological analysis of the semileptonic $B_c \to P$ wave charmonium decay in four steps, organized into four subsections. In subsection \ref{subsec:exp_th_Pwave}, we will discuss the experimental inputs and provide a brief overview of the relevant theory within the NRQCD framework for P-wave charmonium radiative decays. These inputs are essential for extracting the derivative of the radial wave function of the charmonium states at their origin. In the second step discussed in subsection ~\ref{subsec:fit_res_Pwave}, we will extract the charmonia wave function by performing a global analysis to maximize the log-likelihood function. In the third step, presented in subsection~\ref{subsec:BctoPff_shape}, we will provide our estimates for the transition form factors at $q^2 = 0$ and then extrapolate them to the entire physical $q^2$ region using an appropriate extrapolation technique. In the final step, we estimate physical observables in subsection \ref{subsec:BctoP_obs} using the form factors' shape across the entire kinematic region.
 
\subsection{Radiative Decay Modes of the P-wave charmonia}\label{subsec:exp_th_Pwave}
In this subsection, we discuss the radiative decay modes of P-wave charmonia used in our current analysis to extract the derivative of the radial wave functions for the P-wave charmonia. The relevant experimental data have been provided in table~\ref{tab:radexptinpt}. The corresponding theoretical expressions and related discussions are provided in a point-wise manner in what follows.
\begin{table}[t]
	\begin{center}
	\renewcommand*{\arraystretch}{1.8}
	\resizebox{0.85\textwidth}{!}{
	\begin{tabular}{|c|c|c|c|}
		\hline
		\textbf{Observables} & \textbf{Measurement(BR)}~\cite{Zyla:2020zbs} & \textbf{Decay Constant} & \textbf{LQCD Estimate~~(in MeV)} \\ \hline \hline
		$\mathcal{B}(\chi_{c0}\rightarrow\gamma\gamma)$  & $2.04(10)\times 10^{-4}$ & $f_{J/\psi}$& $405(6)$~\cite{Donald:2012ga}  \\
		$\mathcal{B}(\chi_{c1}\rightarrow \rho \gamma)$  & $2.16(17)\times 10^{-4}$& $f_{\eta_c}$ & $394.7(24)$~\cite{Davies:2010ip} \\
		$\mathcal{B}(\chi_{c1}\rightarrow \phi \gamma)$  & $2.4(5)\times 10^{-5}$&& \\
		$\mathcal{B}(\chi_{c1}\rightarrow J/\psi \gamma)$  & $34.3(13)\times 10^{-2}$&& \\
		$\mathcal{B}(h_c\rightarrow \eta_{c} \gamma)$ & $60.0(40)\times 10^{-2}$ & & \\
        $\mathcal{B}(J/\psi \rightarrow e^+ e^-)$  & $5.971(32) \times 10^{-2}$ & & \\
        $\mathcal{B}(\eta_{c} \rightarrow \gamma \gamma)$ & $1.61(12)\times 10^{-4}$&&\\
		\hline
	\end{tabular}
}
	\caption{Experimental measurements for radiative decays of $h_c$, $\chi_{cJ}(J=0,1)$ charmonia and Lattice estimates for decay constant for $J/\psi$, $\eta_c$.}
	\label{tab:radexptinpt}
\end{center}
\end{table}
\begin{enumerate}
    \item \underline{\textbf{Radiative $\chi_{c1}\rightarrow V(V=\rho,\phi)~\gamma $ Decay}}:
 The $\chi_{c1} \to V \gamma$ transition with V representing a light vector meson $\rho$ and $\phi$ is a radiative decay of charmonium involving the emission of a photon and the production of a light vector meson. This process is an annihilation process in which quark-anti-quark pairs annihilate into gluons, which can hadronize into a light vector meson and a photon. For the decay width estimates of the respective decay process, we primarily used the expression provided in ref.~\cite{Kivel:2017nrw}. The amplitude for radiative decay $\chi_{c1}\to \gamma V (V=\rho,\phi)$ has been computed within the QCD factorization framework in this article. The authors use NRQCD in order to separate the heavy from the light degrees of freedom. The corresponding decay width is expressed as:
\begin{eqnarray}
\label{eq.chic1_decaywidth}
\Gamma(\chi_{c1}\to V\gamma)=\frac{1}{12 \pi}\frac{E^5_{\gamma}}{M_1^4}\Bigl[|A_{1V}^{||}|^2+|A_{1V}^{\perp}|^2\Bigr].
\end{eqnarray}
To calculate the longitudinal amplitude $(A_{1V}^{||})$, we consider only the color-singlet contribution in the hard-factorization framework has been considered, as the color-octet contribution is suppressed by a power of $v^2$ in the hard region. The analytical expression for the corresponding longitudinal amplitude is:
\begin{align}
\label{eq:chic1amplitude_para}
A_{1V}^{||}=& -i \langle\mathcal{O}(^3P_1)\rangle \frac{f_v M_1^2}{m^6} Q_V \sqrt{4 \pi\alpha}~\alpha_{s}^2(\mu_h) \frac{N_c^2-1}{2 N_c^2}\int_{0}^{1} dx \phi_V^{||}(x) T_1(x).
\end{align}
In the above equation, the NRQCD matrix element $\langle\mathcal{O}(\chi_{c1})\rangle$ is related to the derivative of the quarkonium radial wave function, $\psi^{'R}_{\chi_{c1}}(0)$ at the origin $r=0$ as:
\begin{equation}
	\langle\mathcal{O}(\chi_{c1})\rangle =\sqrt{2N_c}\sqrt{2 M_J} \sqrt{\frac{3}{4 \pi}} \psi^{\prime R}_{\chi_{c1}}(0) (1+\delta_\text{C} + \delta_\text{NC}),
\end{equation}
where $\delta_\text{C}$ and $\delta_\text{NC}$ are the terms that contribute to the wave function corresponding to the Coulombic and Non-Coulombic correction terms. For a detailed discussion see the ref.~\cite{Chung:2021efj}. And these correction terms are taken as nuisance parameters in our analysis. We use the prior estimates of both these correction terms with $10\%$ error as provided in ref~\cite{Chung:2021efj}:
\begin{equation}~\label{eq:coulombicnc}
	\delta_C=0.266 \pm 0.027 ~~~~~~~~~~~~~~~~~ \delta_\text{NC}= 0.493 \pm 0.049.
\end{equation}
The longitudinal amplitude contains $\phi_V$, is the twist-2 Light-Cone Distribution Amplitude (LCDA) for the final state vector meson and the Hard kernel ($T_1(x)$). The Hard kernel that describes short-distance physics is given by the appropriate partonic configuration with large momentum, which can be computed systematically within perturbative QCD, while long-distance contributions are associated with the matrix elements of the NRQCD operator. Here in eq.~\ref{eq:chic1amplitude_para} the hard kernel is expressed as
\begin{equation}
T_1(x)= ReT_1(x) +i Im\,T_1(X),
\end{equation}
and has been calculated and given in ref.~\cite{Kivel:2017nrw}.
The twist-2 distribution amplitude of the light vector meson $\phi^{\parallel}_V(x)$ has the form 
\begin{equation}
    \phi_V(x, \mu)=6x(1-x)\{1+a_2^V C_2^{3/2}(2x-1)\},
\end{equation}
where the LCDA coefficient is $a_2^V$. Using the explicit expression for the coefficient functions $\phi^{||}_V(x)$ and $T_1(x)$, one can easily obtain the convolution integral as:
\begin{equation}
    \int_0^1 dx \, \phi_V(x) T_1(x)= A_i +a_2^V (\mu_h) B_i,     
\end{equation}
Using explicit expressions for the hard kernel functions, one has
\begin{align}
    A_1=& \int_0^1 dx \, 6x \bar{x}\, T_1(x)=1.32 +5.46 \,i , \\
    B_1=& \int_0^1 dx \, 6x \bar{x} C_2^{3/2} (2x-1) \, T_1(x)= 7.00 + 4.79\,i.
\end{align}
With the given twist-2 LCDA coefficient, one can able to calculate the convolution integral given in eq.~\ref{eq:chic1amplitude_para}. The twist-2 LCDA coefficients are detailed in ref.~\cite{Braun:2016wnx, Ball:2007rt}:
\begin{align}
&a^\parallel_{1\rho}=0~~~~~~~~~~~~~~~~~~~~~~~~~~~~~~a^\parallel_{1\phi}=0\\
&a^\parallel_{2\rho}=0.15\pm 0.07~~~~~~~~~~~~~~~~~a^\parallel_{2\phi}=0.18\pm 0.08.
\end{align}
The amplitudes $A_{1V}^\perp$ associated with the decay amplitude defined in eq.~\ref{eq.chic1_decaywidth}, where the final state is the transverse light meson. Only color singlet contributions to amplitudes have been computed in the literature. To calculate the hard kernel and therefore the amplitude for this decay, one has to consider twist-3 LCDAs to model three-particle wave functions. The details of the calculation are presented in ref.~\cite{Kivel:2017nrw}. The transverse decay amplitude is expressed as:
\begin{equation}
\label{eq:amplitude_perp}
{\small
A_{1V}^{\perp}=  i \langle\mathcal{O}(^3P_1)\rangle \frac{f_V m_V}{m^5}\frac{M_1^2}{m^2} \sqrt{4\pi \alpha} \frac{\pi \alpha}{N_c} \int D \alpha_i \biggl\{\delta_{I0} 9e_Q \frac{G(\alpha_i)}{\alpha_1 \alpha_2 \alpha_3^2} \frac{Q_V}{4}\bigg(\frac{\alpha_1-\alpha_2}{\alpha_1 \alpha_2 \alpha_3^2} V(\alpha_i) +\frac{1-\alpha_3}{\alpha_1 \alpha_2 \alpha_3^2} A(\alpha_i)\bigg)\biggr\}
},
\end{equation}
Here, $e_Q$ implied the electric charge of the heavy quark. The quark charges are
\begin{equation}
    Q_{\rho}=\frac{1}{2}(e_u-e_d)=\frac{1}{2}, ~~~~~~~~~~~~~~~~~ Q_{\phi}=e_s=-\frac{1}{3}\, 
\end{equation}
The abbreviation $\delta_{I0} = 1$ for $\phi$ with isospin $I = 0$, and $\delta_{I0} = 0$ for the $\rho \, (I= 1)$ meson with
\begin{equation}
\int D \alpha_if(\alpha_i)=\int d\alpha_1 \int d \alpha_2 \int d\alpha_3 ~\delta(1-\alpha_1-\alpha_2-\alpha_3) f(\alpha_1,\alpha_2,\alpha_3).
\end{equation} 
The functions $V(\alpha_i)$, $A(\alpha_i)$, and $G(\alpha_i)$ are twist-3 distribution amplitudes that describe the overlap with the transverse light mesons and have the form~\cite{Ball:1996tb, Ball:1998ff, Ball:1998sk}:
\begin{align}
    A(\alpha_i)=&360 \, \zeta_3 \alpha_1 \alpha_2 \alpha_3\bigg(1+\omega_3^A\frac{1}{2}(7\alpha_3-3)\bigg),\\
    V(\alpha_i)=&5040 \, \zeta_3 \omega_3^V \alpha_1 \alpha_2 \alpha_3^2(\alpha_2-\alpha_1),\\
    G(\alpha_i)=&5040 \, \zeta_3 \omega_3^G \alpha_1^2 \alpha_2^2 \alpha_3^2,
\end{align}
The parameters, $\zeta_3$, $\omega_3^{A,V,G}$, being the non-perturbative parameters of the twist 3 DAs, have been taken from~\cite{Kivel:2017nrw}, their numerical values which we have considered in this analysis are given as
\begin{align}
	\label{eq:twist3_DA}
	&\zeta_{3\rho}=0.030\pm 0.010 ~~~~~~~~~~~~~~~~~~\zeta_{3\phi}=0.024\pm 0.008 \nn \\
	&\omega^A_{3\rho}=-3.0 \pm 1.4 ~~~~~~~~~~~~~~~~~~~~\omega^A_{3\phi}=-2.6 \pm 1.3 \nn \\
	& \omega^V_{3\rho}=5.0 \pm 2.4 ~~~~~~~~~~~~~~~~~~~~~~\omega^V_{3\phi}=5.3 \pm 3.0 \nn \\
	& ~~~~~~~~~~~~~~~~~~~~~~~~~~~~~~~~~~~~~~~~~~\omega^{G}_{3\phi}=0.1 \pm 0.01.
\end{align}
In eq.~\ref{eq:chic1amplitude_para} and \ref{eq:amplitude_perp}, $f_V$ represents the decay constant of the light vector meson having the values
\begin{equation}
\label{eq:twist2_DA}
f_\rho=0.221~\text{GeV}~\text{\cite{Zyla:2020zbs}} ,~~~~~~~~~~~~~~~~~f_{\phi}=0.161~\text{GeV}~\text{\cite{Zyla:2020zbs}},
\end{equation}
$m_V$ represents the mass of the light vector mesons, having values 
\begin{equation}
    m_{\rho}= 0.775~\text{GeV}~\text{\cite{Zyla:2020zbs}} , ~~~~~~~~~~~~~~~~~m_{\phi}=1.019~\text{GeV}~\text{\cite{Zyla:2020zbs}} ,
\end{equation}
And then the twist-3 DAs have the following normalization
\begin{equation}
\int D \alpha_i A(\alpha_i)= \zeta_3, ~~~\, \, \int D \alpha_i (\alpha_2-\alpha_1) V(\alpha_i)= \zeta_3 \omega^V_3, ~~~ \, \, \int D \alpha_i \, G(\alpha_i)= \zeta_3 \, \omega_3^G.
\end{equation}
The coefficients of twist-3 DAs are taken from ref.~\cite{Kivel:2017nrw} for all the model parameters considered in our analysis.

\item \underline{\bf Radiative $\chi_{c1} \rightarrow J/\psi \gamma$ and $h_{c}\rightarrow \eta_{c} \gamma$ Decay}
This transition is an electric dipole(E1) transition, with the P-wave charmonium $\chi_{c1}(h_c)$ decaying into an S-wave charmonium $J/\psi(\eta_c)$ with the emission of a photon. Being an $E1$ transition, it involves a change in orbital angular momentum quantum number, $\Delta L=1$, while the spin quantum number, $\Delta S=0$.  The development of EFTs for heavy quarkonium helped to establish a model independent framework for the study of these radiative transitions. A systematic study of M1 transitions within the potential NRQCD (pNRQCD) framework has been performed~\cite{PhysRevD.85.094005}. Using this framework, a perturbative evaluation of the decay rates has been performed. The decay rate for these E1 transitions depends on the overlap between the initial $\chi_{c1} (\stateinequ{3}{P}{1})$ and final $J/\psi(\stateinequ{3}{S}{1})$ state, and $h_c (\stateinequ{1}{P}{1})$ and $\eta_c (\stateinequ{1}{S}{0}) $, respectively. Then the decay rate for this $E1$ transition from a $\stateinequ{3}{P}{1} \to \stateinequ{3}{S}{1}$(spin-triplet transition) and $\stateinequ{1}{P}{1} \to \stateinequ{1}{S}{0}$(spin-singlet transition) is given by~\cite{QuarkoniumWorkingGroup:2004kpm}:
\begin{eqnarray}
\Gamma(i \rightarrow f + \gamma)= \, {4 \over 3} \, \alpha \, e_{c}^2 
(2 J_f+1) \mathcal{S}_{if} \, \omega_{f}^3 \, |I_{SP}|^2~(1+\delta^{\prime}_{SP} ),
\label{eq:decayrat_SP}
\end{eqnarray}
where $\omega_{f}=\frac{M_i^2-M_f^2}{2 M_i}$ representing the energy of the emitted photon with $M_i(M_f)$ is the initial(final) quarkonium state mass. And $e_{c}=\frac{2}{3}$ representing the charge of the charm quark. The statistical factor $\mathcal{S}_{if}=\frac{1}{3}$ and $\frac{1}{9}$ for transitions between spin-singlet and spin-triplet states respectively~\cite{QuarkoniumWorkingGroup:2004kpm}. The term $\delta^{\prime}_{SP}$ in the transition decay rate represents the relativistic corrections~$\mathcal{O}(v^2)$ to the leading E1 decay rate, and is the non-perturbative contribution to the decay rate. The relativistic corrections to the decay rate can be incorporated systematically as an expansion in the powers of the quark-anti-quark relative velocity $v$. the relativistic~$\mathcal{O}(v^2)$ corrections to the decay rates have been calculated using pNRQCD explicitly. Likewise, these corrections appear mainly from two sources~\cite{PhysRevD.85.094005}:
\begin{itemize} 
    \item Contributions from higher-order operators in the pNRQCD Lagrangian, which introduce additional corrections to the LO decay rate.  
    \item Relativistic corrections to the $Q\bar{Q}$ potential lead to a correction in the decay rate.
\end{itemize}
The size of these corrections has been discussed in detail previously and can be found in Ref.~\cite{MartinezNeira:2017prz}. In that work, they reported sizable negative corrections to the LO result, amounting to approximately $10-15\%$ and $40-50\%$, respectively, from the two sources discussed above. The overlap integral, $I_{SP}$ in eq.~\ref{eq:decayrat_SP} is the effective radial overlap integral which is defined as:
\begin{align}
\text{for} \, \, & \chi_{c1} \to J/\psi \gamma: \quad & I_{SP}&=\langle \chi_{c1} |r |J/\psi \rangle = \int_0^\infty r^3 \, \psi^{\prime R}_{\chi_{c1}}(r) \, \psi^R_{J/\psi}(r) \, dr~\, , \nn \\
\text{for} \,\, &h_{c} \to \eta_c \gamma:  \quad &I_{SP}&=\langle h_c |r | \eta_c \rangle = \int_0^\infty r^3 \, \psi^{\prime R}_{h_c}(r) \, \psi^R_{\eta_c}(r) \, dr~ \, .
\label{eq:Ips_overlap_int}
\end{align}
 Here $\psi_{S}^R(r)$ represents the normalized radial solution of the Schrodinger wave function for the $S$ wave, while  $\psi_{P}^{\prime R}(r)$ denotes the normalized derivative of the radial solution for the $P$ wave charmonia states. In this work, we consider the quark-antiquark binding potential to be a simple Coulombic potential. For a Coulombic potential, the normalized radial wave functions have the form
\begin{align}\label{eq:wavfun}
\psi_S^R(r)=&A~exp\bigg(-\frac{\lambda_s}{2} r\bigg),\\
\psi_P^R(r)=& A^{\prime} r~ exp\bigg(-\frac{\lambda_p}{2} r\bigg),
\end{align}
Here the normalization constants are denoted by A and $A^\prime$. The normalization condition allows us to constrain the shape parameters $\lambda_s$, and $\lambda_p$, which are related to the normalization constants as follows:
\begin{equation}
A=\psi_S^R(0)=  \sqrt{\frac{\lambda_s^3}{2}}; \quad \quad \quad \quad \ \ \  A^\prime={\psi^{' R}_P}(0)=\sqrt{\frac{\lambda_p^5}{24}}.
\end{equation}
Here, ${\psi^{R}_S}(0)$, ${\psi^{'R}_P}(0)$ are the radial wave function and derivative of the radial wave function respectively, both defined at the origin $r=0$ for S wave and P wave charmonia respectively. We constrain these shape parameters from the experimental measurements. 

Hence, in addition to the derivatives of the $P$-wave functions, the decay rates $\Gamma(\chi_{c1}\to J/\psi \gamma)$ and $\Gamma(h_{c}\to \eta_c \gamma)$ are also sensitive to the wave functions at the origin, $\psi_{J/\psi}(0)$ and $\eta_c(0)$. These quantities are extracted using the available experimental inputs on the branching fractions $\mathcal{B}(\eta_{c}\rightarrow \gamma\gamma)$ and $\mathcal{B}(J/\psi \rightarrow e^+e^-)$, together with lattice QCD determinations of the decay constants $f_{J/\psi}$ and $f_{\eta_c}$. The input we used to extract the charmonium wave function of the S-wave is presented in table~\ref{tab:radexptinpt}. In the following, we discuss the relevant details.

\begin{itemize}
    \item \underline{\bf Radiative $\eta_{c}\rightarrow \gamma\gamma$ Decay}
\end{itemize}
	Radiative decay of the $\eta_c$ meson is considered the cleanest channel, as it is free from gluon contamination in the final state. This allows us to extract the non-perturbative information about the pseudo-scalar $\eta_c$ meson from this decay mode. This decay mode has been studied extensively in the recent years. Within the NRQCD framework, this decay mode has been studied with great precision. The analytic expression for the decay width corresponding to the di-photon decay of $\eta_c (\stateinequ{1}{S}{0})$ heavy quarkonium at order $\mathcal{O}(v^4)$, where $v$ denotes the relative velocity of the heavy quarks in the heavy quarkonium, is given by~\cite{Bodwin:2002cfe, Guo:2011tz}
	\begin{eqnarray} \label{eq:etac2gamma}
	\Gamma(\eta_c(\stateinequ{1}{S}{0}) \rightarrow \gamma \gamma)&=&
	\frac{F_{\gamma\gamma}(\stateinequ{1}{S}{0})}{m_c^2}\langle
	\mathcal{O}(\stateinequ{1}{S}{0}) \rangle_{\eta_c}
	+\frac{G_{\gamma\gamma}(\stateinequ{1}{S}{0})}{m_c^4}\langle \mathcal{P}(\stateinequ{1}{S}{0}) \rangle_{\eta_c} + \nn \\ &&\frac{H^1_{\gamma\gamma}(\stateinequ{1}{S}{0})}{m_c^6
      }\langle \mathcal{Q}^1_1(\stateinequ{1}{S}{0}) \rangle_{\eta_c} +\frac{H^2_{\gamma\gamma}(\stateinequ{1}{S}{0})}{m_c^6
      }\langle \mathcal{Q}^2_1(\stateinequ{1}{S}{0}) \rangle_{\eta_c},
	\end{eqnarray}
	where 
	\begin{align}
	\langle\mathcal{P}(\stateinequ{1}{S}{0})\rangle_{\textrm{LO}}=
	\mathbf{q}^2 \langle\mathcal{O}(\stateinequ{1}{S}{0})\rangle_{\textrm{LO}}.
	\end{align}
	The leading order long-distance matrix element (LDME) is related to the wave function at the origin as
	\begin{eqnarray} \label{wfodefinition}
	\langle \mathcal{O}(\stateinequ{1}{S}{0})
	\rangle_{\eta_c}&=& 2 N_c |\psi(0)_{\eta_c}|^2.
	\end{eqnarray}
	Here $\mathbf{q}$ is half the relative three-momentum of the heavy quark and anti-quark that the heavy quarkonium consists of. The $F_{\gamma\gamma}(\stateinequ{1}{S}{0})$, $G_{\gamma\gamma}(\stateinequ{1}{S}{0})$, $H^1_{\gamma\gamma}(\stateinequ{1}{S}{0})$, and $H^2_{\gamma\gamma}(\stateinequ{1}{S}{0})$ are the short distance coefficients obtained by matching the expression for the decay width of a heavy quarkonium decaying into light hadrons computed in perturbative QCD to that computed in perturbative NRQCD and are given by
	\begin{align}
	F_{\gamma \gamma}(\stateinequ{1}{S}{0})=& 2\pi \alpha^2 e_c^4 \bigl(1+\frac{\alpha_s}{\pi} \frac{\pi^2-20}{3}\bigr)
	, \\
	G_{\gamma \gamma}(\stateinequ{1}{S}{0})=& 2\pi \alpha^2 e_c^4
	[-\frac{4}{3}+
	\frac{\alpha_s}{\pi}\frac{1}{27}\bigl(48\ln\frac{\mu_{R}^2}{m_c^2}-96\ln 2-15\pi^2+196\bigr)
	],\\
      H^1_{\gamma\gamma}+H^2_{\gamma\gamma}=&\frac{136 \pi}{45} e_c^4 \alpha^2
	\end{align}
    Using the charge of the charm quark $e_c=\frac{2}{3}$, mass for the charm quark $m_c= 1.34\pm 0.27$, and the renormalization scale $\mu_{R}=2 m_c$, we estimated the contribution to the LO decay rate order by order from eq.~\ref{eq:etac2gamma}. We find that the $\mathcal{O}(\alpha_s)$ corrections reduces the decay rate by about $43\%$, while the LO relativistic correction  at $\mathcal{O}(v^2)$ lower it further by approximately $27\%$. Where as the correction at $\mathcal{O}(\alpha_s v^2)$ enhances the rate by about $4\%$, and the next-to leading-order (NLO) relativistic correction contributes an additional positive shift of roughly $6\%$. There may still be missing contribution, denoted as $\delta^\text{corr.}_{\eta_c}$,  arising from higher-order operators in the NRQCD Lagrangian that are not yet fully determined. In our analysis, we conservatively assume that such unknown pieces could affect the LO decay rate by as much as $10\%$.
    
    To constrain the $\eta_c$ wave function, we also utilized the decay constant for this charmonium state, as provided by LQCD, which is shown in table~\ref{tab:radexptinpt}. Discussion of this decay constant is given in our previous work ref.~\cite{Biswas:2023bqz}.
	\begin{itemize}
	    \item \underline{\bf Pure Leptonic $J/\psi\rightarrow e^+ e^- $ Decay}
	\end{itemize}
The pure leptonic decay mode of the $J/\psi$ meson. It is considered one of the cleanest channels for $J/\psi$ probes. The branching fraction for the $J/\psi$ meson in this decay mode is the largest that contributes to its total decay width, allowing us to extract the non-perturbative information about the $J/\psi$ meson. This decay mode has also been studied extensively in recent years. Within the NRQCD framework, this decay mode has been studied with great precision. The analytic expression for the decay width corresponding to the electron-positron decay of a $J/\psi(\stateinequ{3}{S}{1})$ heavy quarkonium at order $\mathcal{O}(v^4)$, where $v$ denotes the relative velocity of the heavy quarks in the heavy quarkonium, is given by~\cite{Bodwin:2002cfe, Lee:2018aoz}:
   \begin{align}
	\Gamma[ J/\psi \to e^+ e^-]=& 
	\frac{2 e_c^2 \pi \alpha^2 \langle O_1 \rangle_{J/\psi}}{ 3  m_c^2} \times \nn \\ & \left[1 - \frac{8}{3} \, \frac{\alpha_s}{ \pi}-\frac{1}{6} \langle v^2 \rangle_{J/\psi}+\frac{\alpha_s}{3 \pi}\bigg(\frac{8}{9}+\frac{8}{3} \ln\frac{\mu_{R}^2}{m_Q^2}\bigg)\langle v^2 \rangle_{J/\psi}+\frac{29}{18}\langle v^4 \rangle_{J/\psi}\right]^2.
	\label{eq:gam_psi}
	\end{align}
    Using the $e_c=\frac{2}{3}$, $m_c=1.34 \pm 0.27$ and the renormalization scale $\mu_{R}=2 m_c$, we calculate the contributions to the LO decay rate order by order. We find that the $\mathcal{O}(\alpha_s)$ corrections reduce the decay rate by about $34\%$, interfering destructively with the LO term. The relativistic correction at $\mathcal{O}(v^2)$ also contributes negatively, lowering the rate by approximately $3\%$. Together, these corrections significantly reduce the resulting decay rate. On the other hand, the $\mathcal{O}(\alpha_s v^2)$ corrections enhance the decay rate by about $4\%$, while the $\mathcal{O}(v^4)$ correction contributes an additional positive shift of roughly $7\%$. Both of these correction terms compensate for the earlier reduction. 
    
    There may still be missing contribution, denoted as $\delta^\text{corr.}_{J/\psi}$, arising from higher order operators in the NRQCD Lagrangian that are not yet fully determined. In our analysis, we conservatively estimate that, these unknown pieces could affect the LO decay rate by up to $10\%$. We also used the decay constant for the $J/\psi$ wave function $\psi^R_{J/\psi}(0)$, as provided by LQCD, which is detailed in table~\ref{tab:radexptinpt}. Further discussion of this decay constant can be found in our previous work ref.~\cite{Biswas:2023bqz}.
	\item \underline{\bf Radiative $\chi_{c0}\rightarrow \gamma \gamma$ Decay}
\end{enumerate}
Radiative decay of the $\chi_{c0}$ meson is considered one of the cleanest channels, as it is free from gluonic contamination in the final state. This allows us to extract the non-perturbative information of the scalar $\chi_{c0}$ meson. In recent years, within the NRQCD factorization framework, this radiative decay of $\chi_{c0}$ has been analyzed with increasing precision. The analytic expression for the decay width corresponding to the di-photon decay of a general P-wave heavy quarkonium of order $\mathcal{O}(\alpha_s^2)$, where $\alpha_s$ denotes the strong coupling constant between two heavy quarks in the heavy quarkonium. The decay width corresponding to the di-photon decay for $\chi_{c0}$ is given by~\cite{Xu:2014zra, Chung:2021efj, Sang:2015uxg}:
\begin{equation}
\Gamma[\chi_{c0}\to\gamma\gamma]=\frac{12 \pi Q_c^4 \alpha^2}{m_{\chi_{c0}}}\frac{\langle\mathcal{O}^{\chi_{c0}}\rangle }{m_c^3}\Bigl[1+\alpha_s c^{(1)}_{\gamma \gamma}-\left(1.83+2.36~\alpha_s \right)v^2_{\chi_{c0}} +\alpha^2_s c^{(2)}_{\gamma \gamma}\Bigr]
\end{equation}
with the short distance coefficients,
\begin{eqnarray}
C^{(1)}_{\gamma \gamma}&=&\frac{C_F}{\pi}\Bigl(\frac{\pi^2}{4}-\frac{7}{3} \Bigr)\nonumber\\
C^{(2)}_{\gamma \gamma}&=&\frac{C_F}{\pi^2} \frac{\beta_0}{4}\Bigl(\frac{\pi^2}{4}-\frac{7}{3} \Bigr) \ln{\frac{\mu^2_R}{m_c^2}}-\gamma^{(2)} \ln{\frac{\mu_\Lambda}{m_c}}+ c^{(2),\text{fin}}_{\gamma\gamma},
\end{eqnarray}
where loop coefficient of the QCD $\beta$ function $\beta_0=\frac{25}{3}$, $\alpha_s$ is the strong coupling constant. $c^{(2),\text{fin}}_{\gamma\gamma}$ includes the contribution from the heavy quark loop and is known numerically as~\cite{Sang:2015uxg}
\begin{equation}
    c^{(2),\text{fin}}_{\gamma\gamma}= -2.39666 + 0.11534 \,i \, .
\end{equation}
And $\mu_R$, $\mu_\Lambda$ are the renormalization and the factorization scale respectively, and the fine structure constant $\alpha= \frac{1}{137}$ and charm quark charge $Q_c=\frac{2}{3}$. We choose the renormalization scale, $\mu_R=2 m_c$, and factorization scale, $\mu_\Lambda= m_c$ in our analysis. The NRQCD matrix element $\langle\mathcal{O}^{\chi_{c0}}\rangle$ is related to the derivative of the quarkonium radial wave function $\psi^{' R}_{\chi_{c0}}(0)$ at the origin r=0 as:
\begin{equation}
    \langle\mathcal{O}^{\chi_{c0}}\rangle= \frac{6 N_c}{4 \pi}|\psi^{' R}_{\chi_{c0}}(0)|^2\bigl (1+\delta_{C}+\delta_{NC})^2 \, ,
\end{equation}
where $\delta_c$ and $\delta_{NC}$ are the Coulombic and non-Coulombic correction terms as discussed previously in eq.~\ref{eq:coulombicnc} and we consider them as nuisance parameters in this analysis. 

Using $m_c= 1.34 \pm 0.27$, $\alpha(2m_c)= 0.245 \pm 0.025$, we evaluate the contributions to the LO decay rate and the available corrections at different order. The $\mathcal{O}(\alpha_s)$ term provides a modest enhancement of about $1\%$ relative to LO. In contrast the relativistic correction at $\mathcal{O}(v^2)$ is sizable and negative, reducing the rate by nearly $37\%$. When combined, these effects substantially reduce the resulting decay rate. Additionally, the correction term at $\mathcal{O}(\alpha_s v^2)$ further lowers the rate by roughly $12\%$. Finally, the NNLO contribution ($\mathcal{O}(\alpha_s^2)$) introduces a compensating positive shift of about $15\%$. 

There may still exist missing contributions, collectively denoted as $\delta^\text{corr.}_{\chi_{c0}}$, which originate from higher-order operators in the NRQCD Lagrangian and remain presently undetermined. In order to account for these unknown effects, we adopt a conservative approach by allowing for a potential impact of up to $20\%$ on the decay rate. Such an estimate effectively translates into an uncertainty in the determination of the underlying wave functions. Incorporating this uncertainty is crucial for ensuring the robustness of the analysis, as it accounts for residual theoretical limitations associated with the truncation of the NRQCD expansion. This conservative treatment prevents an overestimation of the precision of the extracted parameters and provides a more reliable assessment of theoretical uncertainties, thereby strengthening the credibility and stability of the phenomenological conclusions drawn from the study.
%
%
\subsection{Extraction of $\psi^{' R}_M(0): M= \chi_{c0}, \chi_{c1}, h_c$}\label{subsec:fit_res_Pwave} 
The transition matrix elements for the decays of the $B_c$ meson to P-wave charmonia are parameterized in terms of the corresponding form factors as defined in eqs.~\ref{eq:Bctochic0ff}, \ref{eq:Bctohcchic1ff}. These form factors can be written in terms of the respective wave functions and/or derivatives of the wave functions of the initial and final states at the origin, as discussed in the NRQCD theory formalism in subsection ~\ref{subsec:NRQCD_theoP}. A reliable estimate for these wave functions is thus imperative for calculating any exclusive process involving these mesons. There are model-dependent approaches to describe the charmonia wave functions~\cite{Eichten:1995ch, Eichten:1978tg, Eichten:2019hbb}. In this subsection, we will extract the wave functions of the $\chi_{c0}$, $\chi_{c1}$, and $h_c$ states, as well as those of $J/\psi$ and $\eta_c$, using a data-driven analysis, avoiding current model-dependent estimates as inputs in our analysis. 

We determine the parameters by maximizing the log-likelihood, which corresponds to minimizing the $\chi^2$ function. The likelihood is constructed using the available experimental inputs, which are listed in table~\ref{tab:radexptinpt}, namely: 
$\mathcal{B}(\chi_{c0}\rightarrow\gamma\gamma)$, 
$\mathcal{B}(\chi_{c1}\rightarrow \rho \gamma)$, 
$\mathcal{B}(\chi_{c1}\rightarrow \phi \gamma)$,
$\mathcal{B}(\chi_{c1}\rightarrow J/\psi \gamma)$, 
$\mathcal{B}(h_c\rightarrow\eta_{c}\gamma)$ and 
$\mathcal{B}(\eta_{c}\rightarrow \gamma \gamma)$, 
$\mathcal{B}(J/\psi \rightarrow e^+ e^-)$ along with decay constant for $J/\psi$, and $\eta_c$ meson. The radial part of the Schrodinger wave functions at the origin $\psi^{R}_{J/\psi}(0)$ and $\psi^{R}_{\eta_c}(0)$, together with the derivative of the radial wave functions for the decaying charmonia $\psi^{'R}_{\chi_{c0}}(0)$, $\psi^{'R}_{\chi_{c1}}(0)$ and $\psi^{'R}_{h_c}(0)$ are treated as fit parameters. The radial wave-function $\psi^R_{J/\psi}(0)$ and $\psi^R_{\eta_{c}}(0)$ enters the decay $\chi_{c1} \rightarrow J/\psi \gamma$ and $h_c\to \eta_c \gamma$, respectively overlap with $\psi^{'R}_{\chi_{c1}}(0)$ and $\psi^{'R}_{h_c}(0)$. To extract $\psi^R_{J/\psi}(0)$ and $\psi^R_{\eta_{c}}(0)$ wave functions, we use $\mathcal{B}(J/\psi \rightarrow e^+ e^-)$ and $\mathcal{B}(\eta_{c} \rightarrow \gamma \gamma)$ measurements along with lattice inputs for decay constant $f_{J/\psi}$, $f_{\eta_c}$. These priors then allow us to determine $\psi^{' R}_{\chi_{c1}}(0)$ and $\psi^{' R}_{h_c}(0)$. The remaining parameter $\psi^{' R}_{\chi_{c0}}(0)$, is extracted from the decay mode $\chi_{c0} \to \gamma \gamma$.

\paragraph{\underline{Setup for the Analysis}:}
Our primary goal is to simultaneously extract the derivatives of the wave functions for P wave charmonium states $\psi^{'R}_{\chi_{c0}}(0)$, $\psi^{'R}_{\chi_{c1}}(0)$ and $\psi^{'R}_{h_c}(0)$ together with the S-wave charmonium wave functions $\psi^R_{J/\psi}(0)$ and $\psi^R_{\eta_{c}}(0)$. The known and the unknown parameters associated with the analysis are organized as follows:


\begin{itemize}
	\item \textbf{Free Parameters:} $\psi^{'R}_{\chi_{c0}}(0),~\psi^{' R}_{\chi_{c1}}(0),~\psi^{'R}_{h_c}(0)$, $\psi^{R}_{J/\psi}(0)$, $\psi^{R}_{\eta_c}(0)$.
	\item \textbf{Nuisance Parameters:} $m_c$, $\alpha_s(2 m_c)$, $\delta^\text{corr.}_{J/\psi}$, $\delta^\text{corr.}_{\eta_c}$, $\delta^\text{corr.}_{\chi_{c0}}$, $\delta_{C}$, $\delta_{NC}$, $\delta^{\prime}_{SP}$.
	\item \textbf{Other theory parameters:} $\zeta_{3 \rho}$, $\omega_{3^\rho}^A$, $\omega_{3 \rho}^V$, $\zeta_{3 \phi}$, $\omega_{3 \phi}^A$, $\omega_{3\phi}^V$,  $\omega_{3 \phi}^G$, $a_{2 \rho}^{\parallel}$, $a_{2 \phi}^{\parallel}$.
\end{itemize}

For the nuisance parameters, we adopt Gaussian priors characterized by their central values $\mu$ and uncertainties $\sigma$. The charm quark mass and strong coupling constant are taken as $m_c =1.34 \pm 0.27$ and $\alpha_s(2 m_c)=0.245 \pm 0.025$, respectively. Missing-piece corrections are included with priors $\delta^\text{corr.}_{\chi_{c0}}= 0 \pm 0.2$, $\delta^\text{corr.}_{J/\psi}= 0 \pm 0.1$, $\delta^\text{corr.}_{\eta_c}= 0 \pm 0.1$. Coulombic and non-Coulombic effects are incorporated through $\delta_C=0.266 \pm 0.027$ and $\delta_{NC}=0.493\pm 0.049$. In addition, the correction term in the E1 transition, $\delta^{\prime}_{SP}$, is treated as a nuisance parameter with the value~$\delta^\prime_\text{SP}=-0.624$~\cite{Dey:2025xdx}, to which we assign a $10\%$ uncertainty. With all these different pieces now in place, we define the $\chi^2$ function as~\footnote{All the fits and subsequent numerical analyses have been carried out using a~\textit{Mathematica}\textsuperscript{\textcopyright} package~\cite{OptEx}.}
\begin{equation}
	\chi^2=\sum_{i,j}^{\text{data}} (\mathcal{O}_i^{exp}-\mathcal{O}_i^{th})^T V_{i,j}^{-1} (\mathcal{O}_j^{exp}-\mathcal{O}_j^{th})+\chi^2_{nuis.}.
	\label{eq:chi2}
\end{equation}
Here, $\mathcal{O}_i^{th}$ represent the analytical expressions for the decay modes discussed earlier and $\mathcal{O}_i^{exp}$ are the corresponding experimental measurements presented in table~\ref{tab:radexptinpt}. Numerical information regarding nuisance parameters are included by the addition of the $\chi^2_{nuis.}$ term which is defined as: 
\begin{equation}
	\chi^2_{nuis.}=\sum_{i,j}^{\text{parameters}} (\mathcal{I}_i^{P}-\mathcal{V}_i^{P})^T \big(V^{nuis.}\big)_{i,j}^{-1} (\mathcal{I}_j^{P}-\mathcal{V}_j^{P}).
\end{equation}
Where, $\mathcal{I}_k^{P}$ and $\mathcal{V}_k^{P}$ are the $k^{th}$ input parameters and their values, respectively.


The inputs for the remaining theory parameters, along with their central values and associated uncertainties, are taken from the first enumerated item $\left(\chi_{\mathrm{c1}}\to \rho(\phi)\gamma\right)$ in  subsection~\ref{subsec:exp_th_Pwave}. In particular, this includes the twist-2 and twist-3 LCDA parameters defined in eqs.~\ref{eq:twist2_DA} and \ref{eq:twist3_DA}, respectively. The corresponding uncertainties are propagated and treated as theoretical errors, which are combined in quadrature with the experimental uncertainties and subsequently incorporated into the covariance matrix. As a result, the theory expressions depend only on the remaining free and nuisance parameters of interest. In the $\chi^2$ analysis, the parameters whose uncertainties have been propagated are fixed to their central values, while the fit is performed over the remaining parameters. Consequently, the theory expressions for the respective channels are given below:
\begin{eqnarray}
10^{6} \times \mathcal{B}(\chi_{c1} \to \rho \gamma)=&& \frac{|\psi^{' R}_{\chi_{c1}}(0)|^2 (1+\delta_C +\delta_{NC})^2}{m_c^{14} \,  \Gamma_{\chi_{c1}}}\left[4378.084 \, \alpha_s^2(2 m_c)+ 32360.534 \, m_c^2 \, \alpha_s^4(2 m_c)\right] \, ,\nn \\
10^{6} \times \mathcal{B}(\chi_{c1} \to \phi \gamma)=&& \frac{|\psi^{' R}_{\chi_{c1}}(0)|^2 (1+\delta_C +\delta_{NC})^2}{m_c^{14} \, \Gamma_{\chi_{c1}}}\left[49.349 \, \alpha_s^2(2 m_c) + 6950.854 \, m_c^2 \, \alpha_s^4(2 m_c)\right] \nn.
\end{eqnarray}  
\begin{table}[t]
	\centering
		\begin{tabular}{|c|c|c|c|}
			\hline
			\textbf{Parameters} & \textbf{Fit Results} & \textbf{Parameters} & \textbf{Fit Results}\\		
			\hline\hline
			$\psi^{'R}_{\chi_{c0}}(0)$ & $0.143(45)$  & $m_c$ & $1.404(75)$ \\
			$\psi^{'R}_{\chi_{c1}}(0)$ & $0.115(49)$ & $\alpha_s(2 m_c)$    & $0.249(24)$ \\
			$\psi^{'R}_{h_c}(0)$ & $0.122(65)$ & $\delta^\text{corr.}_{J/\psi}$ & $-0.002(53)$ \\
			$\psi^R_{J/\psi}(0)$ & $0.845(35)$ & $\delta^\text{corr.}_{\eta_c}$ & $-0.102(26)$ \\
			$\psi^R_{\eta_c}(0)$ & $1.034(26)$ & $\delta^\text{corr.}_{\chi_{c0}}$ & $0.00(20)$ \\			
			& &$\delta_{C}$         & $0.266(27)$ \\
			& &$\delta_{NC}$        & $0.493(49)$ \\
			& &$\delta^{\prime}_{SP}$ & $-0.635(35)$ \\
			\hline
			$\chi^2/$\text{dof} & $1.705/4$ & & \\
			\text{p-Value} & $0.94$ & & \\
			\hline
		\end{tabular}
	\caption{Fit results for the derivatives of the wave functions for the P wave charmonia states $\chi_{c0}$, $\chi_{c1}$, and $h_c$ are determined through a simultaneous extraction of the S wave charmonium wave function at the origin from a fit to the data given in table \ref{tab:radexptinpt}.}
	\label{tab:fitresultall}
\end{table} 
The theory error obtained for the individual decay modes are sizable: approximately  $71\%$ for $\chi_{c1} \to \rho \gamma$ and  $37\%$ for $\chi_{c1} \to \phi \gamma$. Using these theory errors together with covariance matrix, we construct the likelihood function. With this likelihood function, we simultaneously extract the unknown free parameters.

Our fit results are summarized in table~\ref{tab:fitresultall}. The parameter $\psi^{' R}_{\chi_{c0}}(0)$ carries an uncertainty of about $31\%$ while the $\psi^{' R}_{\chi_{c1}}(0)$ parameter has $43\%$ error. On the other hand, the $\psi^{' R}_{h_{c}}(0)$ parameter is less constrained, with an uncertainty of approximately $53\%$. For the S-wave states, $\psi^{R}_{J/\psi}(0)$ exhibits an uncertainty of about $4\%$ and $\psi^{R}_{\eta_c}(0)$ has $3\%$ error. Note that, the values of $\psi^R_{J/\psi}(0)$ and $\psi^R_{\eta_c}(0)$ remain consistent with our earlier determinations reported in ref.~\cite{Biswas:2023bqz}. These non-perturbative parameters could be validated once the lattice QCD estimates become available. Our determination of the $\delta^{\prime}_{SP}$ parameter yields a large negative value, i.e. $\sim -64\%$ consistent with the numerical analysis reported in Ref.~\cite{MartinezNeira:2017prz}. A more precise and robust extraction of these S and P wave charmonium wave functions will be possible as additional experimental data are available in the near future. In the next subsection onward, we will use the extracted parameters as inputs and conduct our analysis.
\subsection{Shape of the Form Factors}~\label{subsec:BctoPff_shape}
Having extracted the radial wave function, we are now in the position to compute the $B_c \to P$ wave charmonium form factors. We first evaluate the relevant form factors at $q^2=0$ and subsequently extrapolate them to the maximum physical $q^2$ region using an appropriate extrapolation technique. In this subsection, we present our estimates of these form factors and discuss their shape for the $B_c \to P$ wave charmonium transitions, which provide the basis for predicting the observables in the corresponding decay modes.
\paragraph{\underline{Form Factors at $q^2=0$:}}
\begin{table*}[t]
	\centering
	\setlength\tabcolsep{6 pt}
	\label{}
	\begin{center}
		\resizebox{1.0 \textwidth}{!}{
				\begin{tabular}{|c|c|c|c|c|c|c|}
                   \hline
		    \textbf{Decay Channel}&\textbf{Form Factor}&\multicolumn{2}{c|}{\bf NRQCD (This Work)} & 
             \textbf{PQCD} &\textbf{QCDSR} & \textbf{LFQM}\\
                \cline{3-4}
		      &  & LO+NLO & LO+NLO+RC & \cite{Dey:2025xdx} & \cite{Azizi:2009ny} & 
                \cite{Wang:2009mi}\\
				\hline
				$B_c \to \chi_{c0}$ &$ f_{0}(0)=f_{+}(0)$
				& $1.22(71)$ & $1.45(82)$ & $0.431(70)$ & $0.673(20)$ & $0.470(30)$ \\
				\hline
			\end{tabular}
		}
		\caption{Form factor estimates for $B_c \to \chi_{c0}$ at maximum recoil $(q^2=0)$ using the fit results in table~\ref{tab:fitresultall} and $\psi^R_{B_c}(0)$ from table VII of ref.~\cite{Biswas:2023bqz}. We compare our estimates with those from other theoretical models (columns 5, 6 and 7) available in the literature. Theoretical expressions in NRQCD at LO along with the corresponding relativistic corrections have been obtained from ref.~\cite{Zhu:2017lwi}, whereas the same at NLO are available in ref.~\cite{Chen:2021vmb}.}
		\label{tab:Bcchic0FFprd}
	\end{center}
\end{table*}
Using the fit results from table~\ref{tab:fitresultall} and the value for the radial wave function of the $B_c$ meson i.e. $\psi^R_{B_c}(0)= 0.916\pm 0.124$ from table VII of ref.~\cite{Biswas:2023bqz}, we provide numerical estimates for the form factors described in eqs.~\ref{eq:Bctohcchic1ff} and~\ref{eq:Bctochic0ff} at maximum recoil ($q^2=0$) within the framework of the NRQCD effective theory in table~\ref{tab:Bcchic0FFprd} (for the $B_c \to \chi_{c0}$ form factors) and table~\ref{tab:BcPwaveFFprd} (for the $B_c\to\chi_{c1},h_c$ form factors)~\footnote{The averaging procedure has been described in details in our publication regarding $B_c$ decays to the S-wave charmonia~\cite{Biswas:2023bqz}. The interested reader may look at Section 3, page 18 of the said reference.}. The analytical expressions for the form factors in the NRQCD factorization framework to LO, as well as the corresponding relativistic corrections (upto $\mathcal{O}(v^2)$) were obtained from ref.~\cite{Zhu:2017lwi}. We have exploited the QCD relation at maximum recoil between $A_{0,1,2}$ for $B_c\to \chi_{c1},h_c$ in order to extract $A_2(0)$ corresponding to these transitions. The QCD relation $f_+(q^2=0)= f_0(q^2=0)$ for the $B_c\to \chi_{c0}$ transition is also satisfied. Till date the NLO in $\alpha_s$ corrections are available only for $B_c \to \chi_{c0}$ and the corresponding analytic expression has been obtained from ref.~\cite{Chen:2021vmb}. 
\begin{table*}[t]
	\centering
	\begin{center}
		\begin{tabular}{|c|c|c|c|c|c|}
		\hline
		\textbf{Decay}&\textbf{Form}&\textbf{PQCD} &\textbf{QCDSR} &\textbf{LFQM} &\textbf{NRQCD}\\
		\textbf{Mode}&\textbf{Factor}& \cite{Dey:2025xdx} & \cite{Azizi:2009ny} & \cite{Wang:2009mi} & \textbf{(This Work)}\\
		\hline\hline
            &$V(0)$&$1.15(27)$ & $0.46(14)$ &$0.249(4)$&$0.18(11)(3)$\\
		$B_{c}\rightarrow h_{c}$&$A_{0}(0)$&$0.322(61)$ &$0.030(10)$&$0.64(10)$&$1.35(93)(21)$\\
		&$A_{1}(0)$&$0.214(53)$ &$0.084(25)$&$0.140(14)$&$0.104(67)(18)$\\
		&$A_{2}(0)$&$-0.082(36)$& $0.214(71)$&$-1.14(21)$&$-3.1(21)(5)$\\
		\hline
            &$V(0)$&$1.03(17)$ &$0.46(14)$&$1.274(71)$&$2.5(16)(4)$\\
           $B_{c} \rightarrow \chi_{c1}$&$A_{0}(0)$&$0.166(26)$ &$0.030(10)$&$0.130(10)$&$0.086(45)(16)$\\
		&$A_{1}(0)$& $0.165(34)$ &$0.085(25)$&$0.240(6)$&$0.53(30)(8)$\\
		& $A_{2}(0)$&$0.191(57)$ &$0.212(71)$&$0.531(35)$&$1.65(93)(25)$\\
		\hline
	\end{tabular}
		\caption{Form factor estimates for $B_c \to \chi_{c1}, h_c$ at maximum recoil $(q^2=0)$ using fit results from tab.~\ref{tab:fitresultall} and $\psi^R_{B_c}(0)= 0.916 \pm 0.124$ from table VII of ref.~\cite{Biswas:2023bqz}. We have compared our estimates with those from other models, namely the Covariant Light Front Quark Model (CLFQM~\cite{Wang:2009mi}), QCD Sum Rules (QCDSR~\cite{Azizi:2009ny}), and Perturbative QCD~\cite{Dey:2025xdx}. Analytical expressions for the form factors under the NRQCD framework have been obtained from ref.~\cite{Zhu:2017lwi}.}
		\label{tab:BcPwaveFFprd}
	\end{center}
\end{table*}

Since no NLO estimates are currently available for the $B_c \to \chi_{c1}(h_c)$ transition, we assign a conservative $20\%$ prior uncertainty to the LO form factor. This choice is motivated by the observation that the $B_c \to \chi_{c0}$ form factor receives NLO corrections of comparable size relative to its LO contribution. The resulting $20\%$ uncertainty is reflected in the second parenthesis of Table~\ref{tab:BcPwaveFFprd}. To obtain the total error, this contribution must be combined in the quadrature with the other source of uncertainty mentioned in the first parenthesis. In our subsequent analysis, we adopt this total error.

This additional uncertainty is intended to account conservatively for unknown higher-order perturbative effects and other missing dynamical contributions not captured at LO. In this sense, it effectively parameterise our lack of precise theoretical control while avoiding an overestimation of the predictive accuracy. Consequently, the adopted uncertainty band is sufficiently broad to accommodate potential future NLO corrections, thereby ensuring that our predictions remain robust and phenomenologically relevant even when more refined calculations become available.

The form factor prediction exhibit sizable errors at $q^2=0$, primarily due to the large uncertainties in the radial wave functions at $r=0$. For the $B_c\to h_c$ transition, the form factor error ranges between $66\%$ to $70\%$, while for $B_c \to \chi_{c1}$, they lie in the range of approximately $54\%$ to $68\%$. In the case of the $B_c \to \chi_{c0}$ decay mode, the form factor $f_+(0)$ carries an uncertainty of about $55\%$.

Having determined the form factor at $q^2=0$, we next extrapolate them to the entire physical $q^2$ region. Within the NRQCD factorization framework, the predictions are most reliable in the low $q^2$ region. Therefore, the form factor results obtained at maximum recoil are valid only at this point. To estimate the relevant observables, knowledge of the form factors in the full $q^2$ region is required. To model the shape of the form factors across the allowed kinematic range, we employ the pole-expansion method, following the prescriptions of Ref.~\cite{Beyer:1998ka, Melikhov:2000yu}. The relevant form factors $V$, $A_0$, $A_1$, $A_2$, $f_+$, $f_0$ for the $B_c \to \chi_{c0,1}(h_c)$ transitions are expressed as
\begin{equation}
f_i(q^2)=\frac{f_i(0)}{\big(1-\frac{q^2}{m_{pole}^2}\big)\big(1-\alpha_i\frac{q^2}{m_{pole}^2}+\beta \frac{q^4}{m_{pole}^4}\big)}.
\label{eq:pole_expn1}
\end{equation}
Here, the numerator corresponds to the value of the corresponding form factor at maximum recoil ($q^2=0$), available from our form factor estimates as presented in tables \ref{tab:Bcchic0FFprd} and \ref{tab:BcPwaveFFprd}. Note that the slopes of the form factor are strongly influenced by the masses of the corresponding low-lying resonances. The pole masses employed in our analysis, summarized in table~\ref{tab:polmass}, all lie below the pair-production threshold. The parameter $\alpha_i$, which represents the leading order coefficient in $q^2$, is taken to be different for each form factor. And because of less number of data, the subleading parameters $\beta$, which only affect the behavior of the form factors near the higher end of the allowed physical region, are assumed to be same to all the form factors. In eqs.~\ref{eq:pole_expn1}, $\alpha_i$ and $\beta$ are the free parameters that, in the most general case, are to be determined from fits to available lattice or other (model-dependent) form factor data. However, for $B_c$ decays to P-wave charmonium, data is scarce in general while lattice data is completely absent thus far. Therefore, in order to extract these parameters and hence the shape of the form factors we have employed a spin symmetry. Under this symmetry, an initial state (like the $B_c$ meson) decaying to different final states with the same total spin $J$ have the same values for the shape parameters ~\cite{Hernandez:2006gt,Ivanov:2005fd,Dey:2025xdx}. The difference between form factors describing the transition of any particular initial state to different final states with the same $J$ will hence stem solely from their corresponding normalization which are the values of the corresponding form factors at the maximum recoil (i.e. the numerators in eqs.~\ref{eq:pole_expn1}). 

In ref.~\cite{Biswas:2023bqz}, we have estimated the shapes of the form factor in $B_c \to J/\psi$ decays using lattice and in $B_c \to \eta_c$ decays by simultaneous use of lattice and heavy quark spin symmetry (HQSS). We use those inputs to constrain the parameters $\alpha_i$ and $\beta$ for $B_c\to J/\psi(\eta_c)$ respectively. In order to do this, we generate synthetic data points at $q^2= 6, 8, 10$ $GeV^2$ for both $B_c\to J/\psi, \eta_c$ form factors and did a log-likelihood analysis using the expression given in eqs.~\ref{eq:pole_expn1}. Their values at $q^2=0$ are taken from the results obtained in section III (table VII) of ref.~\cite{Biswas:2023bqz} and used as nuisance parameters in the fit. The corresponding results are displayed in table~\ref{tab:poleparamfit}.
\begin{table}[t]
	\centering
	\begin{tabular}{|*{4}{c|}}
		\hline 
		\textbf{Free Parameters} &  \multicolumn{1}{c|}{\bf Fit Results}  & \textbf{Nuisance Parameters}&  \multicolumn{1}{c|}{\bf Fit Results}   \\
		\hline \hline
		$\alpha_V$  &  $1.76(13)$& $V^{J/\psi}\text{(0)}$  &  $0.744(40)$\\
		$\alpha_{A_1}$  &  $1.54(13)$& $A_1^{J/\psi}\text{(0)}$  &  $0.456(15)$\\
		$\alpha_{A_2}$  &  $1.70(19)$& $A_2^{J/\psi}\text{(0)}$ & $0.400(39)$\\
		$\alpha_{A_0}$  &  $1.67(12)$& $A_0^{J/\psi}\text{(0)}$ & $0.489(18)$\\
		$\alpha_{f_+}$  &  $1.79(15)$& $f_+^{\eta_c}\text{(0)=}f_0^{\eta_c}\text{(0)}$  & $0.408(39)$\\
		$\alpha_{f_0}$  &  $1.39(19)$& &\\
		$\beta$  &  $1.30(23)$ & & \\
		\hline 
		$\text{DOF}$  &  $11$  &&\\
		$\text{p-Value}$  &  $0.39$ &&\\
		\hline
	\end{tabular}
	\caption{Fit results for the pole expansion parameters in eqs.~\ref{eq:pole_expn1} for $B_c\rightarrow J/\psi(\eta_{c})$ form factors from a fit to synthetic data points generated at $q^2= 6, 8, 10$ $GeV^2$ for each of the corresponding form factors (see ref.~\cite{Biswas:2023bqz}).}
	\label{tab:poleparamfit}
\end{table} 
To check the predictability of our fit, we regenerate the $q^2$ distributions of the $B_c \to \eta_c$ and $B_c \to J/\psi$ form factors, and find that they are consistent within $1\sigma$ with the lattice-provided form factor results.
We then use the corresponding $\alpha_i$ and $\beta$ for the $B_c\to J/\psi$ transition to estimate the $B_c \to \chi_{c1}, h_c$ form factors since all of these final state charmonia have the same total spin $J=1$. The $\alpha_i$ and $\beta$ estimates for $B_c\to\eta_c$ transitions can similarly be used to arrive at the $q^2$ dependence for the $B_c\to \chi_{c0}$ form factors since both of these are $J=0$ final states. Using the results given in table~\ref{tab:poleparamfit} along with the other results presented in tables~\ref{tab:Bcchic0FFprd} and \ref{tab:BcPwaveFFprd},  we obtain the shape of the form factors describing the $B_c\to h_c, \chi_{c0,1}$ transitions over the respective allowed $q^2$ regions. 

\paragraph{\underline{$z$-Series Expansion}:}
In order to further refine and constrain these shapes within a model-independent framework, we use the results of the above analysis as inputs to a $z$-expansion analysis. Specifically, we employ the Boyd-Caprini-Lellouch (BCL) parametrisation~\cite{Caprini} which provides a systematically improvable and model-independent framework to describe the $q^2$ dependence of hadronic form factors, based on analyticity, crossing symmetry. 

The BCL parametrisation provides a significantly more reliable description of the $q^{2}$ dependence of hadronic form factors compared to traditional pole-based expansions. In simple pole models, the functional form is largely dictated by the assumption of dominance by a few resonant states, which makes the resulting $q^{2}$ shape inherently model-dependent and limits the control over systematic uncertainties. In contrast, the BCL framework exploits the analyticity and unitarity properties of form factors in QCD by mapping the complex $q^{2}$ plane onto a small domain in a conformal variable, $z(q^{2})$. This transformation compresses the entire physical kinematic region into $|z|\lesssim 0.02$, allowing the form factor to be expressed as a rapidly convergent power series in $z$. As a result, only a few expansion coefficients are required to accurately describe the full $q^{2}$ range, leading to improved numerical stability and reduced truncation uncertainties. In this framework, the inclusion of pseudo-data points derived from our pole-based analysis enables a meaningful determination of the BCL coefficients while maintaining consistency with known theoretical inputs. This model-independent and systematically improvable framework enables a more precise and robust determination of the form factor shape, thereby reducing the uncertainty compared to pole-based parametrisation.


Using the pole-expansion results to generate pseudo-data points at $q^2=2,\,4,\,6~\mathrm{GeV}^2$ for each form factor. These synthetic data points are then used to fit the corresponding coefficients in the BCL expansion, thereby providing a controlled description of the form-factor shapes. The resulting pseudo-data are listed in tables~\ref{tab:Bc2chic0syndat} and \ref{tab:Bc2hcc1syndat}, which can be directly compared with more precise lattice determinations in the near future\footnote{Note that the currently available $\mathcal{O}(v^2)$ corrections to the form factors are positive and numerically significant ($\sim 30\%$ of the leading-order value).}.

Under the BCL parameterization, form factors are written as a polynomial series in powers of $q^2$.
\begin{table}[t]
	\centering
	\begin{tabular}{|*{4}{c|}}
		\hline 
		\textbf{Form Factors}  &  $q^2\text{=2}$  &  $q^2\text{=4}$  &  $q^2\text{=6}$ 
		\\
		\hline \hline
		$\text{$f_+$}$  & $\text{1.67(94)}$  &  $\text{1.9(11)}$  &  $\text{2.2(12)}$  \\ 
		$\text{$f_0$}$  & $\text{1.62(90)}$  &  $\text{1.80(99)}$  &  $\text{2.0(11)}$ \\ 
		\hline
	\end{tabular}
	\caption{Synthetic data for $B_c \to \chi_{c0}$ form factors at different $q^2$ (in \text{$GeV^2$})at different $q^2$ (in \text{$GeV^2$}) are generated using eq.~\ref{eq:pole_expn1} with parameters input taken from table~\ref{tab:poleparamfit}.}
	\label{tab:Bc2chic0syndat}
\end{table}
\begin{table}[t!]
	\centering
	\begin{tabular}{|*{7}{c|}}
		\hline 
		&\multicolumn{3}{c|}{$\chi _{c1}$ }&\multicolumn{3}{c|}{$h_c$ }  \\
		\cline{2-7}
		$\textbf{Form Factors}$  &  $q^2\text{=2}$  &  $q^2\text{=4}$  &  $q^2\text{=6}$ &  $q^2\text{=2}$  &  $q^2\text{=4}$  &  $q^2\text{=6}$ 
		\\
		\hline \hline
		$\text{$A_0$}$  &  $\text{0.098(54)}$ &  $\text{0.113(62)}$  &  $\text{0.130(71)}$ & $\text{1.5(11)}$ & $\text{1.8(12)}$ &  $\text{2.1(14)}$ \\ 
		$\text{$A_1$}$  &  $\text{0.59(34)}$  &  $\text{0.66(38)}$  &  $\text{0.74(43)}$  &  $\text{0.117(77)}$  &  $\text{0.131(86)}$  &  $\text{0.147(96)}$\\
		$\text{$A_2$}$  &  $\text{1.9(11)}$  &  $\text{2.1(12)}$  &  $\text{2.4(14)}$ &  $\text{-3.5(25)}$  &  $\text{-3.9(28)}$  &  $\text{-4.5(32)}$ \\
		$\text{V}$  &  $\text{2.9(19)}$  &  $\text{3.4(22)}$  &  $\text{3.9(25)}$ 
         & $\text{0.20(13)}$  &  $\text{0.23(15)}$  &  $\text{0.27(18)}$ \\
		\hline
	\end{tabular}
	\caption{Synthetic data for $B_c \to \chi_{c1}(h_c)$ transition form factors at different $q^2$ (in \text{$GeV^2$}) are generated using eq.~\ref{eq:pole_expn1} with parameters input taken from table~\ref{tab:poleparamfit}.}
	\label{tab:Bc2hcc1syndat}
\end{table}
\begin{equation}
f_i(q^2) = \frac{1}{P(q^2)} \sum_{k=0}^{n} a_i^k z^k( q^2,t_0)
\label{eq:BCLexpanP}
\end{equation}
with $i=A_0, A_1, A_2, V, +, 0$ ;
and
\begin{eqnarray}
z(q^2,t_0) &=& \frac{ \sqrt{t_+ - q^2} - \sqrt{t_+ - t_0}}{\sqrt{t_+ - q^2} + \sqrt{t_+ - t_0}}, \nonumber\\
P(q^2)&=&\prod_{m_{pole}} z(q^2, M_{pole}^2).
\label{eq:zexp}
\end{eqnarray}
In the above, $t_{-}=(m_{B_c}-m_{\chi_{c0,1}(h_c)})^2$, $t_+=(m_{B_c}+m_{D^*})^2$ and
$t_0=t_+(1-\sqrt{1-\frac{t_-}{t_+}})$. The choice of $t_0$ determines the value of $q^2$ at which $z=0$. Here we choose $t_0=t_-$. All information on the poles corresponding to each form factors have been presented in table~\ref{tab:polmass}. 

The objective here is to extract the BCL expansion coefficients $a_i$'s. To determine these coefficient, we performed a log-likelihood analysis. Using eq.~\ref{eq:BCLexpanP} together with the synthetic data presented in tables~\ref{tab:Bc2chic0syndat} and \ref{tab:Bc2hcc1syndat} for the $B_c \to \chi_{c0}$ and $B_c \to \chi_{c1}(h_c)$ transitions respectively, we construct the $\chi^2$ function. Minimization of this function yields the relevant BCL coefficients for each decay mode. The fit results are presented in appendix~\ref{app:info}, specifically in tables~\ref{tab:BchcBCLfit},~\ref{tab:Bcchic1BCLfit} and ~\ref{tab:Bcchic0RCBCLfit}. 

The convergence of the above $z$-series is verified by the hierarchy $a_0 > a_1 z(q^2) > a_2 z(q^2)^2$. In this work, we truncated the BCL expansion at linear order, i.e., up to the first power in $z$. To account for the uncertainty associated with neglected higher-order terms, we introduce an additional truncation error.  In particular, we estimate the possible contribution of the next quadratic term while neglecting relatively small higher-order terms beyond that, following the prescribed of ref.~\cite{Caprini}. This error is propagate as a systematic uncertainty in the form factors. The truncation uncertainty is defined as 
\begin{equation}\label{eq:trunc_errorP}
	\delta f_i(q^2)=\frac{a_{n+1}^\text{max}|z(q^2)^2|}{P(q^2)},
\end{equation}
where, $a_{n+1}^\text{max}$ denotes the maximum value of the coefficient $a_{n+1}$. Since we truncate at $n=1$, the relevant parameters is $a_2^\text{max}$. For this coefficient, we adopt a conservative estimate by assuming the quadratic term contributes about $50\%$ of the linear term. Although the convergence property of the $z$-series requires $a_1 z(q^2) >> a_2 z(q^2)^2$, our prescription deliberately enlarges the estimate to avoid underestimating the systematic error and thereby ensure a conservative error budget. 

We calculate the maximum possible truncation error at the physical endpoints of the kinematic region , where $|z_\text{max}|=0.011$, $0.019$ and $=0.020$ for the $B_c \to \chi_{c0}$, $B_c \to \chi_{c1}$ and $B_c \to h_c$ transitions, respectively. At $q^2_\text{max}$, we obtain the corresponding values of $a_2^\text{max}$ for each channels. Substituting these into eq.~\ref{eq:trunc_errorP}, we estimate the numerator at $|z_\text{max}|$, where the truncation error is largest. The resulting uncertainty $\delta f_i$ is found to be $\sim$ 2-5\% relative to the LO coefficient $a_0^{f_i}$ for all channels. Since the values of $|z_\text{max}|$ are small, we found that the quadratic term has only a minor effect on the form factor shape. Consequently, the systematically estimated truncation errors are negligible compared to the leading order contribution and do not affect the extracted form factor uncertainty. For this reason, we do not include them in the phenomenological results discussed later in this section.

\begin{figure}[t!]
	\centering
	\includegraphics[scale=0.35]{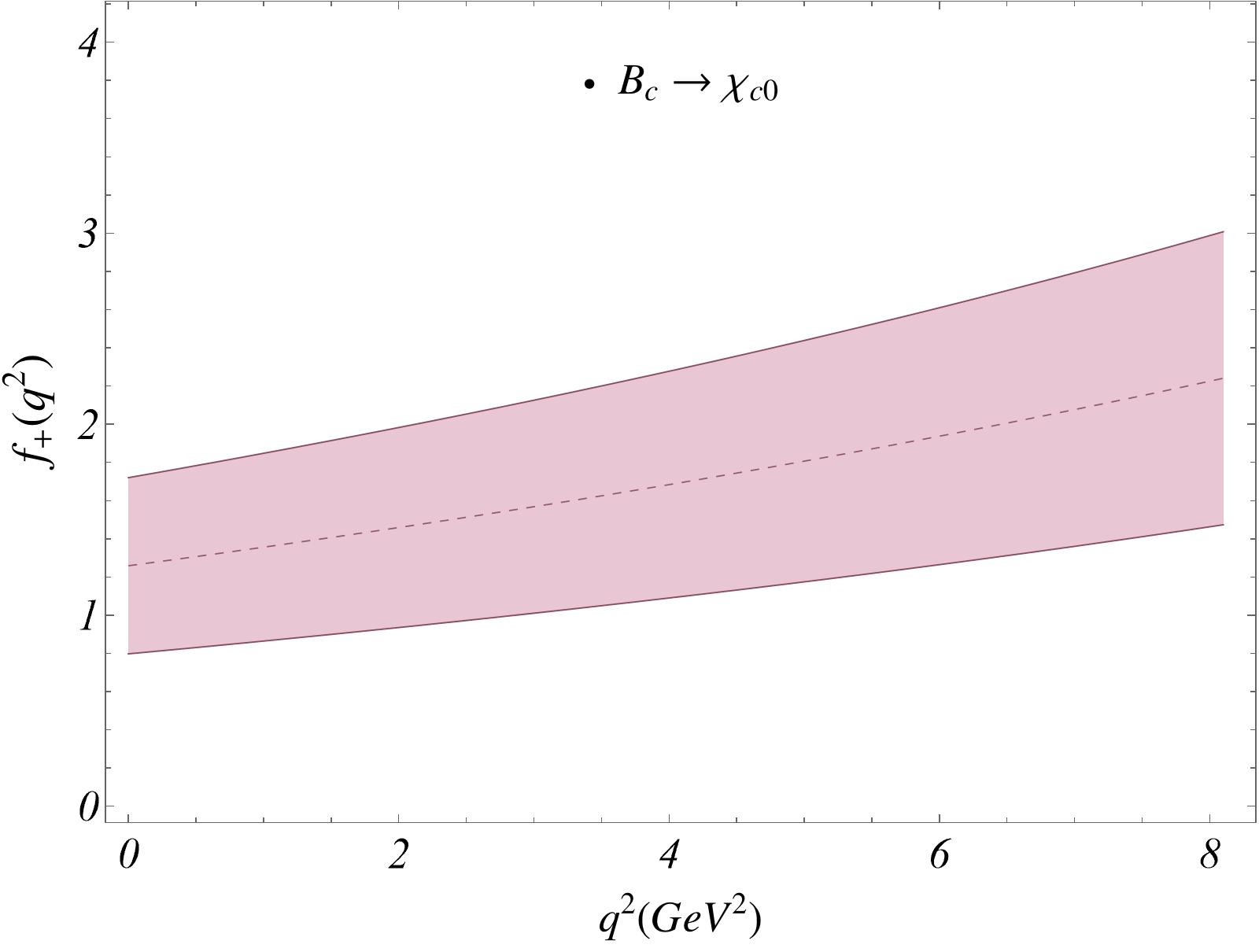}
    \includegraphics[scale=0.35]{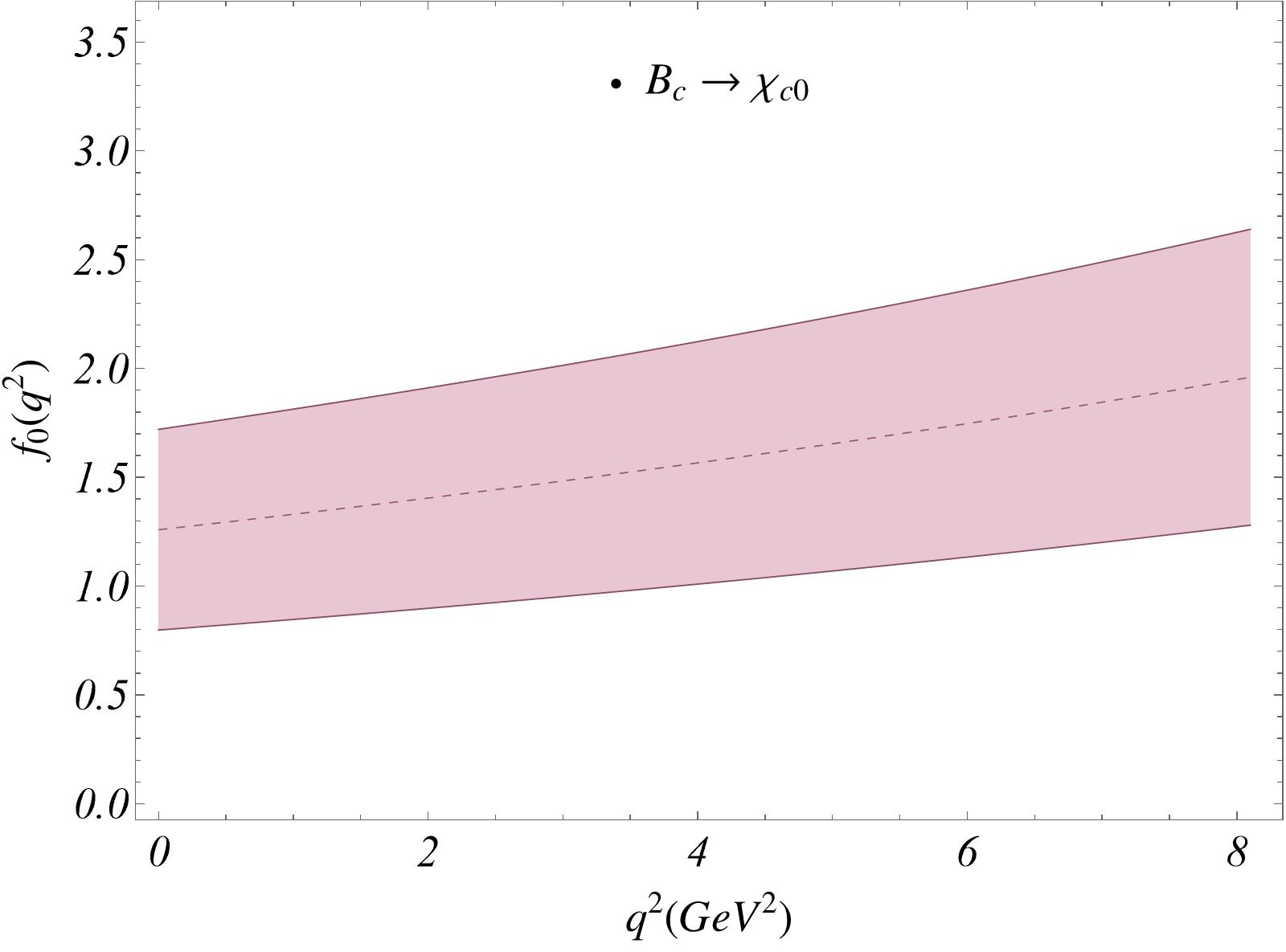}
	\caption{$1\sigma$ bands for the $B_c\to\chi_{c0}$ form factors under the BCL parametrization using the fit results from tables~\ref{tab:Bcchic0RCBCLfit}.}
	\label{fig:ffchic0}
\end{figure}

\begin{figure}[t!]
	\centering
    \includegraphics[scale=0.35]{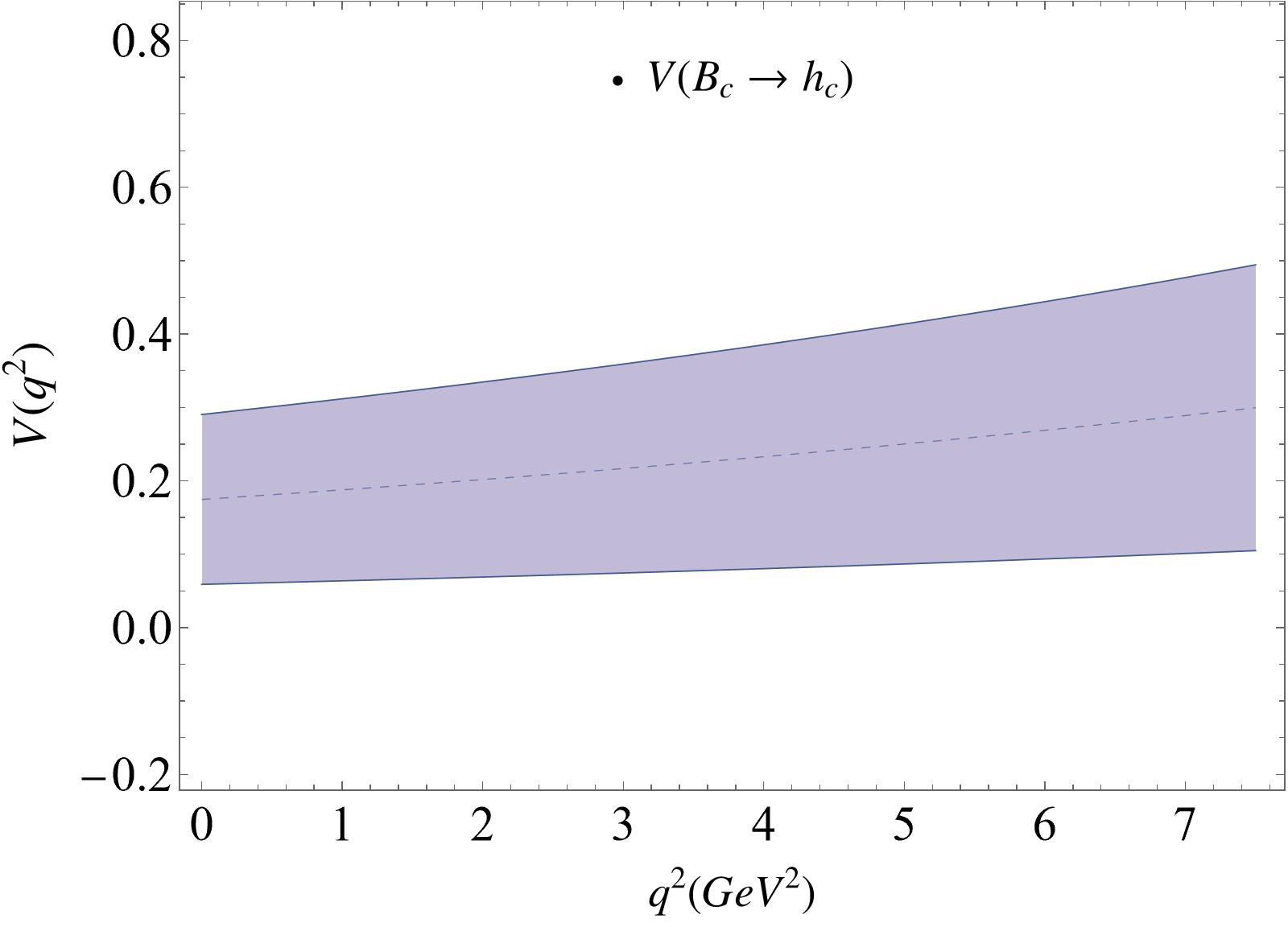}
	\includegraphics[scale=0.35]{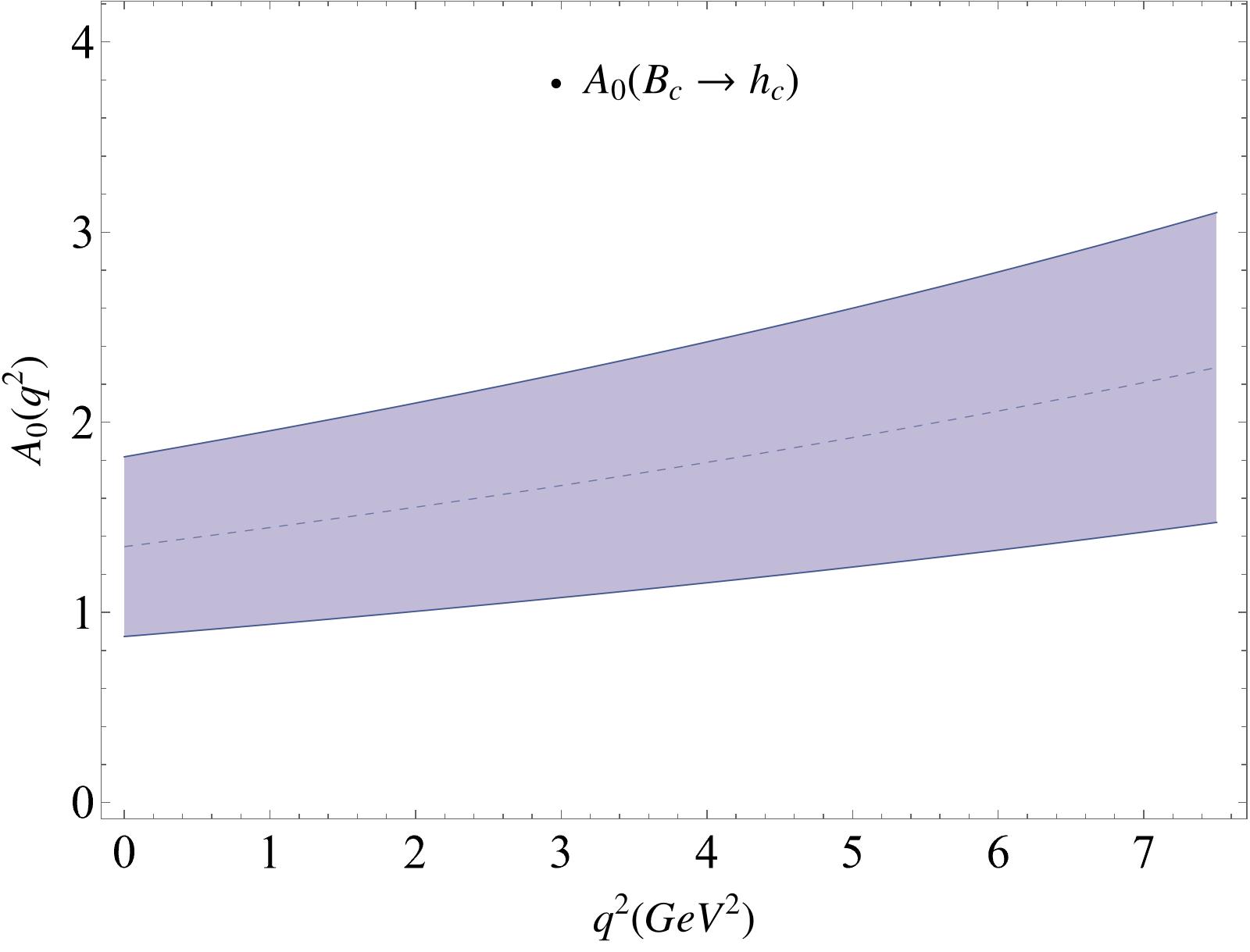}\\
	\includegraphics[scale=0.35]{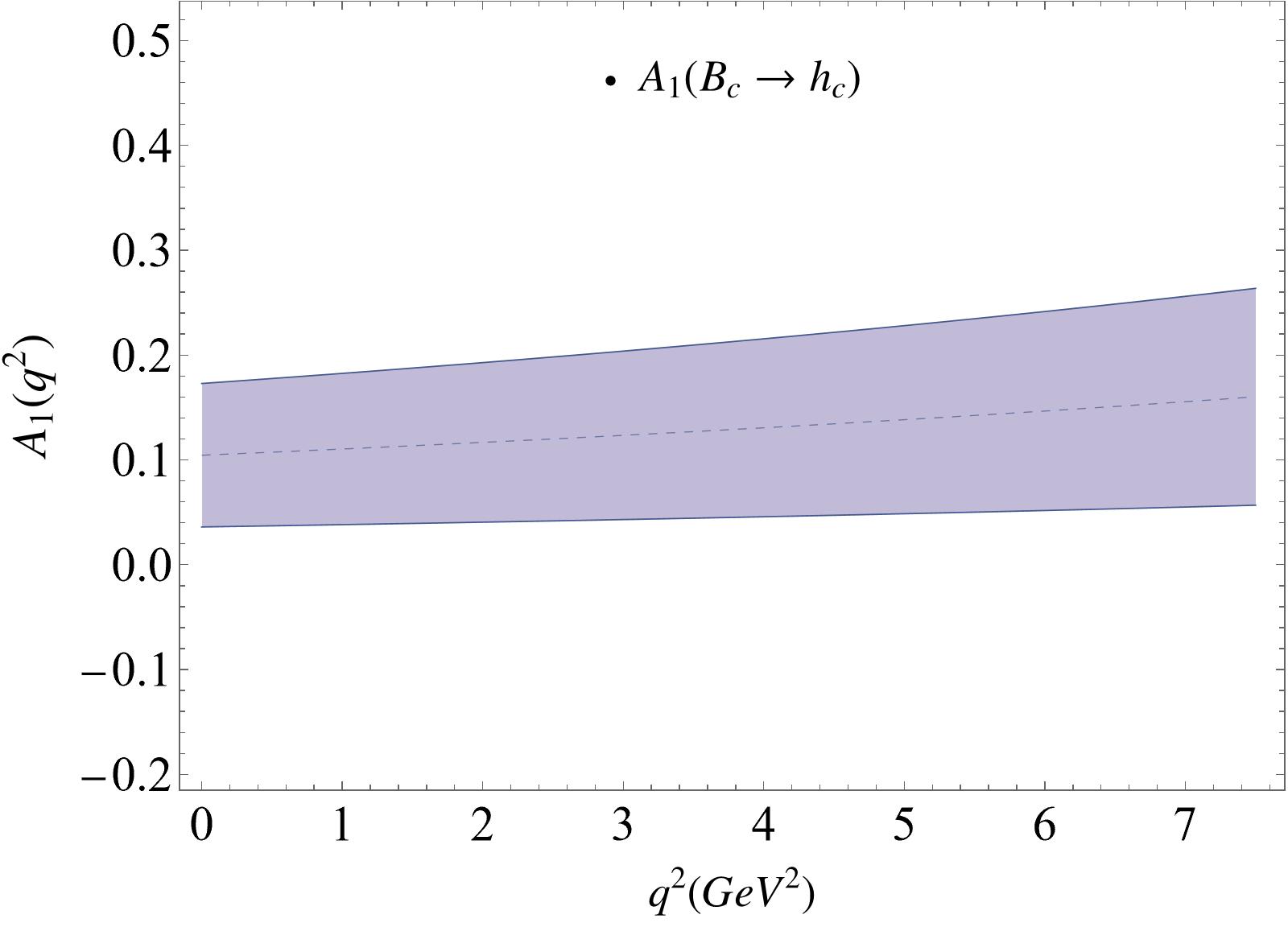}
	\includegraphics[scale=0.35]{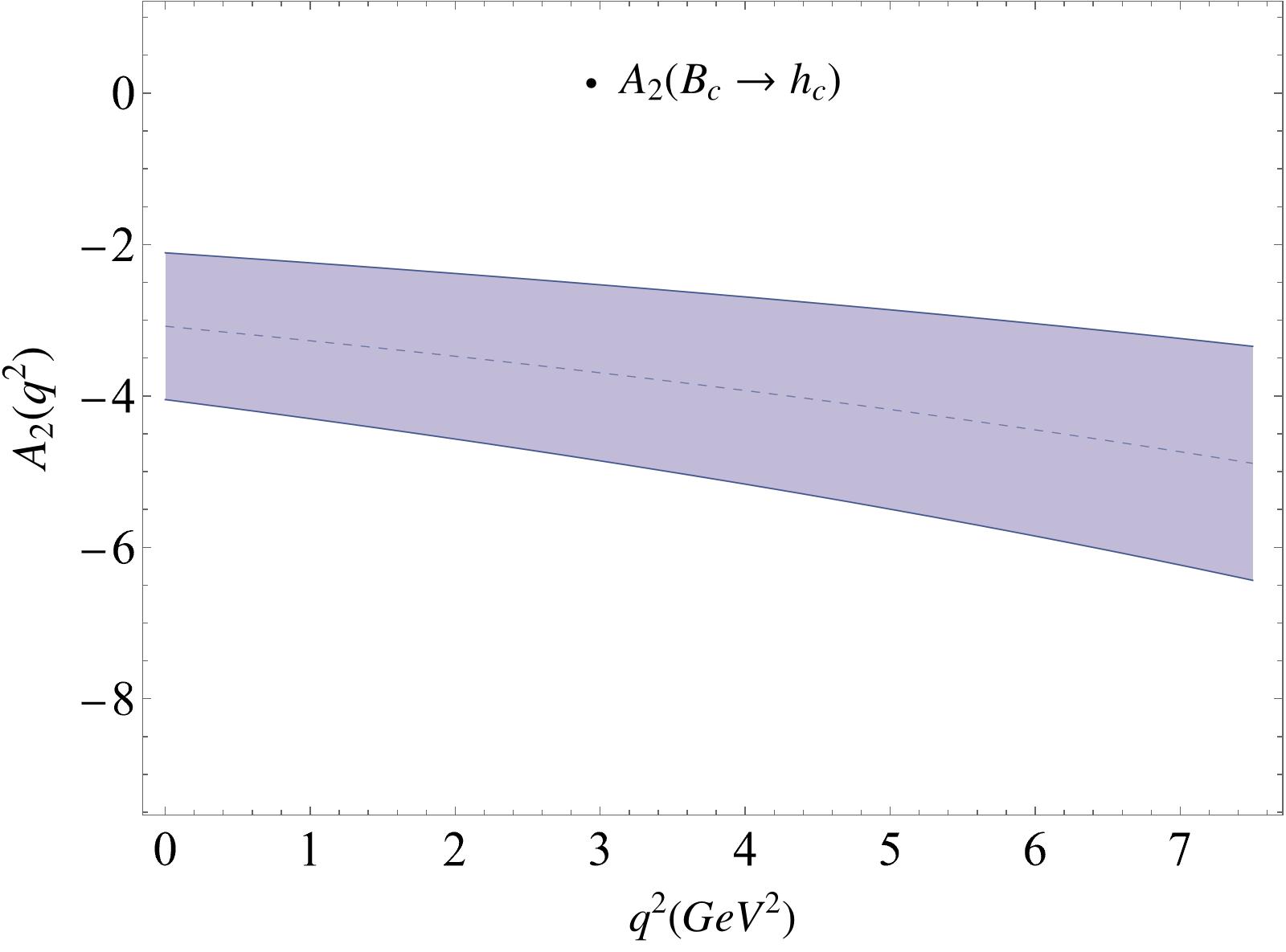}
	\caption{1$\sigma$ bands for the $B_c \to h_c$ form factors under the BCL parametrization using the fit results provided in table~\ref{tab:BchcBCLfit}.}
	\label{fig:ffBchc}
\end{figure}

\begin{figure}[t]
	\centering
        \includegraphics[scale=0.35]{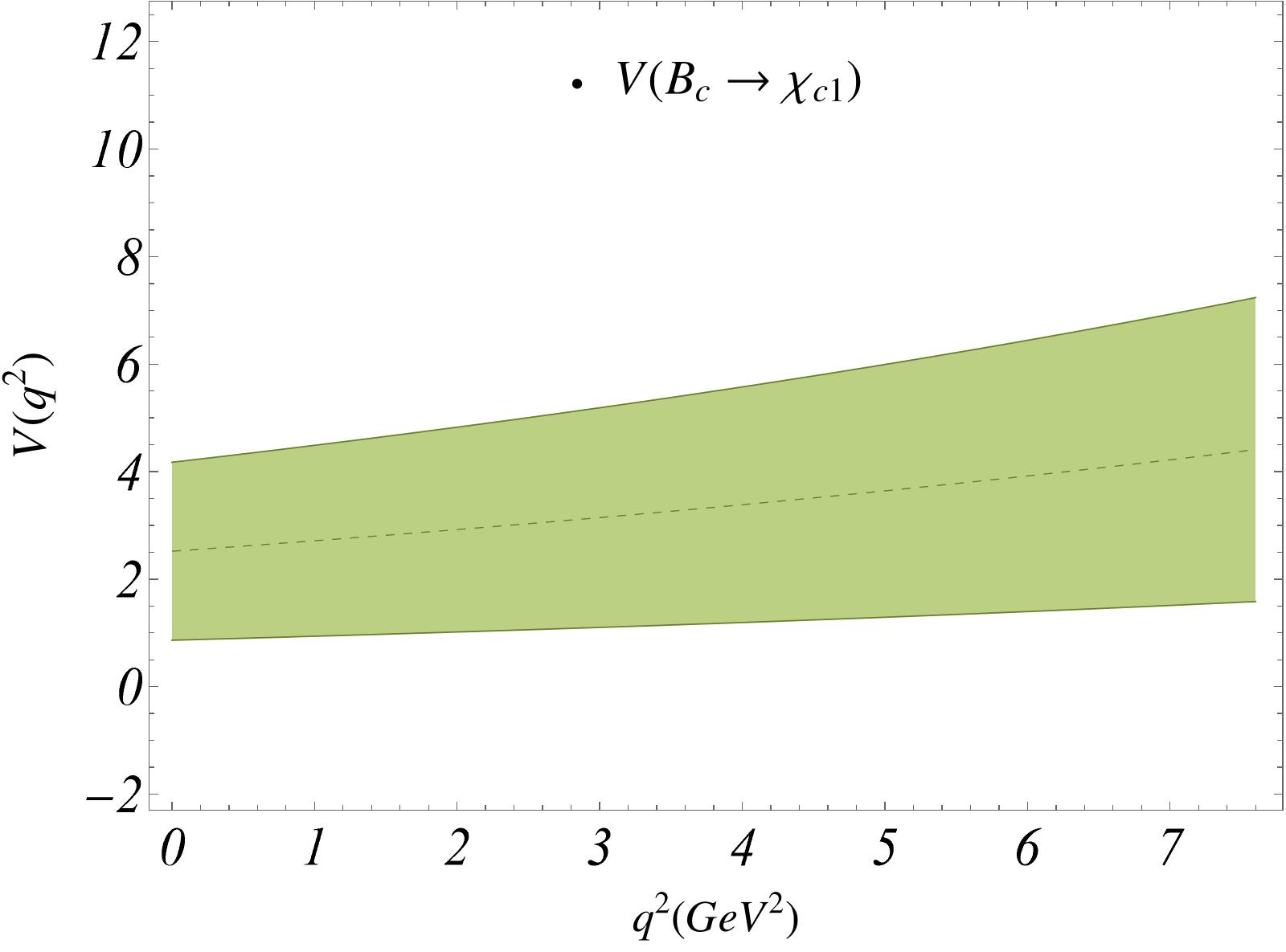}
	\includegraphics[scale=0.35]{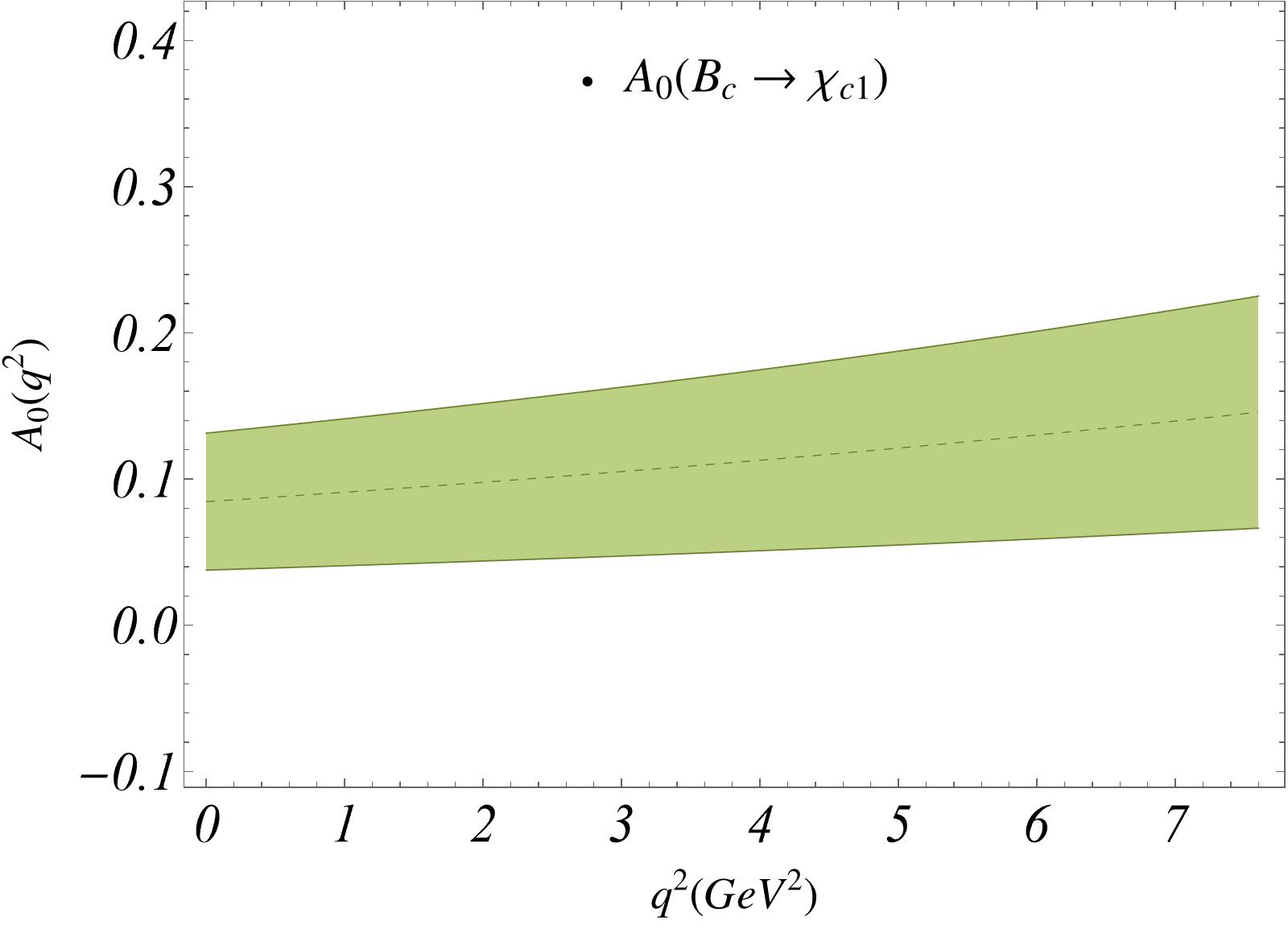}\\
	\includegraphics[scale=0.35]{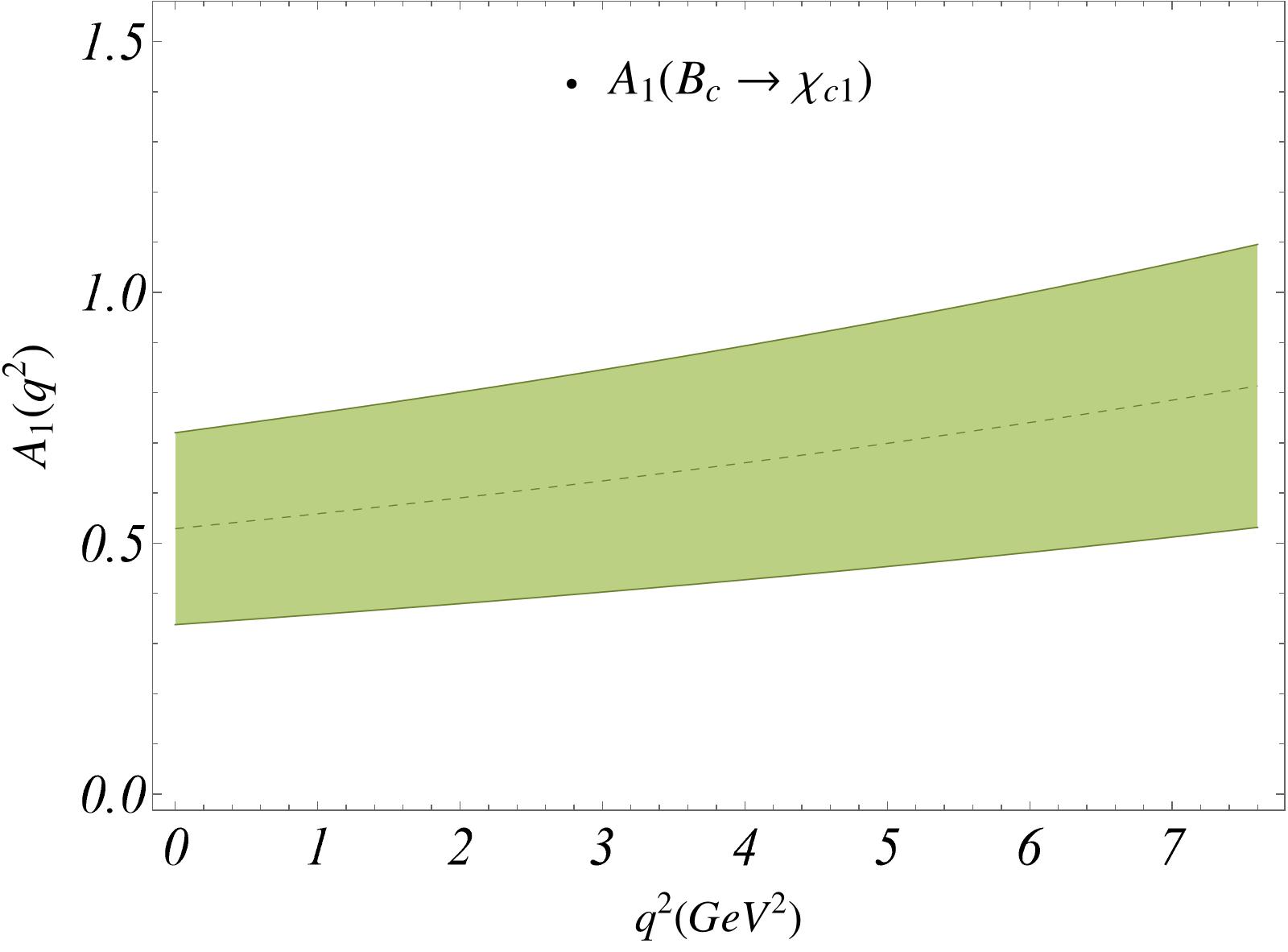}
	\includegraphics[scale=0.35]{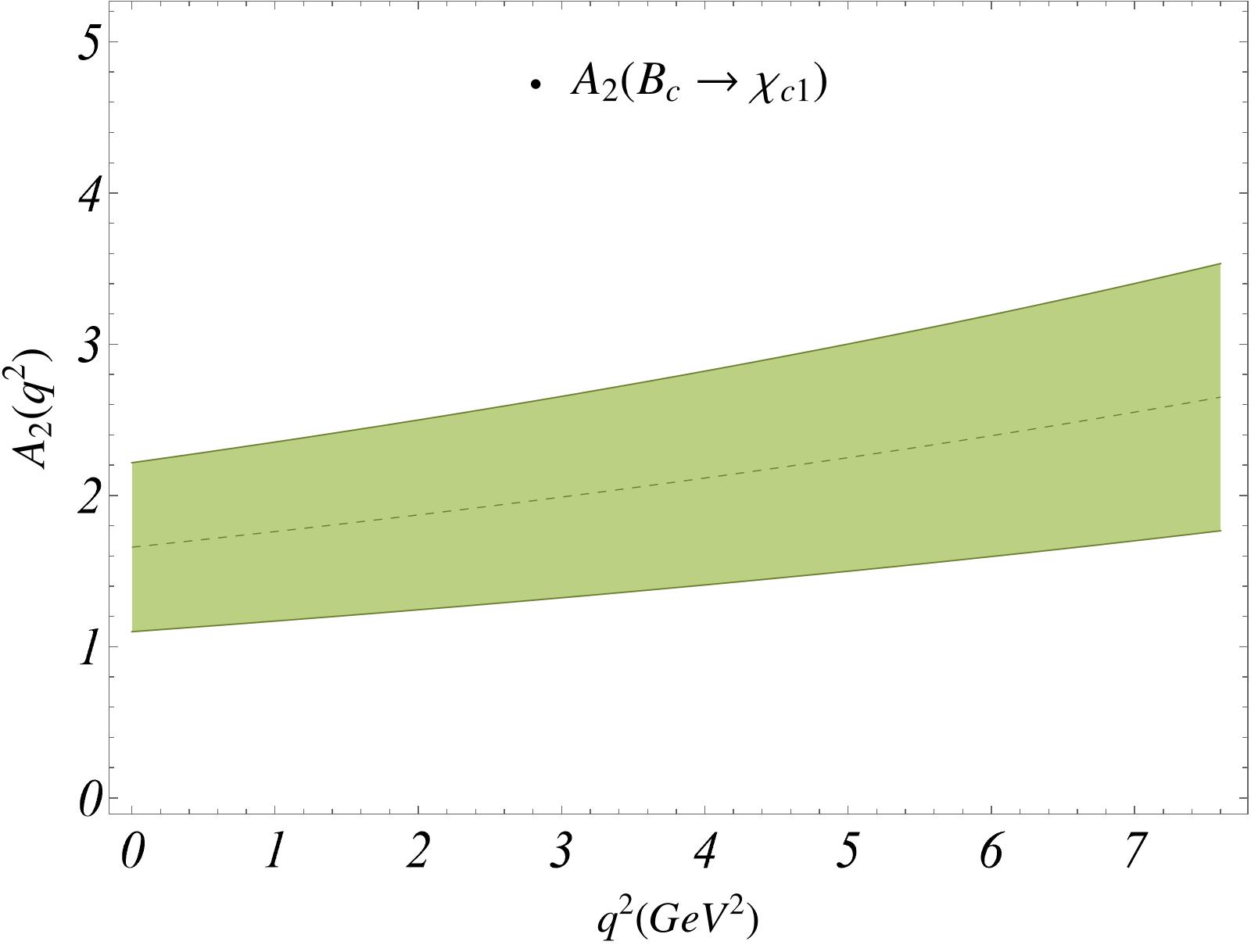}
	\caption{1$\sigma$ bands for the $B_c\to\chi_{c1}$ form factors under the BCL parametrization using the fit results provided in table~\ref{tab:Bcchic1BCLfit}.}
	\label{fig:ffBcchic1}
\end{figure}

 From the fitted coefficients including their central values, uncertainties and correlations, we obtain the shape of the form factors. The $q^2$ dependence of the $B_c \to \chi_{c0}$ form factors $f_{+,0}(q^2)$ under the BCL parametrization is displayed in Fig.~\ref{fig:ffchic0}. And the respective BCL parametrized form factors for the $B_c\to h_c$ and $B_c\to \chi_{c1}$ transitions are shown in figs.~\ref{fig:ffBchc} and~\ref{fig:ffBcchic1}. 

As discussed earlier in this section, the physical observables are intrinsically dependent on the transition form factors relevant to the decay processes. Once the form factors are determined across the full physical $q^2$ range, we can provide quantitative estimates of the associated observables. This extrapolation from maximum recoil to the zero-recoil $q^2$ region ensures that our predictions retain all the information about the dynamics of the semileptonic decay. In the following section, we will present a detailed analysis of these results for relevant $B_c \to P$ wave charmonium decays.
\subsection{Observables Prediction}~\label{subsec:BctoP_obs}
In this section, we present predictions for the branching fractions of each decay mode. Based on the estimated shapes of the transition form factors for $B_c \to P$ wave charmonium, we provide branching fraction predictions both normalized by the factor $|V_{cb}|^2$ in table~\ref{tab:brfracpredPnorm}. 
\begin{table}[t]
	\centering
		\begin{tabular}{|*{4}{c|}}
			\hline 
			\textbf{Modes} & $\frac{1}{|V_{cb}|^2} \mathcal{B}(B_c \to \chi_{c0}\ell \bar{\nu}_{\ell})$ &  \text{$\frac{1}{|V_{cb}|^2} \mathcal{B}(B_c \to \chi_{c1} \ell \bar{\nu}_{\ell})$}&  \text{$\frac{1}{|V_{cb}|^2} \mathcal{B}(B_c \to h_c \ell \bar{\nu}_{\ell})$}\\
			\hline \hline
			$\tau$ & $4.0 \pm 2.8$ & $1.6 \pm 1.0$ & $0.94 \pm 0.71$  \\
			\cline{1-4}
			$\mu$ & $21.0 \pm 15.0$ & $11.0 \pm 7.5$ & $14.0 \pm 10.0$ \\
			\hline
		\end{tabular}
	\caption{SM predictions for normalized semileptonic Branching fraction with $|V_{cb}|$ of $B_c \to h_c, \chi_{cJ}(J=0,1)$ decay with $|V_{cb}| = 0.0403(5)$ obtained from ref.~\cite{Ray:2023xjn}.}
	\label{tab:brfracpredPnorm}
\end{table}
The decay rates exhibit substantial uncertainties, reflecting the large error in the transition form factors. The predicted branching fractions exhibit large relative uncertainties at the level of $\sim 60$ to 75\%, which are largely independent of the lepton flavour and primarily driven by poorly constrained hadronic form factors. The slightly larger uncertainties in the $h_c$ channel further reflect the limited theoretical control over the corresponding transition amplitudes. These sizable errors limit the precision reach of absolute predictions and motivate the use of ratios or global analyses where hadronic uncertainties partially cancel.
Armed with the functional dependencies governing the shape of the form factors over the full kinematic region, we provide numerical estimates for LFU ratios, is defined as
\begin{equation}
R(H)=\frac{\mathcal{B}(B_c\rightarrow H \tau \bar{\nu}_{\tau})}{\mathcal{B}(B_c\rightarrow H \mu \bar{\nu}_{\mu})},
\label{eq:RH}
\end{equation}
which compare the decay rates into $\tau$ and $\mu$ mode in the final sate. These ratios constitute theoretically clean probes of lepton flavour universality, as a large part of the hadronic uncertainties cancels between the numerator and denominator. This cancellation arises from the strong positive correlations between $\mathcal{B}(B_c \to H \tau \nu)$ and $\mathcal{B}(B_c \to H \mu \nu)$, since both observables depend on the same underlying form factors and input parameters. Consequently, correlated variations tend to shift the two branching fractions in the same direction, leading to a significant reduction in the propagated uncertainty of the ratio. The observables $R(H)$ are therefore primarily sensitive to lepton-mass effects and provide robust probes of possible new physics contributions. Their numerical predictions offer complementary tests of the SM in the heavy-quark sector, while any significant deviation from the expected values could signal the presence of physics beyond the SM. The corresponding estimates are presented in Table~\ref{tab:ObsRPwave}.
\begin{table}[h!]
	\centering
	\begin{tabular}{|c|c|c|c|}
		\hline
		\textbf{Observables} &\multicolumn{1}{c|}{\bf Our Prediction} & \textbf{~z-series+NRQCD}\cite{Wang:2018duy}& \textbf{pQCD}\cite{Dey:2025xdx}\\		
		\hline\hline
		$R(\chi _{c0})$  & $0.185(3)$ & $0.110(10)$ & $0.169(11)$ \\
		$R(\chi_{c1})$  &  $0.147(26)$ &  $0.100(10)$ & $0.126(2)$ \\
		$R(h_c)$ &  $0.068(2)$ & $0.06 ^{+0.03}_{-0.01}$ & $0.113(3)$\\
		\hline 
	\end{tabular}
	\caption{Our predictions for the observable $R(H)$ defined in eq.~\ref{eq:RH} with $H=\chi_{c0}$, $\chi_{c1}$, $h_{c}$ and compared to the predictions from other works (refs.~\cite{Wang:2018duy, Dey:2025xdx}).}
	\label{tab:ObsRPwave}
\end{table}

Consequently, $R(\chi_{c0})$ and $R(h_c)$ are predicted with percent level precision, making them theoretically clean observables. On the other hand, $R(\chi_{c1})$ retains a relatively larger uncertainty, reflecting an enhanced sensitivity to form-factor shapes and helicity structures in the axial-vector channel, where the cancellation of uncertainties is less effective. This hierarchy highlights the role of LFU ratios as precision probes of new physics, while also indicating residual theoretical limitations in channels with more involved dynamical structures.

We have compared our predictions to other model-dependent results reported in refs.~\cite{Chang:2001pm, Azizi:2009ny, Wang:2009mi, Zhu:2017lwi, Rui:2018kqr} and the predictions based on the model-independent unitarity bounds on the form factors and the HQET relations between them~\cite{Wang:2018duy}. The authors of ref.~\cite{Wang:2018duy} use $z$-expansion to extrapolate the form factors to the whole kinematically allowed region. However, they truncate the series at the first order in the expansion parameter. Therefore, their predictions for the form factors suffer from large, unconsidered truncation errors. Note that the estimates for the derivatives of the wave functions for the P wave charmonia are obtained from different QCD models in model-dependent analyses~\cite{Eichten:1995ch, Eichten:1978tg, Eichten:2019hbb}, whereas our predictions are purely data driven. All these predictions can be tested in future experiments.
\section{Exclusive Non-Leptonic Decays of $B_c \to h_c(\chi_{c0,1})$}~\label{sec:BctoP_nonlep}
Non-leptonic decays of the heavy mesons are instrumental in understanding the nature of low-energy QCD. The non-leptonic transitions of the $B_c$ meson can be classified into three separate categories: 
\begin{itemize}
	\item decays of the $\bar{b}$ quark with the c quark as the spectator 
	\item c quark decays with the $\bar{b}$ quark as spectator and 
	\item annihilation of the constituent quarks c and $\bar{b}$ into non-leptonic final states. 
\end{itemize}

We study exclusive non-leptonic two-body $B_c$ decays under the factorization approximation with the leading order non-factorizable corrections calculated in NRQCD \cite{Qiao:2012hp,Zhu:2017lqu,Zhu:2017lwi,Chen:2021vmb}. A few such non-leptonic decays have been measured experimentally~\cite{Zyla:2020zbs}. In evaluating the BRs for non-leptonic decays of the $B_c$ meson, we have used the transition form factors evaluated in the NRQCD effective theory \cite{Qiao:2012hp, Zhu:2017lqu, Zhu:2017lwi,Chen:2021vmb} and the corresponding relevant mesonic decay constant from PDG~\cite{Zyla:2020zbs}. Our estimate can be compared with BR predictions to those from other theoretical/phenomenological studies from refs.~\cite{Chang:2001pm,Ivanov:2006ni,Hernandez:2006gt,Ebert:2010zu,Rui:2017pre,Losacco:2023uvp,Wang:2011jt}.
\begin{figure}[t]
	\centering
	\begin{tikzpicture}[baseline={(current bounding box.center)}]
	\begin{feynman}
	\vertex(a);
	\vertex[right=2cm of a](a1);
	\node[right=2cm of a,  dot, style=black];
	\node[above right=1cm and 1.55cm of a1, dot, style=black];
	\vertex[right= 2cm of a1] (a2);
	\vertex[above right=1cm and 1.55cm of a1] (a4);
	\vertex[above right=1cm and 2.6cm of a1] (a5);           
	\vertex[above right=2cm and 2cm of a1] (a6);           
	\vertex[right=0.75cm of a] (a3);
	\vertex[below=2cm of a] (b);
	\vertex[right=4cm of b] (c1);
	\vertex[right=0.75cm of b] (c2);
	\diagram* {
		(a) -- [anti fermion ,arrow size=1pt,edge label={\(\overline{b}\)}] (a1) -- [anti fermion ,arrow size=1pt,edge label={\(\overline{c}\)}] (a2) ,(b) --[fermion ,arrow size=1pt,edge label'={\(c\)}] (c1),(a)--[fill=gray!15,plain,bend left](b),(a)--[fill=gray!15,plain,bend right,edge label'={\(B_c^+\)}](b),(a2)--[fill=gray!15,plain,bend left,edge label={\((c\overline{c})\)}](c1),(a2)--[fill=gray!15,plain,bend right](c1), (a4)[dot] --[boson ,arrow size=1pt, edge label'={\(\rm W^+\)}] (a1),(a4) [dot] --[anti fermion ,arrow size=1pt, edge label'={\(\rm \overline{q}\)}] (a5),(a6) --[anti fermion ,arrow size=1pt, edge label'={\(\rm u\)}] (a4),(a5)--[fill=gray!15,plain,bend left](a6),(a5)--[fill=gray!15,plain,bend right,edge label'={\(P,V\)}](a6),};
	\end{feynman}
	\end{tikzpicture} 
	\caption{Tree level Feynman diagram for non-leptonic $B_c \to (c\bar{c})(P, V)$ decays, where P and V stand for a light pseudo-scalar meson and a vector meson and (c$\bar{c}$) denotes a (P wave) charmonium.}
	\label{fig:feynnonlep}
\end{figure}
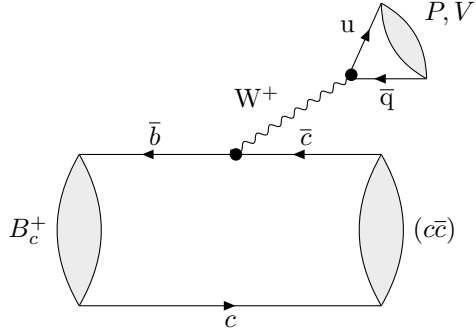 
\subsection{Formalism}
The theoretical description of non-leptonic decays involves the matrix elements of local four-fermion operators. The effective CKM favored Hamiltonian for the $b\to c \bar{u}d$ transition can be written as follows:
\begin{equation}\label{eq:heffnl}
\mathcal{H}_{eff}=\frac{G_F}{\sqrt{2}}V^*_{ud} V_{cb} \big(C_1 (\mu) Q_1(\mu)+C_2(\mu) Q_2({\mu})\big),
\end{equation}
Here, $V_{ud}$, $V_{cb}$ are the CKM matrix elements and $C_i({\mu})$s are the Wilson coefficients (WCs) which take into account the short-distance effects. The WCs $C_{1,2}(\mu)$ are evaluated perturbatively at the W scale and are then  evolved down to the renormalization scale $\mu \approx m_b$ by the renormalization group equations. The effects of soft gluons below the scale $\mu$ with the virtualities remain in the hadronic matrix elements of the local four-fermion operators $Q_i$. The four-fermion effective operators $Q_{1,2}(\mu)$ are defined as
\begin{align}
Q_1 &= \bar{d}_{\alpha}\gamma^{\mu} (1-\gamma_{5})u_{\alpha} \bar{c}_{\beta}\gamma_\mu(1-\gamma_{5})b_{\beta}, \nonumber \\
Q_2 &= \bar{d}_{\alpha}\gamma^{\mu} (1-\gamma_{5})u_{\beta} \bar{c}_{\beta}\gamma_\mu(1-\gamma_{5})b_{\alpha}, 
\end{align}
where $\alpha$ and $\beta$ are color indices, and the Einstein summation convention over repeated indices are understood. The Fierz rearrangement
\begin{eqnarray}
T^{A}_{\alpha \beta} T^{A}_{\rho \lambda}=-\frac{1}{6} \delta_{\alpha \beta} \delta_{\rho \lambda}+\frac{1}{2} \delta_{\alpha \lambda} \delta_{\rho \beta},
\end{eqnarray} 
can be used to change the above basis to 
\begin{eqnarray}
Q_0=Q_1, ~~~~Q_8=-\frac{1}{6}Q_1+\frac{1}{2}Q_2,
\end{eqnarray}
with the following WCs~\cite{Buchalla:1995vs}
\begin{eqnarray}
C_{0}=C_1+C_2/3= \frac{2}{3}C_{+}+\frac{1}{3}C_{-}\,,~~C_{8}=2 C_2 = C_{+}-C_{-}\,,
\end{eqnarray}
where
\begin{eqnarray}
C_{\pm}=\left[\frac{\alpha_s(M_W)}{\alpha_s(\mu)}\right]^{
	\frac{\gamma_\pm} {2\beta_0}}\,,~~\gamma_\pm=\pm 6\frac{N_c\mp
	1}{N_c}\,,~~\beta_0=\frac{11N_c-2n_f}{3}\,.
\end{eqnarray}
In this article, we focus only on the CKM favored processes and do not consider non-leptonic $B_c$ decays into $D^{(*)},$ $D_s^{(*)}$ mesons, as these decays are strongly CKM suppressed. We also limit our analysis to charmonium final states with light pseudoscalar/vector mesons ($\pi$, $K^{(*)}$, $\rho$). The corresponding Feynman diagram is shown in fig.~\ref{fig:feynnonlep}.
Under the naive-factorization assumption, the amplitude can be expressed as:
\begin{equation}
\langle M_{c\bar{c}} M| \mathcal{H}_{eff}|B_c\rangle \propto C_i \langle M_{c\bar{c}}| (\bar{b}c)_{V-A}|B_c\rangle \times \langle M| (\bar{q_1} q_2)_{V-A}|0\rangle \approx C_i f_M F^{B_c \to M_{c\bar{c}}},
\end{equation}
where $f_M$ and $F^{B_c \to M_{c\bar{c}}}$ are the light meson decay constant and the $B_c$ to charmonium form factor, respectively. In the above equation, the matrix elements of the weak current between the vacuum and a pseudo-scalar (P) or a vector (V)  meson have been parameterized by the decay constant $(f_P,f_V)$ defined as:
\begin{equation*}
\langle P(p_{\mu})|{(\bar{q_1} q_2)}_{V-A}|0\rangle =- i f_P p_{\mu},~~~\\\nonumber
\langle V|{(\bar{q_1} q_2)}_{V-A}|0\rangle = i f_V m_V \epsilon^*_{\mu},
\end{equation*}
where  $m_V$ and $\epsilon_{\mu}$ are the mass and polarization vector of the vector meson. The matrix element $\langle\chi_{c0}| (\bar{b}c)_{V-A}|B_c\rangle$ are defined as:
\begin{align}\label{eq:qcdffdef}
\langle M(p)|\bar{c}\gamma^{\mu}b|B_c(P)\rangle =&-i\bigg[f_{+}(q^2)(P^{\mu}+p^{\mu}-\frac{m_{B_c}^2-m_{M}^2}{q^2} q^{\mu})+f_{0}(q^2)\frac{m_{B_c}^2-m_{N}^2}{q^2} q^{\mu}\bigg],\nonumber
\end{align} 
for $B_c$ decays to $M=\chi_{c0}$, while:
\begin{align}
\langle M(p,\epsilon^*))|\bar{c}\gamma^{\mu}b|B_c(P)\rangle=&-i\bigg[2m_{M}A_0^{M}\frac{\epsilon^*\cdot q}{q^2}q^{\mu}-A_2^{M}(q^2)\frac{\epsilon^*\cdot q}{m_{B_c}+m_{M}}(P^{\mu}+p^{\mu}-\frac{m_{B_c}^2-m_{M}^2}{q^2}q^{\mu})\nonumber
\\& +(m_{B_c}+m_{M})A_1^{M}(q^2)(\epsilon^{*\mu}-\frac{\epsilon^* \cdot q}{q^2}q^{\mu})\bigg], \nn \\
\langle M(p,\epsilon^*))|\bar{c}\gamma^{\mu}\gamma^{5}b|B_c(P)\rangle =&\frac{2 V^{M}(q^2)}{m_{M}+m_{h_c}}\epsilon^{\mu \nu \rho\sigma}\epsilon^*_{\nu}p_{\rho}P_{\sigma}
\end{align}
for $M=h_c, \chi_{c1}$. In general, the matrix element between the vacuum and the meson state can be parameterized by the meson wave function and its decay constant as follows:
\begin{align}\label{eq:decaycons}
\langle 0|q_1(0) \gamma^{\mu} \gamma_5 q_2(0)|P(P)\rangle &=if_P P^{\mu}\int_{0}^{1} dx \phi_P(x),\nonumber\\
\langle 0|q_1(0) \gamma^{\mu}  q_2(0)|V(P),\epsilon_{\lambda=0}\rangle &=if_V M_V\epsilon^{\mu} \int_{0}^{1} dx \phi_{V\parallel}(x),\nonumber\\
\langle 0|q_1(0) \sigma^{\mu \nu}  q_2(0)|V(P),\epsilon_{\lambda=\pm 1}\rangle &=if_V^{\perp} \int_{0}^{1} dx (\epsilon^{\mu} P^{\nu}-\epsilon^{\nu} P^{\mu})~ \phi_{V\perp}(x).
\end{align}	
Hence, following the QCD factorization approach, the most general expression for the decay amplitude for $B_c \to M_1(c\bar{c}) M_2$ decays is given by 
\begin{equation}\label{eq:QCDF}
{\cal A}(B_c \to M_1(c\bar{c}) M_2 ) \sim F^{B \to M_1}(m_{M_2}^2) f_{M_2} \int_{0}^{1}{dx\ T_H(x)\ \phi_{M_2}}(x) + \mathcal{O}(\frac{1}{m_b}) + \mathcal{O}(v^2),
\end{equation} 
where $F^{B \to M_1}(m_{M_2}^2)$ and $f_{M_2}$ are the $B_c\to M_1$ form factor evaluated at $q
^2 = m_{M_2}^2$ ($m_{M_2}$ being the mass of the $M_2$ meson) and the decay constant for $M_2$ meson respectively. $M_1$ is the meson that takes away the spectator quark (the charmonium in our case), $\phi_{M_2}$ is the light-cone-distribution amplitude (LCDA) of the meson $M_2$ and $T_{H}$ are the perturbatively calculable hard-scattering kernels for the various operators in the effective weak Hamiltonian (eq.~\ref{eq:heffnl}). 
In the naive factorization approximation, $T_H$ is independent of $x$, and $\int_{0}^{1}{dx\ \phi_{M_2}}(x) = 1$ is obtained from the normalization. Hence, under this approximation the amplitude becomes proportional to the product of the decay constant and the transition form factor. Once the non-factorizable contributions, i.e. contributions from the diagrams involving gluon exchanges that do not belong to the form factor for the $B_c\to M_1$ transition are added, the hard scattering kernels given in eq.~\ref{eq:QCDF} can be generalized as (ref.\cite{Qiao:2012hp,Chen:2021vmb}),
\begin{table}[t]
	\renewcommand{\arraystretch}{1}
	\centering
	\setlength\tabcolsep{6pt}
	\label{}
	\begin{tabular}{ccc|c|c|}
		\hline
		\multicolumn{3}{c|}{Gegenbauer Coefficients} & Decay constant & Form factors \\
		\cline{1-3} 
		Mesons& Coefficient & Values &  (MeV) \cite{Zyla:2020zbs}  &    \\ 
		\hline
		$\pi$~\cite{RQCD:2019osh}&$a_1$&$0$ & $f_{\pi} = 130.5$ &  $f_0(m_{\pi}^2) = 1.45(83)$   \\
		&$a_2$&$0.116^{+19}_{-20}$ &  & \\
		\hline
		$K$~\cite{RQCD:2019osh}& $a_1$&$0.0525^{+31}_{-33}$ & $f_{K} = 155.72$ & $f_0(m_K^2)= 1.47(84)$  \\
		& $a_2$&$0.106^{+15}_{-16}$ & &  \\
		\hline
		&~~~$a_1^{||,\perp}$&0 &  &    \\
		$\rho$~\cite{Braun:2016wnx}&$a_2^{||}$&$0.132 \pm 0.027$ & $f_{\rho} = 221$ & $f_+(m_{\rho}^2) = 
		1.52(86)$   \\
		&$a_2^{\perp}$&$0.101\pm 0.022$ &  &  \\
		\hline
		&$a_1^{||}$&$0.03\pm 0.02$ &  & \\
		$K^*$~\cite{Ball:2007rt}&$a_1^{\perp}$ & $0.04\pm0.03$ & $f_{K^{*}} = 220$ & $f_+(m_{K^*}^2) = 1.54(87) $ \\
		&$a_2^{||}$&$0.11 \pm 0.09$ & &  \\
		&$a_2^{\perp}$&$0.10\pm 0.08$ & & \\
		\hline
		\hline 
	\end{tabular}
	\caption{Gegenbauer moments for the twist-2 distribution amplitudes of light mesons and the inputs for the decay constants used in the estimate of the branching fractions of the non-leptonic $B_c$ decays.}
	\label{tab:Gigenmoments}
\end{table}
{\small
\begin{equation}\label{eq:NLamplitude1}
\mathcal{A}(B_{c}^{+}\to \chi_{cJ}(h_{c}) P)=
\frac{G_{F}}{\sqrt{2}}V_{ud}^{*}V_{cb} \int_{0}^{1}{dx\ \phi_\pi(x)\Big[C_0(\mu)   
	\big(T_{f,0}(\mu) + T_{nf,0}(x,\mu)\big) \langle Q_{0} \rangle
	+ C_{8}(\mu)   T_{nf,8}(x,\mu) \langle Q_{8}\rangle  \Big]},
\end{equation}
}
\begin{equation}
T_{f,i}(\mu)=\sum_{k=0}^\infty (\frac{\alpha_s}{4\pi})^k
T^{(k)}_{f,i}(\mu)\,,~~~T_{nf,i}(\mu)=\sum_{k=0}^\infty (\frac{\alpha_s}{4\pi})^k
T^{(k)}_{nf,i}(\mu)\,,
\end{equation}
where $P$ is a pseudoscalar meson. In these equations, the hard kernels $T_{f}$ and $T_{nf}$ represent the perturbatively calculable factorizable and non-factorizable contributions to the decay amplitudes respectively. These can further be classified as the contributions to $\langle Q_0 \rangle$ and $\langle Q_8 \rangle$, respectively. In NRQCD, the matrix elements $\langle Q_{0} \rangle$ and $\langle Q_{8} \rangle$ for $B_c \to \chi_{c0} \pi$ decays can be written as \cite{Chen:2021vmb}, 
{\small
\begin{align}\label{eq:matrixBctochic0}
\langle Q_0 \rangle^{B_c\to \chi_{c0} \pi}=& -(M_{B_c}^2 - M_{\chi_{c0}}^2) f_\pi f^{B_c\to \chi_{c0}}_0(q^2 = M_\pi^2)          \nonumber \\  
\langle Q_8 \rangle_x^{B_c\to \chi_{c0} \pi}=&\frac{f_{\pi}}{8\pi}\frac{\psi^{\prime R}_{h_c}(0)\psi^{R}_{B_c}(0)\phi_{\pi}(x)}{\sqrt{8 m_c^3} \sqrt{m_b +m_c}}\frac{\alpha_{s}}{\pi} \times \frac{64 \pi^2}{9(2r-1)^2}\Biggl[\frac{8r^2-14r+7}{(2r-1)(r-1)(x-x_1+i\epsilon)}
+\frac{4(2r^2+3r-3)}{(2r-1)(x-x_2-i\epsilon)} \nonumber\\
&+\frac{2(r-1)}{(2r-3)(x-x_1+i\epsilon)^2}
+\frac{4(r-1)^2}{(2r-3)(x-x_2-i\epsilon)^2}\Biggr].
\end{align}
}
The same matrix elements for $B_c\to h_c(\chi_{c1}) \pi$ decays are obtained as \cite{Chen:2021vmb}\footnote{W.r.t to their S-wave counterparts $\eta_c$ and $J/\psi$, the matrix elements of $\chi_{c0}$ and $h_c(\chi_{c1})$ differ by a sign since they have opposite parity.}
{\small
\begin{align}\label{eq:matrixBctohcchic1}
\langle Q_0 \rangle^{B_c \to h_c(\chi_{c1}) \pi}=&  2 m_{h_c(\chi_{c1})} f_\pi P_{B_c}\cdot\epsilon^*_{h_c(\chi_{c1})} A_0^{B_c\to h_c(\chi_{c1})}(q^2 = M_\pi^2)\nonumber \\
\langle Q_8\rangle_x^{B_c \to  h_c \pi}=&\frac{f_{\pi}}{8\pi}\frac{\psi^{\prime R}_{h_c}(0)\psi^{R}_{B_c}(0)\phi_{\pi}(x)}{\sqrt{8 m_c^3} \sqrt{m_b +m_c}}\frac{\alpha_{s}}{\pi} \times \frac{128\sqrt{3} \pi^2}{9(2r-1)^2}\Biggl[-\frac{1}{2(r-1)(x-x_1+i\epsilon)}
+\frac{2(r-1)}{(x-x_2-i\epsilon)}\nonumber \\
&-\frac{(r-1)}{(2r-3)(x-x_1+i\epsilon)^2}
+\frac{2(r-1)^2}{(2r-3)(x-x_2-i\epsilon)^2}\Biggr],\nonumber \\
\langle Q_8 \rangle_x^{B_c \to \chi_{c1} \pi}=&\frac{f_{\pi}}{8\pi}\frac{\psi^{\prime R}_{\chi_{c1}}(0)\psi^{R}_{B_c}(0)\phi_{\pi}(x)}{\sqrt{8 m_c^3} \sqrt{m_b +m_c}}\frac{\alpha_{s}}{\pi} \times \frac{64\sqrt{6} \pi^2}{9(2r-1)^3}\Biggl[\frac{4r-3}{(r-1)(x-x_1+i\epsilon)}
-\frac{2(4r^2-6r+3)}{(x-x_2-i\epsilon)}\Biggr].
\end{align}
}
In the above, we define $P_{B_c}\cdot\epsilon^*_{ h_c(\chi_{c1})}=\frac{m_{B_c}}{m_{ h_c(\chi_{c1})}}|p|$, $|p|=\frac{\sqrt{[m_{B_c}^2-(m_{ h_c(\chi_{c1})}+m_{p})^2][m_{B_c}^2-(m_{ h_c(\chi_{c1})}-m_{p})^2]}}{2 m_{B_c}}$, $z= \frac{m_c}{m_b}$, $m_{B_c}=(m_b+m_c)$ , $m_{ h_c(\chi_{c1})}= m_{\chi_{c0}}=2m_c$.  Note that at the tree level, the hard kernel $T_{f,0}^0=1$ and independent of $x$. Also, at the one-loop level the $T_{f,0}^1(\mu)$ does not depend on $x$.  For these types of contributions the convolution integral in eqs.~\ref{eq:NLamplitude1} and~\ref{eq:NLamplitude2} could be simply written as $\int_{0}^{1}{dx\ \phi_\pi(x)} = 1$, the normalization condition. Also, at the lowest order for both the decay modes $T_{nf,0}^0=0$ and $T_{nf,8}^0 =1$. The detailed mathematical expressions for the $T_{f,0}^1$, $T_{nf,0}^1$ and $T_{nf,8}^1$ are provided in ref.~\cite{Qiao:2012hp}\footnote{Although in ref.~\cite{Qiao:2012hp} the corresponding expressions are provided for S-wave charmonia, the same for the P-wave charmonia discussed in this article will differ by only an overall negative sign due to the change in parity.}. However, we have not taken into account the non-factorizable contributions $T_{nf,0}^1$ and $T_{nf,8}^1$ in our analysis. The LCDA of light meson can be written as an expansion in Gegenbauer polynomials, defined as:
\begin{equation}\label{eq:LCDApseudo}
\phi_{M} (x,\mu^2)=6x(1-x) \big[ 1+ \sum_{n=1}^{\infty} a_n^M (\mu^2) C_n^{3/2} (2x-1)\big],
\end{equation}
where the Gegenbauer polynomial is given by the coefficient $C_n^{3/2}(2x -1)$ and $a_n^{M}$s are the Gegenbauer moments. 
In the asymptotic limit, the above equation can be written as $\phi_M(x) \approx 6 x (1-x)$. From the definitions given in eq.~\ref{eq:decaycons} one can work out and verify that only the parallel component of the wave functions contributes to the decay amplitude in $B_c \to \chi_{c0} V(V=\rho, K^*)$ decays. The transition amplitude for $B_c \to \chi_{c0} V$ will hence have only the longitudinal components $\mathcal{A}_{||}$ which can be expressed as 
{\small
\begin{align}\label{eq:NLamplitude2}
&\mathcal{A}_{||}(B_c \to \chi_{c0}~ V)= \\
& \frac{G_{F}}{\sqrt{2}}V_{ud}^{*}V_{cb} \int_{0}^{1}dx\ \phi_{V,||}(x)\times\nn \bigl(C_0(\mu)   
\big(T_{f,0}(\mu) + T_{nf,0}(x,\mu)\big) \langle Q_{0} \rangle^{B_c \to \chi_{c0} V}
+ C_{8}(\mu)   T_{nf,8}(x,\mu) \langle Q_{8}\rangle^{B_c \to \chi_{c0} V}  \bigr),
\end{align}
}
with the twist-2 LCDA given by 
\begin{equation}\label{eq:LCDAvector}
\phi_{V,||} (x,\mu)=6x(1-x) \big[ 1+ \sum_{n=1}^{\infty} a_{n,||}(\mu) C_n^{3/2} (2x-1)\big].
\end{equation}
The matrix elements $\langle Q_{8} \rangle^{B_c \to h_c/\chi_{c0}/\chi_{c1} V}$ can be obtained from eq.~\ref{eq:matrixBctochic0} with the replacement $f_P \to f_V$ and $\phi_{p} \to \phi_V$\footnote{One should also keep in mind that for non-leptonic bottom decays where both the final-state hadrons are $J=1$ mesons (i.e. vectors), QCD factorization can only reliably predict the longitudinal components of the BR's since the transverse contributions are next-to-leading order in $\Lambda/m_b$ ($\Lambda$ denoting the QCD scale), and hence suffer from large power corrections. However, in the context of our current article such states are $\chi_{c1}K^*/\rho$ and $h_cK^*/\rho$, for which our predictions are at the leading order in QCD (and obviously also in the heavy-quark limit). Our predictions therefore correspond to the total BRs, summed over all polarization.}. In this analysis, we predict the BRs for the decays $B_c\to \chi_{c1},h_c(\pi,K)$ (table~\ref{tab:chi1hcPpred}), $B_c\to\chi_{c1},h_c(K^*,\rho)$ (table~\ref{tab:chi1hcVpred}) and $B_c\to\chi_{c0}(\pi,K,K^*,\rho)$ (table~\ref{tab:chic0PVpredkin}). The available inputs for the Gegenbauer moments are given in table \ref{tab:Gigenmoments}. Our predictions for these non-leptonic modes involving a P-wave charmonium can be compared with other predictions in the literature: for eg. refs.~\cite{Hernandez:2006gt, Ebert:2010zu, Rui:2017pre}.
\subsection{Results}
In the following we present predictions for the non-leptonic BR's discussed thus far in this section using the P-wave charmonium fit results presented in table~\ref{tab:fitresultall} as input and $\psi^R_{B_c}(0)$ from table~VII of ref.~\cite{Biswas:2023bqz}. We also use the BCL fit results from tables~\ref{tab:BchcBCLfit},~\ref{tab:Bcchic1BCLfit} and \ref{tab:Bcchic0RCBCLfit} for estimating the form factors at the square of the mass of the relevant light meson (i.e. $q^2$= $m^2_{\pi}$, $m^2_{\rho}$, $m^2_{K}$, $m^2_{K^*}$). The corresponding estimates have been provided in tables~\ref{tab:Gigenmoments} and \ref{tab:formfactorinp_nonlep}. Values used for the Gegenbauer moments and decay constants corresponding to each of the relevant light meson can be found in table~\ref{tab:Gigenmoments}.  \\
\begin{table}[t]
	\centering
	\resizebox{1.0\textwidth}{!}{
	\begin{tabular}{|*{6}{c|}}
		\hline
		\textbf{Modes}  &  \multicolumn{2}{c|}{\bf BR Prediction}&  \multicolumn{3}{c|}{\bf Other Methods}\\
		\cline{2-6}
		&\textbf{Naive Fact.}&$\textbf{With corrections}$ & \textbf{NRCQM}& \textbf{RQM} & \textbf{pQCD} \\
		&&\textbf{(fac. + Non-fac. (LO))}&~\cite{Hernandez:2006gt}&~\cite{Ebert:2010zu}&~\cite{Rui:2017pre} \\
		\hline\hline
		$10^{5} \times$$ BR(B_c\to \chi _{\text{c1}} \pi)$ & $1.7(19)$ &$1.6(18)$ &$0.14(1)$&$20$&$51(11)$\\
		$10^{6} \times$$ BR(B_c\to \chi _{\text{c1}} K)$  & $1.3(14)$  &$1.2(14)$&$0.11(1)$&$15$&$38(10)$\\
		\hline
		$10^3\times BR(B_c\to h_c \pi)$  &  $4.2(59)$ &$4.1(59)$&$0.53(7)$&$0.46$&$0.54(11)$\\
		$ 10^4 \times$ $BR(B_c\to h_c K)$  &  $3.2(45)$   &$3.2(45)$&$0.41(6)$&$0.35$&$0.43(8)$\\
		\hline
	\end{tabular}
}
	\caption{Our predictions for the BRs for exclusive non-leptonic two-body decays of the $B_c$ meson into $J=1$ P-wave charmonium and a light pseudo-scalar meson in the factorization approach. We use form factors from table~\ref{tab:formfactorinp_nonlep}. Values for the derivatives of the P-wave charmonium wave functions and the $B_c$ wave function have been obtained from table~\ref{tab:fitresultall} of our current article and table~VII of ref.~\cite{Biswas:2023bqz} respectively. Gegenbauer moments and decay constant corresponding to each light meson can be found in table~\ref{tab:Gigenmoments}. These predictions can be compared with the corresponding predictions from refs.~\cite{Hernandez:2006gt},\cite{Ebert:2010zu},\cite{Rui:2017pre}.}
	\label{tab:chi1hcPpred}
\end{table}
\begin{table}[t]
	\centering
	\begin{tabular}{|*{5}{c|}}
		\hline
		\textbf{Modes}  & \textbf{BR Prediction}&  \multicolumn{3}{c|}{\bf Other Methods}\\
		\cline{3-5}
		& \textbf{(Naive Fac.)}& \textbf{NRCQM}& \textbf{RQM} & \textbf{pQCD} \\
		& &~\cite{Hernandez:2006gt}&~\cite{Ebert:2010zu}&~\cite{Rui:2017pre}\\
		\hline \hline
		$10^{3}\times BR(B_c\to \chi _{\text{c1}}\rho)$  & $0.92(118)$ &$0.10(1)$&$0.15$&$2.8(5)$ \\
		$10^{3}\times BR(B_c\to \chi _{\text{c1}}K^*)$  &  $0.063(80)$ &$0.0073(7)$&$0.010$&$0.18(5)$  \\
		\hline
		$10^{3}\times BR(B_c\to h_c\rho)$ &$10.5(145)$ &$1.3(1)$&$1.0$& $2.3(4)$  \\
		$10^{3}\times BR(B_c\to h_c K^*)$ & $0.53(73)$&$0.071(8)$&$0.070$& $0.13(4)$ \\
		\hline
	\end{tabular}
	\caption{Our predictions for the BRs for exclusive non-leptonic two-body decay of the $B_c$ meson into $J=1$ P wave charmonium and a light vector meson in the naive-factorization approach. We use form factors from table~\ref{tab:formfactorinp_nonlep}. Gegenbauer moments and decay constant corresponding to each light meson can be found in tab.~\ref{tab:Gigenmoments}.}
	\label{tab:chi1hcVpred}
\end{table}
\begin{table}[t]
	\centering
	\resizebox{1.0\textwidth}{!}{
		\begin{tabular}{|*{6}{c|}}
			\hline
			\textbf{Modes}  &  \multicolumn{2}{c|}{$\textbf{BR Prediction}$}&\multicolumn{3}{c|}{$\textbf{Other Methods}$} \\
			\cline{2-6}
			&\multicolumn{1}{c|}{\bf Naive Fac.} & \multicolumn{1}{c|}{\bf With corrections}& \textbf{NRCQM}& \textbf{RQM} & \textbf{pQCD}\\
			&\multicolumn{1}{c|}{ }&\multicolumn{1}{c|}{\bf (fac. + Non-fac. (LO))}&~\cite{Hernandez:2006gt}&~\cite{Ebert:2010zu}&~\cite{Rui:2017pre} \\
			\hline \hline
			$10^{3}\times BR(B_c\to \chi _{\text{c0}}\pi)$ & $5.3(60)$ & $5.2(60)$ &\text{$0.26(3)$}&\text{0.21}&\text{1.6(4)} \\
		   $10^{3}\times BR(B_c\to \chi _{\text{c0}}\rho)$&   $15.7(179)$&$14.4(164)$&$0.67^{+0.06}_{-0.01}$&$0.58$&$5.8(13)$ \\
			$10^{3}\times BR(B_c\to \chi _{\text{c0}} K)$ &   $0.40(46)$ &$0.40(46)$  &\text{$0.020(2)$}&\text{0.016}&\text{0.12(4)}\\
			$10^{3}\times BR(B_c\to \chi _{\text{c0}}K^*)$ & $0.84(96)$& $0.75(85)$&$0.037(5)$&$0.040$&$0.33(8)$\\
			\hline
		\end{tabular}
	}
	\caption{BR predictions for exclusive non-leptonic two-body decays of the $B_c$ to the $J=0$ P-wave charmonium ($\chi_{c0}$) and a light (pseudoscalar or vector) meson considering only (a) factorizable contributions at the next-to-leading-order and (b) factorizable contributions at the next-to-leading-order and non-factorizable contributions at the leading order; both with and without relativistic effects. We use form factors from table~\ref{tab:Gigenmoments}. Charmonium and $B_c$ meson wave functions values have been obtained from table \ref{tab:fitresultall} and table~VII of ref.~\cite{Biswas:2023bqz} respectively. Gegenbauer moments and decay constant corresponding to each light meson can be found in table~\ref{tab:Gigenmoments}.}
    \label{tab:chic0PVpredkin}
\end{table}
A few comments are in order regarding our predictions for the non-leptonic decays:
\begin{itemize}
	\item Considering $B_c$ decays into $J=1$ P wave charmonia, the corresponding NLO calculations are only available for $A_0$ but not for $A_{1,2}$ and $V$. We hence refrain from presenting the (incomplete) NLO estimates in our predictions for these modes (tables~\ref{tab:chi1hcPpred} and~\ref{tab:chi1hcVpred}). However, for the $J=0$ case, NLO calculations do exist at the maximum recoil, where the QCD relation $f_+(0)=f_0(0)$ is analytically valid. Hence, we have been able to provide NLO numbers for our predictions corresponding to $\chi_{c0}$ in the final state in table~\ref{tab:chic0PVpredkin}.
	\item The LO+ NLO predictions with and without the relativistic corrections presented in table~\ref{tab:chic0PVpredkin} under `With correction (fact.+Non-fac.(LO))' include the non-factorizable corrections at LO via the $T_{nf,8}$ kernel. As mentioned previously, we do not take non-factorizable corrections into account at the NLO level. However, we do incorporate them at LO.
	\item  Since the relativistic corrections that are available in the literature have been computed only by one group (ref.~\cite{Zhu:2017lwi}), the robustness of the results require cross checks and validation. As such, in table~\ref{tab:chic0PVpredkin} we provide our results for the NLO factorizable corrections with and without the inclusion of the relativistic corrections, both considering and without considering the contributions from LO non-factorizable diagrams. Interestingly, LHCb measured $\frac{\sigma(B_c^+)}{\sigma(B^+)} \times \mathcal{B}(B_c^+ \to \chi_{c0} \pi^+)=9.8^{+3.4}_{-3.0}(stat.)\pm 0.8(syst.)\times 10^{-6}$~\cite{LHCb:2016utz} in 2016 while studying the $B_c^+ \to K^+ K^- \pi^+$ decays thus reporting the first evidence for the $B_c^+ \to \chi_{c0}(\to K^+K^-)\pi^+$ transition.
\end{itemize}
\section{Predictions for P-wave Charmonium Production Modes}\label{sec:Pwaveprod_Zdecay}
In this section, using our data-driven estimates for the radial wave functions ($\psi^R_{B_c}(0)$, $\psi^R_{J/\psi}(0)$, $\psi^R_{\eta _c}(0)$ (table VII of ref.~\cite{Biswas:2023bqz}) and the derivatives of the radial wave functions, $\psi^{\prime R}_{h _c}(0)$, and $\psi^{\prime R}_{\chi _{cJ}}(0)$ for $J=0,1$ (table~\ref{tab:fitresultall}) of the $B_c$, $J/\psi$, $\eta_c$, $h_c$ ,and $\chi_{cJ}$ mesons respectively, we present an update for the BR's and production cross-sections for P-wave charmonia. 
\subsection{Z Boson Decays into Charmonium Mesons}
Theoretical sensitivity for $Z$ decays to charmonium final states is greater in general than that for $B_c$ decays to such final states since the former are cleaner because they involve fewer strong interaction effects. The corresponding (SM) predictions are hence more precise. These transitions can hence serve as a useful probe for the decay dynamics of charmonium states. A promising candidate for such decay is $Z\rightarrow J/\psi \gamma$, which has been discussed in our article dedicated to the phenomenology of the S-wave charmonia~\cite{Biswas:2023bqz}.
Exclusive radiative Z boson decays into charmonium final states have been studied under the framework of the NRQCD effective theory in refs.~\cite{Likhoded:2017jmx,Luchinsky:2017jab}. We have taken the corresponding expressions from these references and present them here: 
\begin{align}
&\Gamma(Z\rightarrow\gamma\chi_{c0})=\frac{2\alpha c_{V}^{2} e_{c}^{2} f_{\chi_{c0}}^{2}}{3 M_{Z}}\frac{(1-12 r^{2})^{2}}{(1-4 r^{2})} \\
&\Gamma(Z\rightarrow\gamma\chi_{c1})=\frac{2\alpha c_{V}^{2} e_{c}^{2} f_{\chi_{c1}}^{2}}{3 M_{Z}}\frac{(1+4 r^{2})^{2}}{(1-4 r^{2})} \\
&\Gamma(Z\rightarrow\gamma h_{c})=\frac{2\alpha c_{A}^{2} e_{c}^{2} f_{h_{c}}^{2}}{3 M_{Z}} (1-16 r^{4})\\
\end{align}
where $r=\frac{m_{c}}{M_{Z}}$. The NRQCD constants $f_{c\bar{c}}$ are defined as in ref.~\cite{Lansberg:2009xh,Chung:2021efj}
\begin{align}
f_{\chi_{c0}}&=\sqrt{\frac{9 N_c}{2 \pi m_c^3}}\psi_{\chi_{c0}}^{\prime R}(0) \times \nn\\
&\Bigl[1+ \frac{\alpha_s C_F}{\pi^2} \frac{\beta_0}{4}\Bigl(\frac{\pi^2}{4}-\frac{7}{3} \Bigr)+ \frac{\alpha^2_s C_F}{\pi^2} \frac{\beta_0}{4}\Bigl(\frac{\pi^2}{4}-\frac{7}{3} \Bigr) \ln{\frac{\mu^2_R}{m_c^2}}+ c^{(2),\text{fin}}_{\gamma\gamma}\Bigr]^{1/2},\\ 
f_{\chi_{c1}}&=\sqrt{\frac{6 N_c}{\pi m_c^3}}\psi_{\chi_{c1}}^{\prime R}(0) \times\bigl(1 + \alpha_s(\mu_R) c_a^{(1)} + \alpha_s^2(\mu_R) c_a^{(2)} \bigr), \\
f_{h_{c}}&=\sqrt{\frac{3 N_c}{2 \pi m_c^3}}\psi_{h_{c}}^{\prime R}(0),
\label{eq:fNRQCD}
\end{align}
where $c^{(2),\text{fin}}_{\gamma\gamma}$, $c_a^{(1)}$, $c_a^{(2)}$ can be found in ref.~\cite{Chung:2021efj}. As per eqs.~\ref{eq:fNRQCD}, our estimates for the decay constants are: 
\begin{equation}\label{eq:festimate}
f_{\chi_{c0}}=0.164 \pm 0.051,~~~~~~~ f_{\chi_{c1}}=0.167 \pm 0.071,~~~~~~~ f_{h_{c}}=0.144 \pm 0.077.
\end{equation} 
Using the values for the decay constants provided in eq.~\ref{eq:festimate}, we calculate predictions for the transitions listed in table~\ref{tab:zradchar} and compare them against those provided by the authors of ref.~\cite{Luchinsky:2017jab}. However, we would like to mention here that in this case, and in the case of most other processes that we provide estimates for in this section, an objective statistical comparison with the existing literature is currently not possible since most of the corresponding references do not provide estimates for the related uncertainties.
\begin{table}[t]
	\centering
	\begin{tabular}{|c|c|c|}
		\hline 
		\textbf{Decay channel}  &   \text{BR}$\times{10^{-8}}$  &  \text{BR}$\times{10^{-8}}$ \cite{Luchinsky:2017jab}  \\
		\hline \hline
		$\text{Z$\to $ }\chi _{\text{c0}} \gamma$  &  $0.124(78)$  &  $0.014$  \\
		$\text{Z$\to $}\chi _{\text{c1}} \gamma$  &  $0.13(11)$  &  $0.087$  \\
		$\text{Z$\to $}h_c \gamma$  &  $0.63(67)$  &  $0.3$  \\
		\hline
	\end{tabular}
	\caption{Predictions for the BR's for radiative decays of Z bosons into charmonium final states.}
	\label{tab:zradchar}
\end{table} 

Exclusive two-body decays of the Z boson into charmonium final states have been studied under the framework of the NRQCD effective theory in refs.~\cite{Likhoded:2017jmx, Luchinsky:2017jab}. We jot down the relevant modes and the corresponding theoretical expressions in what follows:
\begin{align}
&\Gamma(Z\to \chi_{c0} h_c) =\frac{8192 \pi  \alpha_s^2 \beta ^3 c_V^2 f_{\chi_{c0}}^2 f_{h_c}^2 r^2 \left(3-8 r^2\right)^2}{243 M_Z^3}\\ 
&\Gamma(Z\to \chi_{c1} \chi_{c1}) =\frac{1024 \pi  \alpha_s^2 \beta  c_A^2 f_{\chi_{c1}}^4 r^2 \left(32 r^4-18 r^2+1\right)^2}{243 M_Z^3}\\ 
&\Gamma(Z\to h_c h_c) =\frac{1024 \pi  \alpha_s^2 \beta ^5 c_A^2 f_{h_c}^4 r^2}{243 M_Z^3}\\
&\Gamma(Z\to \eta_c \chi_{c1}) =\frac{2048 \pi  \alpha_s^2 \beta ^3 c_A^2 f_{\eta_c}^2 f_{\chi_{c1}}^2 r^2 \left(1-4 r^2\right)^2}{27 M_Z^3}\\ 
&\Gamma(Z\to J/\psi \chi_{c1}) =\frac{8192 \pi  \alpha_s^2 \beta  c_V^2 f_{\JP}^2 f_{\chi_{c1}}^2 r^2 \left(576 r^6+104 r^4-24 r^2+1\right)}{243 M_Z^3}\\
&\Gamma(Z\to h_c J/\psi) =\frac{256 \pi  \alpha_s^2 \beta ^3 c_A^2 f_{h_c}^2 f_{\JP}^2 \left(4608 r^6-1504 r^4+128 r^2+1\right)}{243 M_Z^3}\\ 
&\Gamma(Z\to h_c \eta_c) =\frac{256 \pi  \alpha_s^2 \beta  c_V^2 f_{h_c}^2 f_{\eta_c}^2 \left(36864 r^8-8704 r^6+832 r^4-40 r^2+1\right)}{243 M_Z^3}\\
&\Gamma(Z\to h_c \chi_{c1}) =\frac{256 \pi  \alpha_s^2 \beta ^3 c_V^2 f_{h_c}^2 f_{\chi_{c1}}^2 \left(13312 r^6-3328 r^4+208 r^2+1\right)}{243 M_Z^3}\\ 
&\Gamma(Z\to J/\psi\chi_{c0}) =\frac{256 \pi  \alpha_s^2 \beta  c_V^2 f_{\JP}^2 f_{\chi_{c0}}^2 \left(36864 r^8+9728 r^6-6848 r^4+728 r^2+1\right)}{243 M_Z^3}\\ 
&\Gamma(Z\to \eta_c \chi_{c0}) =\frac{256 \pi  \alpha_s^2 \beta ^3 c_A^2 f_{\eta_c}^2 f_{\chi_{c0}}^2 \left(1-8 r^2\right)^2}{243 M_Z^3}\\
&\Gamma(Z\to \chi_{c0} \chi_{c1}) =\frac{256 \pi  \alpha_s^2 \beta  c_A^2 f_{\chi_{c0}}^2 f_{\chi_{c1}}^2 }{243 M_Z^3}(524288 r^{10}-454656 r^8+138752 r^6-17648 r^4+840 r^2+1)
\end{align}
where, $r=\frac{m_{c}}{M_{Z}}$. The NRQCD constants $f_{Q}$ are defined as in eq.~\ref{eq:fNRQCD}.

We provide numerical predictions for the BR's of Z boson decays to di-charmonium final states in the NRQCD effective theory in table~\ref{tab:zradchar}. All the decay channels discussed in this sub-section are sensitive to the radial wave function and/or the derivative of the radial wave function for the charmonium mesons and have been predicted using our fit results.
\begin{table}[t]
	\centering
	\begin{tabular}{|c|c|c|}
		\hline 
		\textbf{Decay channel}  &  \textbf{BR}$\times{10^{-12}}$\textbf{(This work)}  &  \textbf{BR}$\times{10^{-12}}$ \cite{Likhoded:2017jmx}  \\
		\hline \hline
		$Z \to \eta _c + \chi _{\text{c0}}$  &  $5.1(32)$  &  \
		$0.47$  \\
		$Z \to \eta _c  + \chi _{\text{c1}}$  &  $0.082(70)$  &  \
		$0.054$  \\
		$Z \to \eta _c +  h_c$  &  $0.60(65)$  &  $0.21$  \\
		$Z \to \text{$J/\psi$} +  \chi_{\text{c0}}$  &  $0.72(46)$  &  $0.083$  \\
		$Z \to \text{$J/\psi$} + \chi _{\text{c1}}$  &  $0.0044(38)$  &  $0.0035$  \\
		$Z \to \text{$J/\psi$} + h_c$  &  $3.2(35)$  &  $1.5$  \\
		$Z \to \chi _{\text{c0}} +  \chi _{\text{c1}}$  &  $0.76(81)$  &  $0.076$  \\
		$Z \to \chi _{\text{c0}} +  h_c$  &  $0.0046(53)$  &  \
		$0.00035$  \\
		$Z \to \chi _{\text{c1}} +  \chi _{\text{c1}}$  &  $0.00058(98)$  &  $0.00039$  \\
		$Z \to \chi _{\text{c1}} +  h_c$  &  $0.080(109)$  &  \
		$0.029$  \\
		$Z \to h_c +  h_c$  &  $0.00032(69)$  &  $0.00009$  \\
		\hline
	\end{tabular}
	\caption{Predictions for BR's of Z-decays into di-charmonium final states.}
	\label{tab:zdoublcharmonium}
\end{table}
\subsection{Double-Charmonium Production from Electron-Positron Annihilation}
We have also studied the production of double charmonia via electron-positron annihilation in the scope of the NRQCD framework in this work. The numerical results for the corresponding cross-sections are presented in table~\ref{tab:epemdoublechar}. The explicit expressions for the cross-sections corresponding to these decay channels are provided and discussed in what follows in this subsection.
%
%
\begin{itemize}
	\item \underline{$e^{+}e^{-}\rightarrow h_{c} \eta_{c} $}
\end{itemize}
In the one spin-triplet S-wave and one spin-triplet P-wave  case, the expression for the cross section for $e^+e^-\rightarrow \gamma^{*}\rightarrow H_1+H_2$ process can be written as:
\begin{eqnarray}
\label{jskc} &&\sigma(e^+(p_1)e^-(p_2)\rightarrow
H_1(p_3)+H_2(p_4))=\nonumber
\\ && \frac{2\pi\alpha^2\alpha_s^2|R_{ns}(0)|^2|R_{mp}'(0)|^2\sqrt{s^2-2s(m_3^2+m_4^2)+(m_3^2-m_4^2)^2}
}{27m_c^2s^2}\int^1_{-1}|\bar{M}|^2 d\cos\theta.
\end{eqnarray}
\begin{eqnarray}
\label{echc} |\bar{M}_{\eta_c
	h_c}|^2&=&2048(8192m_c^{10}+1024m_c^8t+1024m_c^8u+320m_c^6t^2+128m_c^6tu+320m_c^6u^2\nonumber\\&
-&16m_c^4t^3-112m_c^4t^2u-112m_c^4tu^2-16m_c^4u^3+2m_c^2t^4-4m_c^2t^3u-12m_c^2t^2u^2\nonumber\\&
-&4m_c^2tu^3+2m_c^2u^4-t^4u-3t^3u^2-3t^2u^3-tu^4)/((8m_c^2-t-u)^7m_c^2).
\end{eqnarray}

\begin{itemize}
	\item \underline{$e^{+}e^{-}\rightarrow \chi_{cJ}(J=0,1) h_{c}$}
\end{itemize}
In the two p-wave (nP and mP) case, the expression for the cross section corresponding to
$e^+e^-\rightarrow \gamma^{*} \rightarrow H_1(nP)+H_2(mP)$ is
\begin{eqnarray}
\label{jsetc} &&\sigma(e^+(p_1)e^-(p_2)\rightarrow
H_1(p_3)+H_2(p_4))=\nonumber\\
&&\frac{2\pi\alpha^2\alpha_s^2|R_{p}'(0)|^4\sqrt{s^2-2s(m_3^2+m_4^2)+(m_3^2-m_4^2)^2}
}{9m_c^2s^2}\int^1_{-1}|\bar{M}|^2 d\cos\theta.
\end{eqnarray}
For the $\chi_{cJ}h_c$ production $|\bar{M}|^2$ reads
\begin{eqnarray}
\label{kc0hc} |\bar{M}_{\chi_{c0}
	h_c}|^2&=&16384(t^2+u^2-32m_c^4)(16m_c^2 - 3t - 3u)^2/(3(8m_c^2 - t
- u)^7m_c^2),
\end{eqnarray}
\begin{eqnarray}
\label{kc1hc} |\bar{M}_{\chi_{c1} h_c}|^2&=&- 4096(8192m_c^{10} -
3072m_c^8t - 3072m_c^8u + 2048m_c^6t^2 + 3584m_c^6tu \nonumber\\
&+&2048m_c^6u^2 - 16m_c^4t^3 + 144m_c^4t^2u + 144m_c^4tu^2 -
16m_c^4u^3 - 52m_c^2t^4 \nonumber\\
&-& 128 m_c^2t^3u - 152m_c^2t^2u^2 - 128m_c^2tu^3 - 52m_c^2u^4 +
t^4u \nonumber\\&+& 3t^3u^2 + 3t^2u^3 + tu^4) /((8m_c^2 - t -
u)^7m_c^4),
\end{eqnarray}
\begin{itemize}
	\item \underline{$e^{+}e^{-}\rightarrow J/\psi \chi_{cJ}(J=0,1)$}
\end{itemize}
The cross-section for $e^+e^-\rightarrow J/\psi+\chi_{cJ}$ can be written as:
\begin{equation*}
\sigma=\frac{8}{3}\kappa\mathcal{C}_1.
\end{equation*}
The universal factor $\kappa$ is:
\begin{eqnarray}
\kappa_{J/\psi+\chi_{cJ}}=\frac{2\pi \alpha^{2}\alpha_{s}^2|\psi^R_{S}(0)|^2|\psi^{\prime R }_{P}(0)|^2\sqrt{s^2-16m_c^2s}}{27m_c^2s^2}
\end{eqnarray}
where $|\psi^R_{S}(0)|$ and $|\psi^{\prime R }_{P}(0)|$ are the radial wave function and its derivative at the origin respectively.

At the leading order (LO),
\begin{eqnarray}
\mathcal{C}_{1}^{J=0}&=&\frac{16r^2(144r^4+152r^3-428r^2+182r+1)}{3mc^6}\nonumber\\
\mathcal{C}_{1}^{J=1}&=&\frac{128r^3(18r^3+13r^2-12r+2)}{mc^6}
\end{eqnarray}
for J/$\psi$ +$\chi_{cJ}$. The dimensionless variable r is defined as r$\equiv$$\frac{4 m_c^2}{s}$\footnote{There is a typo in the corresponding expression for $\mathcal{C}_{1}^{J=0}$ in ref.~\cite{Sun:2021tma}. The coefficient for $r$ is taken to be 128 in the said reference while it should be 182 as we have written here. This has been confirmed via an email correspondence with the authors of the said reference.}.
\begin{table*}[t]
	\begin{center}
		\begin{tabular}{|c|ccc|}
			\hline 
			$Mode$ &  \multicolumn{3}{c|}{$\sigma$}\\
			 \hline \hline
			& $c_{L}$ & $c_{H}$ & $c$\\
			$J/\psi + \chi_{c0}$ & $-0.2147$ & $-0.4372$ & $8.1947$\\
			$J/\psi + \chi_{c1}$ & $-0.2883$ & $-0.5543$ & $0.7284$\\
			\hline 
		\end{tabular}
		\caption{The coefficients $c_{L}$, $c_{H}$, and $c$ for $e^{+}e^{-} \to J/\psi+\chi_{cJ}(J=0,1)$ at $\sqrt{s}=10.6$ GeV corresponding to $m_c$=$1.4$ GeV.}
		\label{tab:}
	\end{center}
\end{table*}
At the next to leading order (NLO), the corresponding expressions can be written in terms of their respective LO expressions as:
\begin{equation}
\sigma^{NLO}=\sigma^{LO}\big[1+\frac{\alpha_{s}}{\pi}\big(\frac{1}{2}\beta_0 \log\frac{\mu_r^2}{4 m_c^2}+c_Ln_L+c_H n_H+c\big)\big]
\end{equation}
where the dimensionless coefficients $c_L, c_H $ and c for various processes have provided in ref.~\cite{Sun:2021tma}.
The expression for the production cross-section for the associate production of P-wave charmonia has been taken from ref.~\cite{Liu:2004ga}. In our analysis we have also presented predictions for associated production cross sections of P-wave charmonia along with S-wave charmonia. The corresponding expressions have been taken from refs.~\cite{Sun:2021tma,Liu:2004ga}.
\begin{table}[t]
	\centering
	\begin{center}
		\begin{tabular}{|c|c|c|}
			\hline 
			\textbf{Decay Channel} ($\mu=2 m_c$)  &  \textbf{Cross-section (in fb) }  & \
			\textbf{Cross-Section (in fb)}  \\
            &\textbf{This Work} & \\
			\hline \hline
			$\sigma(J/\psi + \chi _{c1})$ &  $0.90(76)$  &  $1.055(426)$ \cite{Sun:2021tma}  \\
			$\sigma(J/\psi + \chi _{c0})$  &  $13.3(83)$  &  $10.81(345)$ \cite{Sun:2021tma} \\
			$\sigma(\eta _c + h_c)$  &  $1.3(15)$  &  $0.79$ \cite{Liu:2004ga}  \\
			$\sigma(\chi _{c0} + h_c)$  &  $0.28(38)$  &  $0.22$ \cite{Liu:2004ga}  \\
			$\sigma(\chi_{c1} + h_c)$  &  $0.86(146)$  &  $\text{1.0}$ \cite{Liu:2004ga}  \\
			\hline
		\end{tabular}
		\caption{Our predictions for the cross sections of double charmonia production from electron-positron annihilation (second column) and comparisons with other estimates for the same in the literature (third column) in units of femtobarns.}
		\label{tab:epemdoublechar}
	\end{center}
\end{table}
Our prediction for $\sigma(e^+ e^-$ $\rightarrow J/\psi + \chi_{c0}$) is $7.9\pm5.2$ fb at $\mu=2 m_c$. It is consistent with the corresponding estimates from both Belle ($6.4 \pm 1.7\pm 1.0$ fb~\cite{Belle:2004abn}) and BaBar ( $10.3 \pm 2.5 ^{+ 1.4}_{-1.8}$ fb~\cite{BaBar:2005nic}) within $1\sigma$.	
\subsection{Associated Production of Charmonium with a Photon from Electron-Positron Annihilation}
This subsection is dedicated to discussions and predictions for associated production of a P-wave charmonium with a photon via electron-positron annihilation. 
\begin{itemize}
	\item \underline{$e^{+}e^{-}\rightarrow \chi_{cJ}\gamma  $}
\end{itemize}
The expression for the cross-section for the production of $\chi_{cJ}$ is written as
\begin{eqnarray}
\sigma_{\chi_{c0}}&=&\frac{(4\pi\alpha)^3Q_c^4}{18\pi m_c^3 s^2}\big[c^{00} + \alpha_{s} c^{10} + (c^{02} + \alpha_{s} c^{12})\langle v^2\rangle\big]\langle0|\mathcal{O}^{\chi_{c0}}|0\rangle\\
\sigma_{\chi_{c1}}&=&\hat{\sigma}^{(0)}\big[1 + \alpha_{s} c^{10} + (c^{02} + \alpha_{s} c^{12})\langle v^2\rangle\big]\langle0|\mathcal{O}^{\chi_{c1}}|0\rangle
\end{eqnarray}
where
\begin{equation}
\hat{\sigma}^{(0)}_{\chi_{c1}}=\frac{(4\pi\alpha)^3Q_c^4(1+r)}{3\pi m_c^3 s^2(1-r)}.
\end{equation}
In the subsequent calculations for the $\chi_{cJ}$ production process, we select the fine structure constant $\alpha=1/137$. The strong-coupling constant is chosen as  $\alpha_s=0.230$ and $\langle v^2\rangle^{\chi_{cJ}}=0.20$. The asymptotic behaviour of the quantities $c^{ij}$ (called radios) in the limit of small distance (or high energy) is depicted in table~\ref{tab:asympradio}.
\begin{table}[t]
	\centering
	\begin{tabular}{c|cl}
		\hline
		\hline
		$\chi_{c0}$	&$\lim\limits_{r\rightarrow0}c^{02}_{\chi_{c0}}$&$=-\frac{13}{10}$	\\\hline
		&$\lim\limits_{r\rightarrow0}c^{10}_{\chi_{c0}}$&$=-\frac{2}{9\pi}[3(1-2\ln2)\ln r + 3\ln2(3\ln2 - 11) +\pi^2]$ \\ &&$\approx0.08\ln r+0.61$	\\\hline
		&$\lim\limits_{r\rightarrow0}c^{12}_{\chi_{c0}}$&$=\frac{1}{135\pi}[9(23-26\ln2)\ln r+3\ln2(117\ln2-443)-637+39\pi^2]$ \\ &&$\approx0.11\ln r-2.37$	\\ \hline\hline
		$\chi_{c1}$	&$\lim\limits_{r\rightarrow0}c^{02}_{\chi_{c1}}$&$=-\frac{11}{10}$	\\\hline
		&$\lim\limits_{r\rightarrow0}c^{10}_{\chi_{c1}}$&$=-\frac{2}{9\pi}[3(3-2\ln2)\ln r + 3\ln2(3\ln2 - 5)+21 +\pi^2]$ \\ &&$\approx-0.34\ln r-1.75$	\\\hline
		&$\lim\limits_{r\rightarrow0}c^{12}_{\chi_{c1}}$&$=\frac{1}{135\pi}[9(39-22\ln2)\ln r+3\ln2(99\ln2-95)-637+33\pi^2]$ \\ &&$\approx0.50\ln r+1.36$	\\ \hline\hline
	\end{tabular}
	\caption{The asymptotic behaviours of the radios in the high-energy limit.}
	\label{tab:asympradio}
\end{table}
The long-distance matrix elements (LO) are obtained from the radial wave functions at the origin in the potential model calculations
\begin{eqnarray}
\langle0|\mathcal{O}^{\chi_{c0}}|0\rangle&=&\frac{6N_c|\psi^{\prime R}_{\chi_{c0}}(0)|^2}{4\pi},
\end{eqnarray}
and
\begin{eqnarray}
\langle0|\mathcal{O}^{\chi_{cJ}}|0\rangle&=&(2J+1)(1+\mathcal{O}(v^2))\langle0|\mathcal{O}^{\chi_{c0}}|0\rangle\nonumber\\
&\approx&(2J+1)\langle0|\mathcal{O}^{\chi_{c0}}|0\rangle.
\end{eqnarray}

We provide our predictions for associated charmonium production with photon in table~\ref{tab:epemradchar}
\begin{table}[t]
	\centering
	\begin{tabular}{|*{3}{c|}}
		\hline 
		$\textbf{Decay channel}$  &  $\textbf{Cross-section(in fb)}$  &  $\textbf{Cross-section(in fb)}$  \\
        &\text{This Work}& \text{Ref no.~\cite{Xu:2014zra}}\\
		\hline \hline
		$\sigma (e^+ e^-\text{$\to $ }\chi _{\text{c0}} + \text{ $\gamma $})$  &  \
		$0.67(42)$  &  $1.60(40)$  \\
		$\sigma(e^+ e^-\to \chi _{\text{c1}} + \text{ $\gamma $})$  &  $6.6(56)$  &  $14.7(40)$  \\
		\hline
	\end{tabular}
	\caption{Our estimates for the associated production of charmonium with a photon from $e^+e^-$ annihilation (second column) and comparison with other estimates in the literature. }
	\label{tab:epemradchar}
\end{table}
\section{Conclusion}\label{sec:conclusions}
In this article we have performed a thorough phenomenological analysis of exclusive $B_c$ decays to P-wave charmonia; i.e.: the $J=1$ $h_c,\chi_{c1}$ and the $J=0$ $\chi_{c0}$ $c\bar{c}$ states. A (non-perturbative) input of prime importance for the theoretical estimation of exclusive decays are the related form factors. The shape of the form factor governs the shape of the decay width and hence the uncertainties for various related observables, both binned and integrated. Ensuring that the form factor is well behaved, at least within the kinematic range relevant for a particular set of exclusive decays is hence of utmost importance. The common practice in the phenomenological community is to obtain the high $q^2$ behaviour of the relevant form factor from lattice QCD analyses while estimating the low $q^2$ dynamics from light-cone sum rules. With the behaviour at the end points of the relevant kinematic region objectively known, one can use them for an extrapolation procedure (subject to other kinematic, QCD and analytic constraints) involving a polynomial to represent the functional behaviour of the form factor throughout the relevant kinematic range. However, this ideal situation is obviously the result of robust and extensive studies, which is not the case for $B_c$ decays to P-wave charmonia currently. Hence, in order to have an interpretation of $B_c\to\chi_{cJ}, h_c$ form factors, one must rely on theoretical models. To that end, we work within the framework of the Non-relativistic QCD (NRQCD) model.

We use the pole-expansion method described in eqs.~\ref{eq:pole_expn1} for the numerical description of the associated form factors. Under this method, the form factor is expressed as a ratio of its estimate at large recoil ($q^2=0$) to a  polynomial in $q^2$. In order to obtain the estimates at large recoil (i.e. the numerator for the form factor corresponding to a particular final state), we carry our a fit of existing experimental data on decays of P-wave charmonia (collected in table~\ref{tab:radexptinpt}) to theoretical expressions for these modes under the NRQCD framework. The denominator is a function of two parameters. Under this parameterization, the only difference between the form factors corresponding to two final states with the same $J$ stems from the numerator: the value at large recoil. In order to fix the parameters in the denominator, we use the results from our analysis of $B_c$ decays to the S-wave charmonia $J/\psi,\eta_c$ that can be obtained from ref.~\cite{Biswas:2023bqz}. We also generate pseudo data points using the pole-expansion parameters and carry out a fit using the BCL parameterization for these form factors. We provide the shapes of the form factors dictated by the BCL parametrization over the entirety of the corresponding relevant kinematic region in figs.~\ref{fig:ffBchc},~\ref{fig:ffBcchic1} and~\ref{fig:ffchic0}. Armed with the numerical estimates of these form factors, we predict values for observables sensitive to lepton-flavor universality (table~\ref{tab:ObsRPwave}).

We then take a step further and use our estimates to predict a few other non-leptonic decay modes of the $B_c$ involving a P-wave charmonium with a light pseudoscalar or vector meson in the final state. We would like to stress here that this is, to the best of our knowledge, the first (or at least one of the earliest) works that is not completely theoretical but also makes use of experimental data in order to constrain the corresponding form factors. We have also provided (hessian) uncertainties for our predictions, thus enabling them to be cross-checked in a robust, objective and statistically meaningful manner with future experimental measurements. We are hopeful that in the future, lattice and LCSR collaborations will start working on these modes enabling a model-independent treatment of these form factors and much more precise SM estimates for many related observables; thus making these decays important probes in the search for NP.
\section*{Acknowledgments}
SS acknowledges the Council of Scientific and Industrial Research (CSIR), Government of India, for the JRF fellowship grant (File No. 09/731(0173)/2019‑EMR‑I). SS further acknowledges the Anusandhan National Research Foundation (ANRF), Government of India, for financial support through research grant no. ANRF/IRG/2024/000256/PS. SS also thanks Utsab Dey and Karthik Jain for their valuable discussions.

\appendix
\section{Supplementary Numerical Information Relevant to the Analysis}~\label{app:info}
In this appendix, we present some supplemental information relevant to our analysis. In tab.~\ref{tab:polmass} we list the pole masses employed for the transition form factors of the $B_c\to\chi_{c1,0}, h_c$ decays below the $BD^*$ threshold. Based on the spin-parity information of each form factor, the corresponding pole mass is incorporated into the expansion given in eqs.~\ref{eq:pole_expn1} and \ref{eq:zexp}.  
\begin{table}[h]
	\centering
	\renewcommand{\arraystretch}{1.8}
	\centering
	\setlength\tabcolsep{30pt}
	\begin{tabular}{|*{4}{c|}}
		\hline 
		\text{$0^+$}/GeV &\text{$0^-$}/GeV &\text{$1^-$}/GeV & \text{$1^+$}/GeV\\
		\hline \hline
		\text{6.712}&\text{6.275}&\text{6.331}&\text{6.745}\\
		& \text{6.872} & \text{6.926}&\text{6.75}\\
		& \text{7.25} & \text{7.02} & \text{7.15}\\
		& &\text{7.28} & \\
		\hline
	\end{tabular}
	\caption{Pole-masses used for the pseudoscalar, vector and axial-vector form factors corresponding to $B_c\to\chi_{c1,0}, h_c$ transitions below the $BD^*$ threshold in eq.~\ref{eq:zexp}.}
	\label{tab:polmass}
\end{table} 

In table~\ref{tab:BCLcoeffwNRQCD} we presents the fit results for the BCL coefficients $a_n^i$ corresponding to the $B_c \to \eta_{c}$ transition form factors obtained from a fit to the synthetic data generated under HQSS displayed in table~\ref{tab:syndatfpf0}. We truncate the series at N=1 for $f_0$ and N=2 for $f_+$, since higher order coefficients are insensitive to the data. And the correlation between the BCL parameters presented in table~\ref{tab:corrBCLcoeff}.
\begin{table}[h!]
	\centering
	\begin{tabular}{|c|c|cccc|}
		\hline
		\text{Parameters} & \text{Fit Results} &$a_1^{f_0}$ & $a_0^{f_+}$ & $a_1^{f_+}$ & $a_2^{f_+}$ \\
		\hline
		$a_1^{f_0}$&$-2.4(11)$ & 1. & 0.944 & 0.994 & 0.867 \\
		$a_0^{f_+}$ & $0.68(9)$ & . & 1. & 0.964 & 0.841 \\
		$a_1^{f_+}$& $-7.01(79)$ & .& . & 1. & 0.836 \\
		$a_2^{f_+}$& $20(14)$ & . & .& . & 1. \\
		\hline
	\end{tabular}
	\caption{Correlations matrix between the BCL coefficients (fit with NRQCD input at $q^2=0$). Corresponding fit results are given in table \ref{tab:BCLcoeffwNRQCD}.}
	\label{tab:corrBCLcoeff}
\end{table}

Our objective is to simultaneously extract the derivatives of the wave functions for P wave charmonium wave functions, $\psi^{'R}_{\chi_{c0}}(0)$, $\psi^{'R}_{\chi_{c1}}(0)$, $\psi^{'R}_{h_c}(0)$ along with the S-wave functions $\psi^R_{J/\psi}(0)$ and $\psi^R_{\eta_{c}}(0)$. A toy analysis was performed using Gaussian priors for nuisance parameters, defined by central values $\mu$ and uncertainties $\sigma$. For the radial wave functions, we adopt $\psi^{R}_{J/\psi}(0)=0.842 \pm 0.045$ and $\psi^{R}_{\eta_c}(0)=1.029 \pm 0.037$ from our earlier study~\cite{Biswas:2023bqz}. The charm quark mass and strong coupling constant are taken as $m_c =1.34 \pm 0.27$ and $\alpha_s(2 m_c)=0.245 \pm 0.025$. Missing-piece corrections are included with priors $\delta^\text{corr.}_{\chi_{c0}}= 0 \pm 0.2$, $\delta^\text{corr.}_{J/\psi}= 0 \pm 0.1$, $\delta^\text{corr.}_{\eta_c}= 0 \pm 0.1$, together with Coulombic and non-Coulombic terms $\delta_C=0.266 \pm 0.027$ and $\delta_{NC}=0.493\pm 0.049$. The twist-2 and twist-3 LCDA parameters are included as specified in eqs.~\ref{eq:twist2_DA}, \ref{eq:twist3_DA}. The E1 transition correction, $\delta^\prime_\text{SP}=-0.624$~\cite{Dey:2025xdx}, is treated as a nuisance parameter with $10\%$ uncertainty. With these inputs, we form the likelihood function and perform a combined analysis with results summarized in the table below.
\begin{table}[t]
	\centering
		\begin{tabular}{|c|c|c|c|}
			\hline
			\textbf{Parameters} & \textbf{Fit Results} & \textbf{Parameters} & \textbf{Fit Results}\\		
			\hline\hline
			$\psi^{'R}_{\chi_{c0}}(0)$ & $0.145$  & $m_c$ & $1.407$ \\
			$\psi^{'R}_{\chi_{c1}}(0)$ & $0.124$ & $\alpha_s(2 m_c)$    & $0.250$ \\
			$\psi^{'R}_{h_c}(0)$ & $0.127$ & $\delta^\text{corr.}_{\chi_{c0}}$ & $0.00$ \\
			$\psi^R_{J/\psi}(0)$ & $0.841$ & $\delta_{C}$ & $0.266$ \\
			$\psi^R_{\eta_c}(0)$ & $1.033$ & $\delta_{NC}$ & $0.493$  \\ 			
			 $\delta^\prime_{SP}$& $-0.630$ & $\omega_{3 \phi}^A $ & $-2.601$\\
			 $\zeta_{3 \rho}$ & $0.024$ & 	$\omega_{3 \phi}^V $ &  $5.263$\\
		   	 $\omega_{3 \rho}^A $ & $-2.597$ & 	$\omega_{3 \phi}^G $ & $0.10$\\
			 $\omega_{3 \rho}^V $ & $5.367$ & 	$a_{2 \rho}^\parallel$& $0.13$\\
			 $\zeta_{3 \phi}$ &  $0.024$ & 	$a_{2 \phi}^\parallel$ & $0.176$\\
			\hline
			$\chi^2/$\text{dof} & $2.919/5$ & & \\
			\text{p-Value} & $0.71$ & & \\
			\hline
		\end{tabular}
	\caption{Fit results for the derivatives of the wave functions for the P wave charmonia states $\chi_{c0}$, $\chi_{c1}$, and $h_c$ are determined through a simultaneous extraction of the S wave charmonium wave function at the origin from a fit to the data given in table \ref{tab:radexptinpt}.}
	\label{tab:toy_analysisfitRes}
\end{table} 

Fit results for the derivatives of the wave functions for the P wave charmonia states $\chi_{c0}$, $\chi_{c1}$, and $h_c$ are determined through a simultaneous extraction of the S wave charmonium wave function at the origin from a fit to the data given in table \ref{tab:radexptinpt}.
\begin{landscape}
	\begin{table}
		\renewcommand*{\arraystretch}{1.2}
		\resizebox{1.6 \textwidth}{!}{
			\begin{tabular}{|c|c|*{13}{c}|}
				\hline
				\text{Parameters} & \text{Fit Results} & $\psi^{' R}_{\chi_{c0}}(0)$ & $\psi^{' R}_{\chi_{c1}}(0)$ & $\psi^{' R}_{h_{c}}(0)$ & $\psi^{R}_{J/\psi}(0)$ & $\psi^{R}_{\eta_c}(0)$ & $m_c$ & $\alpha_s(2 m_c)$ & $\delta^\text{corr.}_{J/\psi}$ & $\delta^\text{corr.}_{\eta_c}$ & $\delta^\text{corr.}_{\chi_{c0}}$ & $\delta_{C}$ & $\delta_{NC}$ & $\delta^\prime_{SP}$ \\
				\hline \hline
				$\psi^{' R}_{\chi_{c0}}(0)$ & $0.143(45)$ & $1.$  &  $0.358$  &  $0.211$  &  $-0.134$  &  $0.$  &  $0.354$  &  \
				$-0.201$  &  $0.219$  &  $0.13$  &  $-0.912$  &  $-0.049$  &  \
				$-0.089$  &  $-0.267$  \\
				$\psi^{' R}_{\chi_{c1}}(0)$ & $0.115(49)$ & $0.358$  &  $1.$  &  $0.54$  &  $-0.285$  &  $0.$  &  $0.794$  &  \
				$-0.481$  &  $0.473$  &  $0.278$  &  $-0.003$  &  $-0.036$  &  \
				$-0.065$  &  $-0.697$  \\
				$\psi^{' R}_{h_{c}}(0)$ & $0.122(65)$ & $0.211$  &  $0.54$  &  $1.$  &  $-0.599$  &  $0.348$  &  $0.503$  & \
				$-0.206$  &  $0.55$  &  $0.017$  &  $-0.001$  &  $-0.029$  &  \
				$-0.052$  &  $-0.77$  \\
				$\psi^{R}_{J/\psi}(0)$ & $0.845(35)$ & $-0.134$  &  $-0.285$  &  $-0.599$  &  $1.$  &  $-0.004$  &  \
				$-0.376$  &  $0.043$  &  $-0.733$  &  $-0.225$  &  $0.001$  &  $0.$  & \
				$0.$  &  $0.778$  \\
				$\psi^{R}_{\eta_c}(0)$ & $1.034(26)$ & $0.$  &  $0.$  &  $0.348$  &  $-0.004$  &  $1.$  &  $0.01$  &  \
				$0.02$  &  $0.016$  &  $-0.587$  &  $0.$  &  $0.$  &  $0.$  &  \
				$-0.002$  \\
				$m_c$ & $1.404(75)$ & $0.354$  &  $0.794$  &  $0.503$  &  $-0.376$  &  $0.01$  &  $1.$  & \
				$-0.093$  &  $0.799$  &  $0.608$  &  $-0.002$  &  $0.$  &  $0.$  &  \
				$-0.65$  \\
				$\alpha_s(2 m_c)$ & $0.249(24)$ & $-0.201$  &  $-0.481$  &  $-0.206$  &  $0.043$  &  $0.02$  &  \
				$-0.093$  &  $1.$  &  $0.298$  &  $0.442$  &  $0.002$  &  $0.$  &  \
				$0.$  &  $0.277$  \\
				$\delta^\text{corr.}_{J/\psi}$ & $-0.002(53)$ &  $0.219$  &  $0.473$  &  $0.55$  &  $-0.733$  &  $0.016$  &  $0.799$ \
				&  $0.298$  &  $1.$  &  $0.673$  &  $-0.001$  &  $0.$  &  $0.$  &  \
				$-0.708$  \\
				$\delta^\text{corr.}_{\eta_c}$ & $-0.102(26)$ & $0.13$  &  $0.278$  &  $0.017$  &  $-0.225$  &  $-0.587$  &  \
				$0.608$  &  $0.442$  &  $0.673$  &  $1.$  &  $-0.001$  &  $0.$  &  \
				$0.$  &  $-0.286$  
				\\
				$\delta^\text{corr.}_{\chi_{c0}}$ & $0.0(20)$ & $-0.912$  &  $-0.003$  &  $-0.001$  &  $0.001$  &  $0.$  &  \
				$-0.002$  &  $0.002$  &  $-0.001$  &  $-0.001$  &  $1.$  &  $0.$  &  \
				$0.001$  &  $0.002$  
				\\
				$\delta_{C}$ & $0.266(27)$ & $-0.049$  &  $-0.036$  &  $-0.029$  &  $0.$  &  $0.$  &  $0.$  &  \
				$0.$  &  $0.$  &  $0.$  &  $0.$  &  $1.$  &  $0.$  &  $0.$ 
				\\
				$\delta_{NC}$ & $0.493(49)$ & $-0.089$  &  $-0.065$  &  $-0.052$  &  $0.$  &  $0.$  &  $0.$  &  \
				$0.$  &  $0.$  &  $0.$  &  $0.001$  &  $0.$  &  $1.$  &  $0.$  
				\\
				$\delta^\prime_{SP}$ & $-0.635(35)$ & $-0.267$  &  $-0.697$  &  $-0.77$  &  $0.778$  &  $-0.002$  &  \
				$-0.65$  &  $0.277$  &  $-0.708$  &  $-0.286$  &  $0.002$  &  $0.$  & \
				$0.$  &  $1.$ \\
				\hline
			\end{tabular}
		}
		\caption{Fit results for the derivatives of the wave functions for the P wave charmonia states $\chi_{c0}$, $\chi_{c1}$, and $h_c$ are determined through a simultaneous extraction of the S wave charmonium wave function at the origin from a fit to the data given in table \ref{tab:radexptinpt}.}
	\end{table}
\end{landscape}

In table~\ref{tab:BchcBCLfit} we presents the fit results for the BCL coefficients $a_i^n$ corresponding to the $B_c\to h_c$ transition form factors along with their correlations obtained from a fit to the synthetic data displayed in table~\ref{tab:Bc2hcc1syndat}. The goodness of fit is $99\%$. These extracted coefficients are then used in describing the shape of the transition form factors across the relevant kinematic range as shown in figs.~\ref{fig:ffBchc}.

\begin{table}[ht]
	\centering
	\begin{tabular}{|c|c|*{7}{c}|}
		\hline  
		Parameters & Fit Results & $a_0^{A^{h_c}_0}$ & $a_1^{A^{h_c}_0}$&$a_0^{A^{h_c}_1}$&$a_1^{A^{h_c}_1}$&$a_0^{A^{h_c}_2}$&$a_0^{V^{h_c}}$&$a_1^{V^{h_c}}$\\ 
		\hline\hline
		$a_0^{A^{h_c}_0}$& $0.21(8)$ & $1.0$  &  $-0.991$  &  $0.995$  &  $-0.984$  &  $-0.996$  &  $-0.028$ \
        &  $0.022$  \\
        $a_1^{A^{h_c}_0}$& $-1.9(7)$ & $.$  &  $1.0$  &  $-0.988$  &  $0 .985$  &  $0.981$  &  $0 .053$  \
        &$-0.041$  \\
        $a_0^{A^{h_c}_1}$& $0.011(7)$ & $.$  &  $.$  &  $1.0$  &  $-0.992$  &  $-0.991$  &  $-0.049$  \
        & $0.038$  \\
        $a_1^{A^{h_c}_1}$& $-0.06(4)$ &$.$  &  $.$  &  $.$  &  $1.0$  &  $0.980$  &  $0.067$  &  $-0.051$  \\
        $a_0^{A^{h_c}_2}$& $-0.3(1)$ &$.$  &  $.$  &  $.$  &  $.$  &  $1.0$  &  $0.021$  &  $-0.016$  \\
        $a_0^{V^{h_c}}$& $0.02(1)$ &$.$  &  $.$  &  $.$  &  $.$  &  $.$  &  $1.0$  &  $-0.995$  \\
        $a_1^{V^{h_c}}$& $-0.16(10)$ &$.$  &  $.$  &  $.$  &  $.$  &  $.$  &  $.$  &  $1.0$  \\
		\hline
	\end{tabular}
	\caption{Fit results for the BCL coefficients $a_i^n$ corresponding to the $B_c\to h_c$ transition form factors (with n=1) for  $A_0$, $A_1$, $A_2$, $V$ and their correlations. The p-Value is $99\%$.}
	\label{tab:BchcBCLfit}
\end{table}

Table~\ref{tab:Bcchic1BCLfit} shows the fitted values of the BCL expansion coefficients $a_i^n$ for the $B_c\to \chi_{c1}$ transition form factors. The coefficients were obtained through a $\chi^2$ analysis to the synthetic data provided in table~\ref{tab:Bc2hcc1syndat}. Alongside the central values and their uncertainties, the table~\ref{tab:Bcchic1BCLfit} includes the correlations matrix, provides a quantitative measure of the robustness of the fit. The fit quality, with a p-value of $99\%$, indicates good agreement between the synthetic data and the BCL parametrization. These extracted coefficients are then used in describing the shape of the transition form factors across the relevant kinematic range as shown in figs.~\ref{fig:ffBcchic1}.

\begin{table}[ht]
	\centering
	\begin{tabular}{|c|c|*{7}{c}|}
		\hline  
		Parameters & Fit Results & $a_0^{A^{\chi_{c1}}_0}$&$a_1^{A^{\chi_{c1}}_0}$&$a_0^{A^{\chi_{c1}}_1}$&$a_1^{A^{\chi_{c1}}_1}$&$a_0^{A^{\chi_{c1}}_2}$&$a_0^{V^{\chi_{c1}}}$&$a_1^{V^{\chi_{c1}}}$\\  
		\hline\hline
		$a_0^{A^{\chi_{c1}}_0}$ & $0.014(7)$  & $1.0$  &  $-0.995$  &  $0 .996$  &  $-0 .945$  &  $0.992$  &  $0.119$ &  $-0.098$  \\
        $a_1^{A^{\chi_{c1}}_0}$ & $-0.12(6)$ & $.$  &  $1.0$  &  $-0.992$  &  $0 .954$  &  $-0.987$  &  $-0.103$ &  $0.086$  \\
        $a_0^{A^{\chi_{c1}}_1}$& $0.05(2)$  & $.$  &  $.$  &  $1.0$  &  $-0.959$  &  $0.997$  &  $0.097$  \
        &  $-0.080$  \\
        $a_1^{A^{\chi_{c1}}_1}$ & $-0.31(9)$ & $.$  &  $.$  &  $.$  &  $1.0$  &  $-0.953$  &  $-0.057$  \
        &  $0.05$  \\
        $a_0^{A^{\chi_{c1}}_2}$& $0.18(6)$  & $.$  &  $.$  &  $.$  &  $.$  &  $1.0$  &  $0.091$  &  $-0.076$\
        \\
        $a_0^{V^{\chi_{c1}}}$ & $0.3(2)$ & $.$  &  $.$  &  $.$  &  $.$  &  $.$  &  $1.0$  &  $-0.993$  \\
        $a_0^{V^{\chi_{c1}}}$ & $-2.0(1)$ &$.$  &  $.$  &  $.$  &  $.$  &  $.$  &  $.$  &  $1.0$  \\
		\hline
	\end{tabular}
	\caption{Fit results for the BCL coefficients $a_i^n$ corresponding to the $B_c\to \chi_{c1}$ transition form factors (with n=1) for  $A_0$, $A_1$, $A_2$, $V$ and their correlations. The p-Value is $99\%$.}
	\label{tab:Bcchic1BCLfit}
\end{table}

Table~\ref{tab:Bcchic0RCBCLfit} shows the fitted values of the BCL expansion coefficients $a_i^n$ for the $B_c\to \chi_{c0}$ transition form factors. The coefficients were obtained through a $\chi^2$ analysis to the synthetic data provided in table~\ref{tab:Bc2chic0syndat}. Alongside the central values and their uncertainties, the table~\ref{tab:Bcchic0RCBCLfit} includes the correlations matrix, provides a quantitative measure of the robustness of the fit. The fit quality, with a p-value of $78\%$, indicates good agreement between the synthetic data and the BCL parametrization. These extracted coefficients are then used in describing the shape of the transition form factors across the relevant kinematic range as shown in figs.~\ref{fig:ffchic0}.
\begin{table}[ht]
	\centering
	\begin{tabular}{|c|c|*{3}{c}|}
		\hline 
	   Parameters & Fit Results &	$a_0^{f_0^{\chi_{c0}}}$          &$a_1^{f_0^{\chi_{c0}}}$&$a_1^{f_+^{\chi_{c1}}}$\\
		\hline\hline
		$a_0^{f_0^{\chi_{c0}}}$& $1.44(51)$ & $1.0$  &  $-0.981$  &  $1.0$  \\
        $a_1^{f_0^{\chi_{c0}}}$&$-15.4(44)$ & $.$  &  $1.0$  &  $-0.979$  \\
        $a_1^{f_+^{\chi_{c1}}}$& $1.53(54)$ & $.$  &  $.$  &  $1.0$  \\
		\hline
	\end{tabular}
	\caption{Fit results for the BCL coefficients $a_i^n$ corresponding to the $B_c\to \chi_{c0}$ transition form factors (with n=1) for  $f_+$, $f_0$(LO+NLO+RC) and their correlations. The p-Value is $78\%$.}
	\label{tab:Bcchic0RCBCLfit}
\end{table}

In this work in section~\ref{sec:BctoP_nonlep}, we estimate the branching fractions for the non-leptonic transition $B_c \to h_c, \chi_{c0,1} M$. Within the nive QCD factorization framework, the general expression for the decay amplitude of $B_c \to M_1(c\bar{c}) M_2$ is given in eq.~\ref{eq:QCDF}, where the relevant form factors $F^{B_c \to M_1}(m_{M_2}^2)$ describe the $B_c \to M_1(c \bar{c}) M_2$ transition. These form factors are evaluated at $q^2=m^2_{M_2}$ and are summarized in the table~\ref{tab:formfactorinp_nonlep}.
\begin{table}[ht]
	\centering
	\begin{tabular}{|*{4}{c|}}
		\hline
		\textbf{Form Factor}&\textbf{Estimates(\text{$GeV^2$})} &\textbf{Form factors}&\textbf{Estimates ($GeV^2$)} \\
		\hline \hline
		$\text{$A_0^{\chi_{c1}}(m^2_{\pi })$}$  &  $\text{0.086(47)}$ & $\text{$A_0^{\chi_{c1}}(m^2_{\rho})$}$  &  $\text{0.089(49)}$ \\
		$\text{$A_0^{\chi_{c1}}(m^2_{K})$}$  &  $\text{0.087(48)}$ & $\text{$A_0^{\chi_{c1}}(m^2_{K^*})$}$  &  $\text{0.090(50)}$ \\
		$\text{\text{$A_0^{h_c}(m^2_{\pi })$}}$  &  $\text{1.35(95)}$ & $\text{$A_0^{h_c}(m^2_{\rho})$}$  &  $\text{1.41(99)}$ \\
		$\text{\text{$A_0^{h_c}(m^2_{K})$}}$  &  $\text{1.37(97)}$ & $\text{$A_0^{h_c}(m^2_{K^*})$}$  &  $\text{1.43(100)}$ \\
		\hline
	\end{tabular}
	\caption{Estimates for the $A_0^{\chi_{c1}}$, $A_0^{h_c}$ form factors at $q^2$= $m^2_{K}$, $m^2_{\pi}$, $m^2_{K^*}$, $m^2_{\rho}$ using fit results from tables~\ref{tab:poleparamfit}.}
	\label{tab:formfactorinp_nonlep}
\end{table}
\bibliographystyle{JHEP}
\bibliography{BcPanylz.bib} 
\end{document}